\documentclass[12pt]{article}
\pdfoutput=1
\usepackage{amssymb,amsmath,dsfont,verbatim,commath}
\usepackage{pdfsync}
\usepackage{colonequals}
\usepackage[normalem]{ulem}
\usepackage{slashed}
\usepackage{tikz}
\usepackage{epsfig}
\usepackage{epstopdf}
\usepackage{latexsym}
\usepackage{graphicx}
\usepackage{booktabs}
\usepackage{bbm}
\usepackage{xspace}
\usepackage{cancel}
\usepackage[utf8]{inputenc}
\usepackage[babel]{csquotes}
\usepackage{enumitem}
\usepackage[numbers,compress]{natbib}
\usepackage[T1]{fontenc}
\usepackage[margin=20pt,small]{caption}
\usepackage{subfig}
\usepackage[toc]{appendix}
\usepackage{color}
\usepackage{float}
\usepackage{multirow}
\usepackage{multicol}
\usepackage{bigstrut}
\usepackage{datetime}
\usepackage{tikz-cd}
\usetikzlibrary{positioning,arrows,calc}
\usetikzlibrary{arrows.meta}
\usetikzlibrary{decorations.pathmorphing}
\usetikzlibrary{decorations.markings}
\usetikzlibrary{shapes.geometric}
\usetikzlibrary{patterns}
\usepackage[
colorlinks=false,
linkcolor=darkblue,  
urlcolor=blue,    
filecolor=blue,
citecolor=red,
linktocpage=true,
pdfstartview=FitV,
bookmarksopen=true,
hidelinks
]{hyperref}
\numberwithin{equation}{section}

\newcommand*{\boxedcolor}{red}
\makeatletter
\renewcommand{\boxed}[1]{\textcolor{\boxedcolor}{%
		\fbox{\normalcolor\m@th$\displaystyle#1$}}}
\makeatother

\newcommand{\D}{\Delta}
\newcommand{\hD}{\hat\D}
\newcommand{\hgamma}{\hat\gamma}
\newcommand{\hchi}{\hat\chi}
\newcommand{\heta}{\hat\eta}
\newcommand{\hr}{{\hat r}}
\newcommand{\hphi}{\hat\phi}

\newcommand{\id}{{\mathds 1}}
\newcommand{\ps}{\mathfrak{s}}
\newcommand{\pf}{\mathfrak{p}}

\newcommand{\aphisq}{a_{\phi^2}}

\newcommand{\hf}{\lambda}

\newcommand{\eps}{\varepsilon}

\newcommand{\xper}{y}
\newcommand{\xpar}{\tau}
\newcommand{\W}{{\mathcal W}}

\definecolor{cardinal}{rgb}{0.6,0,0}
\definecolor{darkgreen}{rgb}{0,0.5,0}
\definecolor{golden}{rgb}{0.92, 0.7, 0}
\definecolor{midnight}{rgb}{0, 0, 0.5}
\definecolor{darkblue}{rgb}{0.2, 0, 0.8}

\tikzset{
  vtx/.style={
    circle,
    draw=blue,
    fill=blue,
    inner sep=1pt
  },
  wcirc/.style={
    circle,
    draw=black,
    fill=white,
    inner sep=2pt
  },
  bcirc/.style={
    circle,
    draw=black,
    fill=black,
    inner sep=1pt
  },
  dcirc/.style={
    circle,
    draw=blue,
    fill=blue,
    inner sep=1pt
  },
  rcirc/.style={
    circle,
    draw=red,
    fill=red,
    inner sep=1pt
  },
  phi/.style={
    thick
  },
  sigma/.style={
    thick,
    dashed
  },
  vl1/.style={
    thick,
    blue
  },
  vl2/.style={
    thick,
    dashed,
    blue
  },
  valign/.style={
    baseline={([yshift=-.55ex]current bounding box.center)}
  }
}

\def\sl{\frak{sl}}

\begin{document}

	\begin{titlepage}
		DESY-23-175
		\bigskip
		\begin{center} 
			{\Large \bf Analytic and numerical bootstrap for the \\ long-range Ising model}

			\bigskip
			\bigskip
			\bigskip
			
			{\bf  Connor Behan$^{1,2}$, Edoardo Lauria$^{3}$, Maria Nocchi$^{1}$, Philine van Vliet$^{4,5}$\\ }
			\bigskip
			\bigskip
			${}^{1}$
			Mathematical Institute, University of Oxford, Andrew Wiles Building, Radcliffe Observatory Quarter, Woodstock Road, Oxford, OX2 6GG, UK\\
			\vskip 3mm
			${}^{2}$
			Instituto de F\'{i}sica Te\'{o}rica, UNESP, ICTP South American Institute for Fundamental Research, Rua Dr Bento Teobaldo Ferraz 271, 01140-070, S\~{a}o Paulo, Brazil
			\vskip 3mm
			${}^{3}$
			LPENS, Département de physique, École Normale Supérieure - PSL\\
			Centre Automatique et Systèmes (CAS), Mines Paris - PSL\\
			Université PSL, Sorbonne Université, CNRS, Inria, 75005 Paris\\
			\vskip 3mm
			${}^{4}$
			Deutsches Elektronen-Synchrotron DESY, Notkestr. 85, 22607 Hamburg, Germany
			\vskip 3mm
			${}^{5}$
			Laboratoire de Physique Théorique de l'École Normale Supérieure, PSL University,\\
			CNRS, Sorbonne Universités, UPMC Univ. Paris 06\\
			24 rue Lhomond, 75231 Paris Cedex 05, France
			\vskip 5mm				
			\texttt{connorbehan@ictp-saifr.org,~edoardo.lauria@minesparis.psl.eu,\\nocchi@maths.ox.ac.uk,~philine.vanvliet@phys.ens.fr} \\
		\end{center}
		
		\bigskip
		\bigskip
		
		\begin{abstract}
			\noindent
			We combine perturbation theory with analytic and numerical bootstrap techniques to study the critical point of the long-range Ising (LRI) model in two and three dimensions. This model interpolates between short-range Ising (SRI) and mean-field behaviour. We use the Lorentzian inversion formula to compute infinitely many three-loop corrections in the two-dimensional LRI near the mean-field end. We further exploit the exact OPE relations that follow from bulk locality of the LRI to compute infinitely many two-loop corrections near the mean-field end, as well as some one-loop corrections near SRI. By including such exact OPE relations in the crossing equations for LRI we set up a very constrained bootstrap problem, which we solve numerically using SDPB. We find a family of sharp kinks for two- and three-dimensional theories which compare favourably to perturbative predictions, as well as some Monte Carlo simulations for the two-dimensional LRI. 
			
		\end{abstract}

	\end{titlepage}

	\setcounter{tocdepth}{2}	
	\tableofcontents

\section{Introduction}
\label{sec:intro}

The importance of studying Conformal Field Theories (CFTs) without a stress tensor can be motivated from various angles. The past decade has seen a surge of interest in Quantum Field Theories (QFTs) that live on a fixed Anti-de Sitter (AdS) background \cite{QFTinAdS,Komatsu:2020sag,Antunes:2021abs,vanRees:2022zmr,Ankur:2023lum,Levine:2023ywq,Meineri:2023mps,Lauria:2023uca}. Due to the lack of dynamical gravity, the CFT which encodes boundary correlation functions has no stress tensor and is therefore called nonlocal.
A closely related development is defect CFT \cite{Liendo:2012hy,Billo:2016cpy,Lemos:2017vnx,Lauria:2018klo,Liendo:2019jpu,Cuomo:2021rkm} dealing with critical systems which have part of their conformal symmetry broken by extended objects. While locality of the bulk places some constraints on the admissible defects, correlators restricted to them will always obey the axioms of a nonlocal CFT. Since the mass of a field in AdS and the co-dimension of a defect can both be varied, it is easy for nonlocal theories to appear in continuous families.

Another motivation comes from the bootstrap philosophy \cite{Rattazzi:2008pe} which naturally leads to some traditional assumptions about QFT being relaxed.
One can notice that the works \cite{ElShowk:2012,ElShowk:2014}, in their efforts to study a local CFT, happened to produce several bounds which were saturated by nonlocal CFTs. In fact, \cite{Dymarsky:2018} is the only non-supersymmetric numerical bootstrap study to date which has treated locality as an input instead of an output. Hence, a complete picture of the solutions found by most numerical bootstrap studies requires an understanding of nonlocal CFTs.

The goal of this paper is to bootstrap the critical long-range Ising (LRI) model in $p$ dimensions, one of the earliest known examples of a nonlocal CFT and likely one of the simplest. By analogy with its short-range version (SRI) which comes from a classical lattice Hamiltonian with only nearest-neighbour interactions, the critical $p$-dimensional LRI model is defined as the critical point of
\begin{equation}
	H = -J \sum_{i,j} \frac{\sigma_i \sigma_j}{|i - j|^{p + \mathfrak{s}}}~, \quad J > 0~,\quad \sigma_i =\pm 1~, \label{lri-hamiltonian}
\end{equation}
which has interactions over an infinite distance \cite{d69}. Initially, it was thought that the interval which led to non-trivial critical behaviour was $\frac{p}{2} \leq \mathfrak{s} \leq 2$ \cite{Fisher:1972zz} but it is actually slightly narrower, as we will see. The presence of long-range interactions parameterized by $\mathfrak{s}$ in the Hamiltonian above means that the critical point of LRI should correspond to a family of nonlocal (and unitary as proven in \cite{af88}) conformal field theories.

Following the birth of the perturbative renormalization group \cite{wf72}, Fisher, Ma and Nickel introduced in \cite{Fisher:1972zz} a continuum description of the LRI based on a quartic deformation of a generalized free field $\hat{\phi}$ with $\D_\phi = \frac{p - \mathfrak{s}}{2}$ and $\mathfrak{s}$ the parameter from \eqref{lri-hamiltonian}. A free field $\phi$ having this scaling dimension must live in $d = p + 2 - \mathfrak{s}$ dimensions and it is a simple exercise to show that we can go from one to the other by considering a trivial defect of co-dimension $q = 2 - \mathfrak{s}$. Writing bulk coordinates as $x^\mu = (\xpar^a, \xper^i)$ with $\mu = 1, \dots d$, $a = 1, \dots p$ and $i = 1, \dots, q$, we can define a family of defect operators by
\begin{equation}
	\widehat{\phi^m}(\tau) = \frac{1}{\sqrt{m!}} :\phi(\tau, 0)^m:~, \label{power-phi}
\end{equation}
where $: \dots :$ denotes normal ordering. In this notation, the nonlocal action\footnote{The overall normalization has been chosen such that $\hat{\phi}$ and all of its powers \eqref{power-phi} in the undeformed theory are unit-normalized.}
\begin{equation}
	S = \mathcal{N}_{\mathfrak{s}} \mathcal{N}_{-\mathfrak{s}} \int d^p\tau_1 d^p\tau_2 \frac{\hat{\phi}(\tau_1) \hat{\phi}(\tau_2)}{|\tau_{12}|^{p + \mathfrak{s}}} + \int d^p\tau \frac{\lambda}{\sqrt{4!}} \widehat{\phi^4}~, \label{flow1}
\end{equation}
provides a description of the LRI based on a single mean field, henceforth referred to as the \emph{mean-field description}.
The interaction in \eqref{flow1} becomes irrelevant below $\mathfrak{s} = \frac{p}{2}$, and CFT data at the fixed point can be expanded in powers of $\varepsilon = 2\mathfrak{s} - p$. Normally $0 < \varepsilon \ll 1$ is needed for reliable results but the expansion can truncate in special cases. In particular, $\hat{\phi}$ and $\widehat{\phi^3}$ are protected operators with exact dimensions
\begin{equation}
	\Delta_{\phi} = \frac{p - \mathfrak{s}}{2}, \quad \Delta_{\phi^3} = \frac{p + \mathfrak{s}}{2}~. \label{non-renorm1}
\end{equation}
For $\hat{\phi}$, this is because local interactions cannot renormalize a nonlocal kinetic term, a rigorous proof of which was given in \cite{lsw17}. For $\widehat{\phi^3}$, \eqref{non-renorm1} can be seen by applying a nonlocal equation of motion \cite{Paulos:2015jfa}. As we will review shortly, both non-renormalization theorems become automatic when we view the LRI as a defect in co-dimension $2 - \mathfrak{s}$.

If \eqref{non-renorm1} held for all $\frac{p}{2} \leq \mathfrak{s} \leq 2$, as originally proposed by \cite{Fisher:1972zz}, the connection between the LRI and SRI models would not be especially strong. Instead, Sak analyzed the weakly irrelevant operator $\hat{\phi} \partial^2 \hat{\phi}$ and concluded that $\D_\phi$ stops changing once it reaches the dimension of the short-range spin field $\D^*_\sigma$ \cite{s73}. Understanding this crossover between two universality classes was a puzzle for many years because the SRI (unlike mean-field theory with $\D_\phi = \frac{p}{4}$) does not contain a marginal local operator. The resolution, found in \cite{Behan:2017dwr,Behan:2017emf}, is that the requisite deformation of the SRI by a nonlocal operator is equivalent to the action
\begin{equation}
	S = S_{\text{SRI}} + \mathcal{N}_{\mathfrak{s}} \mathcal{N}_{-\mathfrak{s}} \int d^p\tau_1 d^p\tau_2 \frac{\hat{\chi}(\tau_1) \hat{\chi}(\tau_2)}{|\tau_{12}|^{p - \mathfrak{s}}} + \int d^p\tau g \sigma \hat{\chi}~, \label{flow2}
\end{equation}
henceforth called the \textit{short-range description}.
Here, $\hat{\chi}$ is a generalized free field of dimension $\D_\chi = \frac{p + \mathfrak{s}}{2}$. This guarantees that $\sigma \hat{\chi}$ becomes irrelevant above $\mathfrak{s} = p - 2\D_\sigma^*$. We will again think of it as a defect mode of a free field $\chi$ in dimension $d = p + 2 + \mathfrak{s}$ and define
\begin{equation}
	\widehat{\chi^m}(\tau) = \frac{1}{\sqrt{m!}} :\chi(\tau, 0)^m:~. \label{power-chi}
\end{equation}
If one considers the $S_{\text{SRI}}$ theory to be known, \eqref{flow2} makes it possible to compute CFT data as an expansion in $\delta = \frac{p - \mathfrak{s}}{2} - \D_\sigma^*$ for $0 < \delta \ll 1$. By the same arguments as before,
\begin{equation}
	\Delta_{\chi} = \frac{p + \mathfrak{s}}{2}~, \quad \Delta_{\sigma} = \frac{p - \mathfrak{s}}{2} \label{non-renorm2}~,
\end{equation}
are non-perturbative statements.

The dual descriptions \eqref{flow1} and \eqref{flow2} can help give a flavour for how observables behave as exact functions of $\mathfrak{s}$. In our study of the critical LRI model, the first half will ensure that more data is available for this purpose. The second half will carry out the numerical bootstrap and show that the LRI at arbitrary values of $\mathfrak{s}$ can be located to high precision. Both parts will make essential use of exact relations between OPE coefficients which involve the basic protected operators. If $\mathcal{O}_1$ is a scalar and $\mathcal{O}_2$ is a spin-$\ell$ tensor of $SO(p)$, they take the form
\begin{align}
	\frac{\lambda_{12\phi}}{\lambda_{12\phi^3}} \propto \frac{\Gamma(\tfrac{\D_\phi + \D_{12} + \ell}{2}) \Gamma(\tfrac{\D_\phi - \D_{12} + \ell}{2})}{\Gamma(\tfrac{\D_{\phi^3} + \D_{12} + \ell}{2}) \Gamma(\tfrac{\D_{\phi^3} - \D_{12} + \ell}{2})}~, \quad \frac{\lambda_{12\sigma}}{\lambda_{12\chi}} \propto \frac{\Gamma(\tfrac{\D_\sigma + \D_{12} + \ell}{2}) \Gamma(\tfrac{\D_\sigma - \D_{12} + \ell}{2})}{\Gamma(\tfrac{\D_\chi + \D_{12} + \ell}{2}) \Gamma(\tfrac{\D_\chi - \D_{12} + \ell}{2})}~, \label{schematic-ope}
\end{align}
where the constant of proportionality is independent of $\mathcal{O}_1$ and $\mathcal{O}_2$.
The spin-0 and general spin versions of \eqref{schematic-ope} were first derived in \cite{Paulos:2015jfa} and \cite{Behan:2018hfx} respectively. Both of these derivations used the nonlocal equation of motion. It was later realized in \cite{Lauria:2020emq} that, as a consequence of bulk locality, these relations hold in any defect CFT where the bulk is free, LRI being a particular example as explained in \cite{Paulos:2015jfa}. Recently, \cite{Levine:2023ywq} showed that they are important for enforcing locality for more general QFTs in AdS. In a perturbative context, we will use the relations to gain an order of perturbation theory which appears to be a new application of them. On the numerical side, we will use the fact that crossing symmetry and unitarity become more powerful when combined with \eqref{schematic-ope}. For the case of boundaries, which are a type of $q = 1$ defect, this was already demonstrated in \cite{Behan:2020nsf,Behan:2021tcn}. The novelty here is that the LRI model requires us to consider many values of $\mathfrak{s}$ and therefore many co-dimensions (which are all fractional) since $q = 2 \pm \mathfrak{s}$ in the two formulations above.

\begin{table}
	\centering
	\begin{tabular}{|l|l|l|}
		\hline
		Observable & Loops & Ref \\
		\hline
		$\D_\phi$ & $\infty$ & \cite{Fisher:1972zz} \\
		$\D_{\phi^2}$ & 3 & \cite{Benedetti:2020rrq} \\
		$\D_{\phi^3}$ & $\infty$ & \cite{Paulos:2015jfa} \\
		$\D_{\phi^4}$ & 3 & \cite{Benedetti:2020rrq} \\
		$\D_{\phi^m}$, $m > 4$ & 2 & \eqref{phi-power-2loop} \\
		$\D_{[\hat{\phi} \hat{\phi}]_{n,\ell}}$, $\ell > 0$ & 2 & \eqref{andt} \\
		$\lambda_{\hat{\phi} \widehat{\phi^m} \widehat{\phi^{m - 1}}}$ & 1 & \eqref{phi-powers-1loop} \\
		$\lambda_{\widehat{\phi^3} \widehat{\phi^m} \widehat{\phi^{m - 1}}}$ & 1 & \eqref{phi-powers-1loop} \\
		\hline
	\end{tabular}
	\begin{tabular}{|l|l|l|}
		\hline
		Observable & Loops & Ref \\
		\hline
		$\D_{\sigma}$ & $\infty$ & \cite{Behan:2017emf} \\
		$\D_{\epsilon}$ & 2 & \cite{Behan:2017emf} \\
		$\D_{\chi}$ & $\infty$ & \cite{Behan:2017emf} \\
		$\D_{\sigma\chi}$ & 2 & \cite{Behan:2017emf} \\
		$\D_T$ & 2 & \cite{Behan:2017emf} \\
		$\lambda_{\sigma \hat{\chi} \mathcal{O}}$, $\mathcal{O} \in \text{SRI}$ & 1 & \eqref{spinning-second} \\
		$\lambda_{\hat{\chi} \hat{\chi} \mathcal{O}}$, $\mathcal{O} \in \text{SRI}$ & 2 & \eqref{spinning-second} \\
		$\lambda_{\hat{\chi}\hat{\chi}[\hat{\chi} \hat{\chi}]_{n,\ell}}$ & 2 & \eqref{chi-corr} \\
		\hline
	\end{tabular}
	\caption{Scaling dimensions and OPE coefficients in the LRI model which are known to some non-trivial order in both $p = 2$ and $p = 3$. Data in the left hand table, computed from \eqref{flow1} with the $\varepsilon$ expansion, have closed form expressions. Data in the right hand table, computed from \eqref{flow2} with the $\delta$ expansion, are sometimes numerical. This difficulty comes from the integrals generated by conformal perturbation theory and the lack of an exact solution for the 3d SRI model.}
	\label{tab:loopcount}
\end{table}
As a preview of our results, Table \ref{tab:loopcount} compares the CFT data available in the literature to the new batch we are able to compute in two and three dimensions. Along with \eqref{power-phi} and \eqref{power-chi}, it uses the double-twist notation
\begin{equation}
	[\Phi \Phi]_{n,\ell} \sim \Phi \partial^{2n + \ell} \Phi~, \label{double-twist}
\end{equation}
to mean the unique normalized primary with $2n$ contracted and $\ell$ uncontracted derivatives appearing in the self-OPE of the generalized free field $\Phi$. In addition to the general dimension results of Table \ref{tab:loopcount}, we also compute anomalous dimensions of $[\hat{\phi} \hat{\phi}]_{n,\ell}$ operators to three loops in $p = 2$ using analytic bootstrap methods \cite{Fitzpatrick:2012yx,Komargodski:2012,Alday:2016,Simmonsduffin:2017,Caron-Huot:2017vep}. This is most successful for the double-twist operators with $n > 0$ whose three-loop anomalous dimensions have compact expressions. In the local Wilson-Fisher fixed points, which \cite{Alday:2017zzv} studied in a similar manner, these operators are absent due to the equation of motion. Extracting CFT data for such infinite families is possible thanks to the Lorentzian inversion formula \cite{Caron-Huot:2017vep} which is easiest to apply when the conformal blocks are known in closed form.

The rest of this paper is organized as follows. In section \ref{sec:exactres} we build upon the construction of the critical LRI as a conformal defect to gain some exact information on the spectrum of LRI at criticality. Most notably, we solve the bulk-defect bootstrap equation for the bulk two-point function and derive the relations above. These results are non-perturbative, and in particular, they are valid regardless of the UV description of LRI.
Sections \ref{sec:perturb} and \ref{sec:inversion} are devoted to perturbation theory. In section \ref{sec:perturb} we review some results from the literature about the flows in equations \eqref{flow1} and \eqref{flow2}, as well as derive the new results for the CFT data in Table \ref{tab:loopcount}. This is done by combining standard Feynman diagrams with the exact OPE relations, the latter allowing us to gain one order in perturbation theory, as we will explain. We then move onto the inversion formula in section \ref{sec:inversion} which computes an infinite amount of additional CFT data in two dimensions. 
In section \ref{sec:numerics} we turn to the numerical bootstrap machinery, where we use the exact OPE relations as an input for the LRI crossing equations and compare favourably to the inversion and diagrammatic calculations. A summary and suggestions for the future are given in section \ref{sec:outlook}.

\section{Long-range Ising model as a defect CFT}
\label{sec:exactres}

The critical point of the $p$-dimensional long-range Ising model can be realized as a co-dimension $q=d-p$ conformal defect in a theory of a $d$-dimensional free massless scalar field $\Phi$, see~\cite{Paulos:2015jfa}. As discussed above, $q$ can be either $2 - \mathfrak{s}$ or $2 + \mathfrak{s}$. The fact that the \emph{bulk} theory is free allows us to derive infinitely many constraints and relations between defect conformal data. This will be the goal of this section. While the LRI spectrum is built out of operators constrained to the $p$-dimensional defect, one has to be mindful of differences between the original LRI and the defect description. The conformal defect description enjoys a manifest $SO(p+1,1)\times SO(q)$ symmetry, while in the original LRI there is no $SO(q)$ symmetry. Hence, physical states of the LRI should have zero transverse spin. In this section, we discuss the bulk two-point function of $\Phi$ and the exact OPE relations derived from bulk locality. We will then interpret results in the context of the LRI where $\Phi$ becomes $\phi$ in one description and $\chi$ in the other.

\subsection{Bulk two-point function of $\Phi$}
We start with reviewing the admissible spectrum of defect modes of $\Phi$, and then solve the bulk-defect crossing symmetry constraint on this two-point function analytically to compute the relevant defect CFT data. We take both bulk and defect operators to be unit-normalized. 

\subsubsection{Defect modes of $\Phi$} 

Consider the bulk-defect OPE of a massless scalar field $\Phi$ of dimension $\D_\Phi$, non-necessarily free. The symmetry allows us to write, schematically:
\begin{align}\label{bOPE}
	\Phi(x)= \sum_{s}\sum_{\psi} b_s^{\Phi,\pf} 	|\xper|^{\D-\D_\Phi}\psi_{s}(\xpar)+\dots~,
\end{align}
where the ellipsis denotes derivatives with respect to the directions parallel to the defect and we employed the notation of the introduction.
In the expression above, the $\psi_s$ are defect local primary operators of scaling dimension $\D$ and transverse spin $s$ (which we will take to be integers) under $SO(q)$ rotations.\footnote{To avoid cluttering, we omitted transverse spin indices. The complete expression for the bulk-defect OPE of a massless scalar can be found e.g. in \cite{Billo:2016cpy}.} When $\Phi$ is a free field, $\square \Phi = 0$ (away from contact points) fixes the scaling dimensions of the defect modes of $\Phi$ to be \cite{Billo:2016cpy}
\begin{align}\label{defSpec}
	\psi_{s}^{(+)}:&\quad \D_s^{(+)}=\Delta_{\Phi}+s~,\nonumber\\
	\psi^{(-)}_{s}:&\quad \D_{s}^{(-)}=\Delta_{\Phi}+2 -q-s~,
\end{align}
where $\D_\Phi = d/2-1$. Notice that $\psi_{s}^{(+)}$ and $\psi_{s}^{(-)}$ form a shadow pair on the defect for all $s$.  We will soon identify these defect modes as either $\psi_0^{(+)}=\hphi$ and $\psi_0^{(-)}=\hphi^3$, or $\psi_0^{(+)}=\hat{\chi}$ and $\psi_0^{(-)}=\sigma$.

When $p$ and $q$ are both integers, the restrictions from unitarity on the values of $s$ for `$-$' modes have been classified in \cite{Lauria:2020emq}. There, special attention is paid to the cases of $p = 1$ and $p + q = 2$. As explained in the introduction, however, we will take $p = 2,3$ and $q$ fractional. In this case we have $s \leq \frac{4 - q}{2}$. If this condition were not met, the second line of \eqref{defSpec} would violate the unitarity bound -- in conflict with the fact that the $p$-dimensional LRI model is reflection positive \cite{af88}. It is worth stressing however that we cannot assume unitarity for the larger theory which lives in the $d$-dimensional bulk.\footnote{Many theories in fractional dimension, including free CFTs, fail to be unitary by the arguments in \cite{Hogervorst:2015akt}. \label{footnotess}}

\subsubsection{Bootstrapping the bulk two-point function for LRI}
With the bulk-defect spectrum of $\Phi$ at hand, our next task will be to construct the full bulk two-point function of $\Phi$. In the defect channel, by $SO(p+1,1)\times SO(q)$ symmetry we have \cite{Billo:2016cpy}:
\begin{align}\label{two_points_def_free}
	\langle \Phi(x_1)\Phi(x_2)\rangle =\frac{1}{(|\xper_1| |\xper_2|)^{\Delta_\Phi}}\sum_{s}\sum_{\pf=\pm}(b_s^{\Phi,\pf})^2 {\hat g}^{(\pf)}_{s}(\hr,\heta)~,
\end{align}
where $b_s^{\Phi,\pm}$ are yet unknown, and the sum is over all the defect modes \eqref{defSpec}. The functions ${\hat g}^{(\pf)}_{s}(\hr,\heta)\equiv \hat{g}_{\D_s^{(\pf)},s}(\hr,\heta)$ are the \emph{defect channel} conformal blocks, which read~\cite{Billo:2016cpy,Lauria:2017wav}
\begin{align}\label{two_points_def}
	\hat{g}_{\D,s}(\hr,\heta)=\hr^{\D} \ _2F_1\left(\frac{p}{2},\D, \D,-\frac{p}{2}+1; \hr^2\right) 
	\frac{s!}{2^{s}\left(\frac{q}{2}-1\right)_s}  
	C_s^{\frac{q}{2}-1}(\heta)~,
\end{align}
with $\hr,\heta,\hchi$ cross-ratios defined as
\begin{align}\label{hetadef}
	\heta=\cos\theta~, \quad \hr= \frac{1}{2} \left(\hchi-\sqrt{\hchi^2-4}\right)~, \quad \hchi = \frac{|\xpar_{12}|^2+|\xper_1|^2+|\xper_2|^2}{|\xper_1| |\xper_2|}~,
\end{align}
and $\theta$ is the transverse angle. In the Euclidean domain $(\hr,\heta) \in [0,1]\times [-1,1]$, while $(\hchi,\cos\theta)\in [2,\infty]\times [-1,1]$.

Let us now take a closer look at the dynamics of the system. In perturbation theory, we can think of the two-point function of $\Phi$ for LRI as a (finite) deformation away from the trivial defect. The two-point function with the trivial defect contains only `$+$' modes, and reads \cite{Gaiotto:2013nva}
\begin{align}\label{two_points_def_freeLtrivial}
	\langle \Phi(x_1)\Phi(x_2)\rangle_{\text{triv}} &=\frac{1}{(x_{12}^2)^{\Delta_\Phi}}
	=\frac{1}{(|\xper_1| |\xper_2|)^{\Delta_\Phi}}\sum_{s\geq 0}(b_{s,\text{triv}}^{\Phi,+})^2 {\hat g}^{(+)}_{s}(\hr,\heta)~,\quad (b_{s,\text{triv}}^{\Phi,+})^2=\frac{2^s \left(\Delta_\Phi\right)_s}{s!}~.
\end{align}
It follows from the diagrammatics that, at each order in perturbation theory with either \eqref{flow1} or \eqref{flow2}, only the singlet part of this trivial two-point function gets corrected.\footnote{See appendix \ref{app:FeynDiagON} for an explicit leading-order check of this fact.}

We are therefore led to conjecture that the actual LRI two-point function should be
\begin{align}\label{ansatz}
	\langle \Phi(x_1)\Phi(x_2)\rangle = &\frac{1}{(|\xper_1| |\xper_2|)^{\Delta_\Phi}}\Bigg((b_0^{\Phi,+})^2 {\hat g}^{(+)}_{0}(\hr,\heta)+(b_0^{\Phi,-})^2 {\hat g}^{(-)}_{0}(\hr,\heta) \nonumber \\
	&+\sum_{s>0}(b_{s,\text{triv}}^{\Phi,+})^2 {\hat g}^{(+)}_{s}(\hr,\heta)\Bigg)~,
\end{align}
for two (yet) unfixed coefficients $(b_0^{\Phi,\pm})^2$. We can simply rewrite this as:
\begin{align}\label{two_points_def_freeLRI}
	\langle \Phi(x_1)\Phi(x_2)\rangle 	&= \frac{1}{(x_{12}^2)^{\Delta_\Phi}} +\frac{\left((b_0^{\Phi,+})^2 {\hat g}^{(+)}_{0}(\hr,\heta)-{\hat g}^{(+)}_{0}(\hr,\heta)+(b_0^{\Phi,-})^2 {\hat g}^{(-)}_{0}(\hr,\heta)\right)}{(|\xper_1|| |\xper_2|)^{\Delta_\Phi}}~.
\end{align}

An alternative justification for the ansatz \eqref{two_points_def_freeLRI} uses the recent theorem of \cite{Lauria:2020emq} as follows.\footnote{We are grateful to Balt van Rees for discussions on this point.} As stated above, the trivial defect two-point function contains only `$+$' modes but the converse is true as well. Indeed, if we allow for `$+$' modes only, the resulting (connected) bulk two-point function of $\Phi$ is simply the unique Klein-Gordon propagator with regular boundary conditions on the defect dictated by the bulk-defect OPE, where the $b_{s}^{\Phi,+}$ are fixed to reproduce the contact term in the bulk. All other connected correlators of $\Phi$ vanish. In order to have non-vanishing correlators we must add at least one `$-$' mode~\cite{Lauria:2020emq}. We can then use $s \leq \frac{4 - q}{2}$ and the assumption that $s \in \mathbb{N}$ to conclude that this must be the `$-$' mode with $s = 0$.

\paragraph{Bulk-defect crossing\\}

\begin{figure}
	\centering
	\subfloat[]{
		\begin{minipage}{0.4\textwidth}
			\includegraphics[width=1\textwidth]{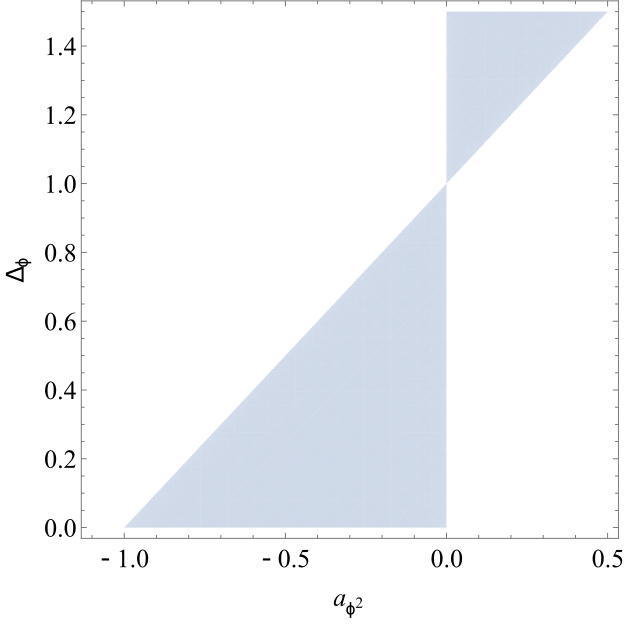} 
	\end{minipage}}
	\hspace{0.5cm}
	\subfloat[]{
		\begin{minipage}{0.4\textwidth}
			\includegraphics[width=1\textwidth]{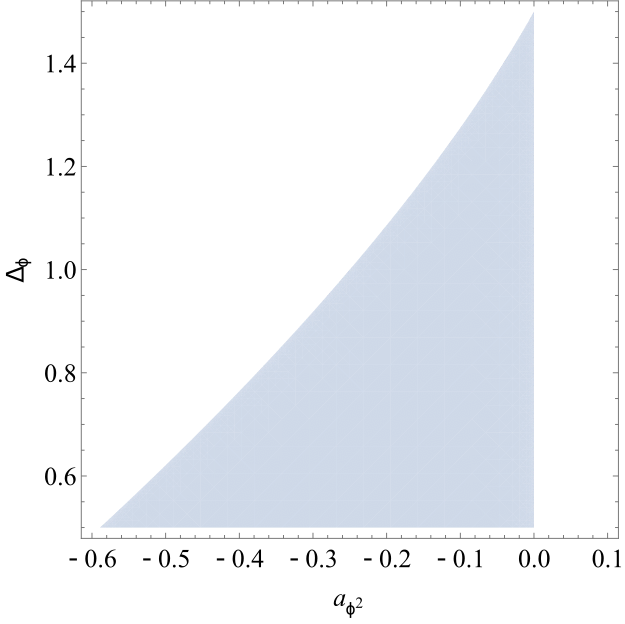}
	\end{minipage}}
	\caption{Allowed region from the bulk-defect bootstrap, in (a) for $p=2$ and in (b) for $p=3$,  upon assuming $(b_0^{\Phi,\pm})^2\geq 0$. Note this is an extra assumption, since unitarity of the bulk theory is not guaranteed -- see footnote \ref{footnotess}. }
	\label{Unitarityregion}
\end{figure}

The bulk two-point function \eqref{two_points_def_freeLRI} must be bulk-defect crossing symmetric. As we have seen, our ansatz is already written in a basis of defect channel conformal blocks, so in order to check crossing we should verify that  \eqref{two_points_def_freeLRI} can be decomposed into bulk channel blocks. Being free, the bulk self-OPE of $\Phi$  features only primary operators $[\Phi\Phi]_{0,\ell}$ of even spin $\ell$ and scaling dimensions $\D_\ell=d+\ell-2$. Using the bulk radial coordinates $(r,\eta)$ defined in \cite{Lauria:2017wav}, we shall therefore write
\begin{align}\label{LRIbulk}
	\langle \Phi(x_1)\Phi(x_2)\rangle &=\frac{\left(r^4-4 \eta^2 r^2+2 r^2+1\right)^{\D_\Phi }}{(|\xper_1| |\xper_2|)^{\D_\Phi} \, \left(4 r\right)^{2 \D_\Phi } }\left[1+\sum_{\ell=\text{even}}\lambda_{\Phi \Phi [\Phi\Phi]_{0,\ell}} a_{[\Phi \Phi]_{0,\ell}} g_{\ell}({r},{\eta})\right]~,
\end{align}
where $a_{[\Phi \Phi]_{0,\ell}}$ are the one-point functions of $[\Phi \Phi]_{0,\ell}$, and the blocks $g_{\ell}({r},{\eta})$ are given by
\begin{align}
	g_{\ell}({r},{\eta})={(4r)}^{\D_\ell} \left[\frac{\left(\frac{1-p-\ell}{2} \right)_{\frac{\ell}{2}}}{\left(\frac{\D_\ell}{2}\right)_{\frac{\ell}{2}}}\, _2F_1\left(-\frac{\ell}{2},\frac{\D_\ell}{2} ;\frac{p+1}{2} ;\eta^2\right)+O(r^2)\right]~,
\end{align}
and higher-order terms can be computed recursively using the results of \cite{Lauria:2017wav}.
Bulk-defect crossing symmetry then dictates:
\begin{align}\label{bpbmsol_d}
	(b_0^{\Phi,-})^2 &= a_{\Phi^2}\frac{\Gamma(p) \Gamma\left(\frac{q-2}{2}\right)}{\Gamma\left(\frac{p}{2}\right) \Gamma\left(\frac{p + q - 2}{2}\right)}~,\quad (b_0^{\Phi,+})^2=1-\frac{\Gamma \left(\frac{p+q-2}{2}\right) \Gamma \left(\frac{4-q}{2} \right)}{\Gamma \left(\frac{q}{2}\right) \Gamma \left(\frac{p-q+2}{2}\right)}(b_0^{\Phi,-})^2,\nonumber\\
	&\lambda_{\Phi \Phi [\Phi\Phi]_{0,\ell}} a_{[\Phi \Phi]_{0,\ell}}=\frac{ \left(\frac{d }{2}-1\right)_\ell \left(\frac{q }{2}-1\right)_{\frac{\ell}{2}}}{2^\ell (1)_{\frac{\ell}{2}} \left(\frac{1}{2} (d +\ell-3)\right)_{\frac{\ell}{2}} \left(\frac{p+1}{2}\right)_{\frac{\ell}{2}}}a_{\Phi^2}~,\quad \ell>0~.
\end{align}
If $(b_0^{\Phi,\pm})^2\geq 0$, which holds in unitary theories, $a_{\Phi^2}$ can be further constrained. In particular for $p/2-1\leq \D_\Phi\leq 3$ the resulting regions are shown in figure~\ref{Unitarityregion}.  However, we should reiterate that unitarity of the bulk theory is not guaranteed. Rather than assuming bulk unitarity, we will simply assume $(b_0^{\Phi,\pm})^2 \geq 0$. Our perturbative predictions are consistent with this assumption, as shown in section~\ref{sec:perturb}.

It should be clear that \eqref{two_points_def_freeLRI} is no longer a valid ansatz for $q = 1$ because boundaries and interfaces do not have transverse spin. This explains why \cite{Liendo:2012hy,Gliozzi:2015qsa} found a different solution to the bootstrap equations, namely:
\begin{align}\label{bsboundary}
	(b_0^{\Phi,+})^2=1 + 2^{d -2}a_{\Phi^2}~, \quad (b_0^{\Phi,-})^2=(d-2)(1 - 2^{d -2}a_{\Phi^2})~.
\end{align}

\subsubsection{The two endpoints of the flow}

Referring to \eqref{flow1} and \eqref{flow2}, we can identify $\Phi = \phi$ when $q = 2 - \mathfrak{s}$ and $\Phi = \chi$ when $q = 2 + \mathfrak{s}$. The trivial defect is then composed of the `$+$' modes, denoted $\hphi$ or $\hat{\chi}$, and the LRI is defined as a defect-localized interaction away from this point. This is either an interaction with the SRI or a self-interaction, depending on the duality frame.

Trivial defects have $b_0^{\Phi,+}=1$ and $b_0^{\Phi,-}=0$ which are equivalent to $a_{\Phi^2} = 0$. After a deformation, upon assuming $(b_0^{\Phi,\pm})^2\geq 0$, the allowed one-point functions in figure~\ref{Unitarityregion} are negative when $\D_\Phi = \D_\phi$ and positive when $\D_\Phi = \D_\chi$. In terms of the small parameters $\eps$ and $\delta$, these deformations affect the co-dimensions as:
\begin{equation}
	q = 2 - \frac{p + \eps}{2}~, \quad q = 2 + p - 2(\delta + \D^*_\sigma)~,
\end{equation}
respectively. In the former case, we have $\aphisq \leq 0$ from the $\hphi^4$ flow while in the latter case, we have $a_{\chi^2} \geq 0$ from the $\sigma \hat{\chi}$ flow.

It is interesting to ask what happens to $\aphisq$ when $a_{\chi^2} = 0$ or vice versa. This is a question about large $\eps$ and $\delta$. In our language, the conjecture of \cite{Behan:2017emf} is that $b_0^{\chi,-} = 0$ when $b_0^{\phi,+} = 0$. 
Instead of the trivial defect, we are therefore dealing with a dual starting point defined by
\begin{align}\label{bminusSRI}
	b_0^{\phi,+}&=0~,\quad (b_0^{\phi,-})^2=\frac{\Gamma \left(\frac{q}{2}\right) \Gamma \left(p-\D_\phi\right)}{\Gamma \left(\D_\phi\right) \Gamma \left(\frac{4-q}{2} \right)} = \frac{\Gamma \left(p - \D_\sigma^* - \delta \right) \Gamma \left(1 - \frac{p}{2} + \delta + \D_\sigma^*  \right)}{\Gamma \left(1 + \frac{p}{2} - \delta - \D^*_\sigma \right) \Gamma \left(\D_\sigma^*+\delta \right)}~.
\end{align}
We can equivalently express the results above in terms of $\aphisq$ via eq.~\eqref{bpbmsol_d} to find
\begin{align}
	\aphisq = \begin{cases}
		\delta-1 +\D_{\sigma}^* =-7/8+\delta~, & p = 2~, \\
		\frac{\sqrt{\pi } \cot (\pi  (\delta +\D_{\sigma}^*)) \Gamma \left(\delta +\D_{\sigma}^*-\frac{1}{2}\right)}{4 \Gamma (\delta +\D_{\sigma}^*-2)}=-0.575408+O(\delta)~, & p = 3~,
	\end{cases} \label{aexpanSRI}
\end{align}
where for $p=2$ we used $\D_{\sigma}^*=1/8$, while for $p=3$ we used $\D_{\sigma}^*=0.5181489$. In order to accommodate both such descriptions of LRI while being consistent with bulk-defect crossing and   $(b_0^{\Phi,\pm})^2\geq 0$, we should allow at least
\begin{align}\label{unitaryintervalaphisq}
	-7/8\leq \aphisq \leq 0~,\quad & p = 2~,\nonumber \\
	-0.575408\leq \aphisq \leq 0~,\quad & p = 3~.
\end{align}
This is the interval in which we will let $\aphisq$ vary when looking at the numerical conformal bootstrap in section~\ref{sec:numerics}. A very similar calculation could be used to estimate $a_{\chi^2}$ at small $\eps$. This result also follows from
\begin{equation}
	\frac{\Gamma(\tfrac{p+2-q}{2})}{\Gamma(\tfrac{2-q}{2})} \left [ a_{\chi^2}^{-1} - \frac{\Gamma(p) \Gamma(\tfrac{q-2}{2})}{\Gamma(\tfrac{p}{2}) \Gamma(\tfrac{p+q-2}{2})} \right ] = \frac{\Gamma(\tfrac{p+q-2}{2})}{\Gamma(\tfrac{q-2}{2})} \left [ a_{\phi^2}^{-1} - \frac{\Gamma(p) \Gamma(\tfrac{2-q}{2})}{\Gamma(\tfrac{p}{2}) \Gamma(\tfrac{p+2-q}{2})} \right ]~, \label{a-relation}
\end{equation}
which we will derive in the next subsection.

\subsection{OPE relations}
In any defect CFT, bulk locality imposes several constraints on the dynamics.  For correlation functions between the bulk field $\Phi$ and two defect insertions $\widehat{\mathcal{O}}$, $\widehat{T}$, and away from contact points, it requires the correlator to be an analytic function of the coordinates. This leads to constraints on defect three-point functions with $\widehat{\mathcal{O}}$, $\widehat{T}$ and the defect modes of $\Phi$ which have been written schematically in \eqref{schematic-ope}. For the LRI specifically, the relations we will use were all derived in \cite{Behan:2018hfx} following the special cases considered in \cite{Paulos:2015jfa}. Here, we will review the modern derivation from \cite{Lauria:2020emq, Behan:2020nsf,Behan:2021tcn} as it applies to a broader class of models. This will also show how the prefactor of \eqref{schematic-ope} is physically related to the one-point function $a_{\Phi^2}$.

\subsubsection{Review of the derivation}
Consider the case where $\widehat{\mathcal{O}}$ is an $SO(p)$ scalar, and $\widehat{{T}}$ is a symmetric and traceless tensor of parallel spin $\ell$. Both  $\widehat{\mathcal{O}}$ and $\widehat{{T}}$  are taken to be scalars under transverse rotations ($s = 0$). Without loss of generality we can place the third operator at infinity and so we investigate:
\begin{align}\label{phiOhOh}
	\langle \Phi(x_1)\widehat{\mathcal{O}}(\xpar_2)\widehat{T}{}^{(l)} (\infty)\rangle~.
\end{align}
In the expression above, to avoid cluttering we have dropped the dependence on the (parallel) spin indices.
We recall that $\Phi$ is a free bulk massless scalar field, so its defect modes are written in \eqref{defSpec} with coefficients $b_s^{\Phi,\pm}$  ($b_s^{\Phi,-}=0$ for $s>0$).
In the defect channel, the correlator above will be written in terms of the OPE coefficients
\begin{align}
	\langle \psi_s^{(\pf)}(\xpar_1)\widehat{\mathcal{O}}(\xpar_2)\widehat{T}{}^{(l)} (\infty)\rangle = \frac{\hf_{{\psi^\pf_s}\widehat{O}\widehat{T}}}{|\xpar_{12}|^{\D_s^{\pf}+\D_{\widehat{\mathcal{O}}}-\D_{\widehat{T}}}}\times (\text{tensor structure})~.
\end{align}
We could now resum the bulk-defect OPE to obtain the complete expression for \eqref{phiOhOh} in terms of the data in the defect channel, but we will not need it in the following. As shown in \cite{Lauria:2020emq}, analyticity of \eqref{phiOhOh} then requires 
\begin{align}\label{ssspinconstr}
	\hf_{{\psi_0^{(-)}}\widehat{\mathcal{O}}\widehat{T}}=-\frac{1}{{R(a_{\Phi^2})}}\frac{\Gamma \left(\frac{q}{2}\right) \Gamma \left(\frac{2\ell + p - q + 2 + 2{\D_{\widehat{T}}}-2\D_{\widehat{\mathcal{O}}} }{4}\right) \Gamma \left(\frac{2\ell + p - q + 2 -2{\D_{\widehat{T}}}+2\D_{\widehat{\mathcal{O}}} }{4}\right)}{\Gamma \left(\frac{4 - q}{2}\right) \Gamma \left(\frac{2\ell + p + q - 2 +2{\D_{\widehat{T}}}-2\D_{\widehat{\mathcal{O}}} }{4} \right) \Gamma \left(\frac{2\ell + p + q - 2 -2{\D_{\widehat{T}}}+2\D_{\widehat{\mathcal{O}}} }{4}\right)} \hf_{{\psi_0^{(+)}}\widehat{\mathcal{O}}\widehat{T}}~,
\end{align}
which is the more precise version of \eqref{schematic-ope}.
We will often consider the case where $\widehat{\mathcal{O}}$ is another zero mode $ \psi_0^{(\pf)}$, so that \eqref{ssspinconstr} gives:
\begin{align}\label{ssspinconstr2}
	\hf_{{\psi_0^{(+)}}\psi^{(+)}_0\widehat{T}}=\kappa_1(\D_{\widehat{T}},\ell)& \hf_{{\psi_0^{(-)}}\psi^{(+)}_0\widehat{T}}~, \quad \hf_{{\psi_0^{(-)}}\psi^{(-)}_0\widehat{T}}=\kappa_2(\D_{\widehat{T}},\ell) \hf_{{\psi_0^{(-)}}\psi^{(+)}_0\widehat{T}}~,
\end{align}
with
\begin{align}\label{kappadef}
	\kappa_1(\D_{\widehat{T}},\ell)&=-R(a_{\Phi^2})\frac{\Gamma \left ( \frac{4 - q}{2} \right ) \Gamma \left(\frac{\ell+{\D_{\widehat{T}}}}{2}\right) \Gamma \left(\frac{\ell+p+q-2-{\D_{\widehat{T}}}}{2} \right)}{\Gamma \left(\frac{q}{2}\right) \Gamma \left(\frac{\ell+p-{\D_{\widehat{T}}}}{2}\right) \Gamma \left(\frac{\ell + 2 - q +{\D_{\widehat{T}}}}{2}\right)}~,\nonumber\\
	\kappa_2(\D_{\widehat{T}},\ell)&=-\frac{1}{R(a_{\Phi^2})}\frac{\Gamma \left ( \frac{q}{2} \right ) \Gamma \left(\frac{\ell+{\D_{\widehat{T}}}}{2}\right) \Gamma \left(\frac{\ell + p - q + 2 -{\D_{\widehat{T}}}}{2} \right)}{\Gamma \left(\frac{4 - q}{2}\right) \Gamma \left(\frac{\ell+p-{\D_{\widehat{T}}}}{2}\right) \Gamma \left(\frac{\ell - 2 + q +{\D_{\widehat{T}}}}{2}\right)}~.
\end{align}
We have defined
\begin{align}
	R(a_{\Phi^2})	\equiv {b_0^{\Phi,-}}/{b_0^{\Phi,+}}~,
\end{align}
where $b_0^{\Phi,\pm}$ and $a_{\Phi^2}$ further satisfy \eqref{bpbmsol_d}.

\subsubsection{Special cases}
For certain special values of the parameters, some of the blocks in the decomposition of \eqref{phiOhOh} are already regular before constraints are introduced. These correspond to poles of the gamma functions in \eqref{ssspinconstr} which cause the exact relations to degenerate. This phenomenon is most interesting when the possibility of operator dimensions infinitesimally close to these special values is disallowed by symmetry. In fractional $q$ models like the LRI, this only happens when
\begin{equation}
	\D_{\widehat{T}}=p + \ell + 2n, \quad \ell \;\; \text{odd}.
\end{equation}
This condition, which guarantees $\kappa_1(\hD_{\widehat{T}},l)=\kappa_2(\hD_{\widehat{T}},l) =0$ and hence $\hf_{\psi^{(+)}_0\psi^{(+)}_0\widehat{T}}=\hf_{\psi^{(-)}_0\psi^{(-)}_0\widehat{T}}=0$, is essential for odd-spin operators because they can only appear in mixed OPEs by Bose symmetry. Importantly, all odd-spin operators in $\psi^{(+)}_0\times \psi^{(-)}_0$ are of this type.

The BCFT setup of \cite{Behan:2020nsf,Behan:2021tcn} also had special cases realized by even-spin operators which could be distinguished from the continuum by Ward identities instead of Bose symmetry. These operators, known as the displacement and its higher-spin cousins, satisfied the following conditions:
\begin{enumerate}
	\item $\D_{\widehat{T}}=2\Delta_0^{(+)}+2n+\ell$ implies $\kappa_1(\D_{\widehat{T}},\ell) =\infty$, $\hf_{\psi^{(+)}_0\psi^{(-)}_0\widehat{T}}=0$ and $\hf_{\psi^{(+)}_0\psi^{(+)}_0\widehat{T}}$ unfixed,
	\item $\D_{\widehat{T}}=2\Delta_0^{(-)}+2n+\ell$ implies $\kappa_2(\D_{\widehat{T}},\ell) =\infty$, $\hf_{\psi^{(+)}_0\psi^{(-)}_0\widehat{T}}=0$ and $\hf_{\psi^{(-)}_0\psi^{(-)}_0\widehat{T}}$ unfixed.
\end{enumerate}
For fractional $q$ on the other hand, we cannot construct any $s = 0$ operators out of the bulk currents which obey Ward identities. Algebraically, this reduced freedom can be seen from the fact that $\Delta_0^{(\pm)}$ do not differ by integers, meaning there is no way to satisfy both conditions at the same time.

\subsubsection{Duality relation}

We can now consider a shadow transformation that exchanges $\D_0^{(+)}$ and $\D_0^{(-)}$. According to \eqref{defSpec}, the one that does the job is $q \leftrightarrow 4 - q$. In terms of LRI parameters, this is equivalent to changing the sign of $\mathfrak{s}$. The first line of \eqref{ssspinconstr} then becomes
\begin{align}
	\hf_{{\psi_0^{(-)}}\psi^{(-)}_0\widehat{T}}=\tilde{\kappa_1}(\D_{\widehat{T}},\ell)& \hf_{{\psi_0^{(-)}}\psi^{(+)}_0\widehat{T}}~, \quad \hf_{{\psi_0^{(+)}}\psi^{(+)}_0\widehat{T}}=\tilde{\kappa}_2(\D_{\widehat{T}},\ell) \hf_{{\psi_0^{(-)}}\psi^{(+)}_0\widehat{T}}~.
\end{align}
The $\tilde{\kappa}_i$, however, are not simply the $\kappa_i$ with $\mathfrak{s} \leftrightarrow -\mathfrak{s}$ applied. This new description of the same LRI has a different bulk so it must use the `$\pm$' modes appropriate to the new co-dimension. This means that if the original set of relations used $R(\aphisq)$, the new set will have $R(a_{\chi^2})$ in its place. Note that the physics has not changed, so it is possible to compare OPE coefficients and find
\begin{align}
	\tilde{\kappa}_1(\D_{\widehat{T}},\ell) = {\kappa}_2(\D_{\widehat{T}},\ell)~, \quad \tilde{\kappa}_2(\D_{\widehat{T}},\ell) = {\kappa}_1(\D_{\widehat{T}},\ell)~.
\end{align}By comparing to eq.~\eqref{kappadef} we see that
\begin{align}\label{dualityconstr}
	R(a_{\phi^2}) R(a_{\chi^2})=1~,
\end{align}
which can be rearranged to give \eqref{a-relation}.

A consequence of this fact is that a co-dimension $2-\ps$ defect with one-point function coefficient $a_{\phi^2}$ gives the same numerical bootstrap bounds as a co-dimension $2+\ps$ defect with one-point function coefficient $a_{\chi^2}$, as far as this small set of correlators is concerned. Applying this duality to the bounds obtained in refs.~\cite{Behan:2020nsf,Behan:2021tcn} should therefore lead to consistent bounds for co-dimension three defects, and it would be interesting to study this further.

\section{Perturbation theory}
\label{sec:perturb}

We can use the exact OPE relations derived in the previous section to study the LRI in perturbation theory. 
As discussed, the LRI admits two dual descriptions: as a GFF coupled to itself -- see eq.~\eqref{flow1} -- and as a GFF coupled to the SRI -- see eq.~\eqref{flow2}. When the interaction in \eqref{lri-hamiltonian} decays sufficiently slowly or sufficiently quickly, one of these flows becomes weakly coupled and hence describes a perturbation of the trivial defect.

As we will review, at the IR fixed point of \eqref{flow1} the critical coupling is $\lambda_*\sim \eps$ with $\eps$ being a small parameter. Observables such as the anomalous dimensions and OPE coefficients of (unit-normalized) LRI operators, as well as the ratio $R$ that appears in the OPE relations, can be expanded in powers of $\eps$ as
\begin{align}
	\D &= \D^{(0)} + \eps\D^{(1)} + \eps^2\D^{(2)} + \eps^3\D^{(3)} + O(\eps^4)~,\nonumber\\
	\hf_{ijk} & =\hf_{ijk}^{(0)} +\eps\hf_{ijk}^{(1)} +\eps^2\hf_{ijk}^{(2)}+\eps^3\hf_{ijk}^{(3)}+O(\eps^4)~,\nonumber\\
	R(\aphisq)& =R^{(0)} +\eps R^{(1)} +\eps^2 R^{(2)} + \eps^3 R^{(3)} + O(\eps^4)~. \label{exp1}
\end{align}

At the IR fixed point of \eqref{flow2} the critical coupling is $g_*^2\sim \delta$ with small $\delta$. Quantities such as the scaling dimensions and the ratio $R$, that are invariant under the $\mathbb{Z}_2$ symmetry that flips the sign of $g_*$, can be expanded in powers of $\delta$ as
\begin{align}
	\D &= \D^{(0)} + \delta \D^{(2)} + \delta^2 \D^{(4)} + O(\delta^3)~,\nonumber \\
	R(\aphisq) &= R^{(0)} + \delta R^{(2)} + \delta^2 R^{(4)} + O(\delta^3)~. \label{exp3}
\end{align}
Depending on their $\mathbb{Z}_2$ charges, OPE coefficients will contain either odd or even powers of $\delta^{1/2}$ (not both) so for them we consider either of:
\begin{align}
	\hf_{ijk} &= \hf_{ijk}^{(0)} + \delta \hf_{ijk}^{(2)} + \delta^2 \hf_{ijk}^{(4)} + O(\delta^3)~,\nonumber\\
	\hf_{ijk} &= \delta^{1/2} \hf_{ijk}^{(1)} + \delta^{3/2} \hf_{ijk}^{(3)} + \delta^{5/2} \hf_{ijk}^{(5)} + O(\delta^{7/2})~. \label{exp2}
\end{align}

Notice that here $\D^{(n)}$ is defined as the shift with respect to the scaling dimension at $\eps = 0$ or $\delta = 0$. Tree-level scaling dimensions can be linear in the small parameter, so $\D^{(n)}$ is in general different from the anomalous dimension $\gamma^{(n)}$, which is naturally defined in perturbation theory around zero coupling with $\eps$ or $\delta$ finite.

\subsection{Review of results in the literature}

In this section, we review some of the available perturbative results for the LRI and its $O(N)$ generalization. Starting from the mean-field end we will consider the following perturbations:
\begin{align}
	S &= \mathcal{N}_{\mathfrak{s}} \mathcal{N}_{-\mathfrak{s}} \int d^p\tau_1 d^p\tau_2 \frac{\hat{\phi}(\tau_1) \hat{\phi}(\tau_2)}{|\tau_{12}|^{p + \mathfrak{s}}} + \int d^p\tau \frac{\lambda}{4!} \hphi^4~, && N = 1~, \nonumber \\
	S &= \mathcal{N}_{\mathfrak{s}} \mathcal{N}_{-\mathfrak{s}} \int d^p\tau_1 d^p\tau_2 \frac{\hat{\phi}(\tau_1) \cdot \hat{\phi}(\tau_2)}{|\tau_{12}|^{p + \mathfrak{s}}} + \int d^p\tau \frac{\lambda}{4} (\hphi \cdot \hphi)^2~, && N > 1~, \label{full-flow1}
\end{align}
where $(\cdot )$ denotes the $O(N)$ scalar product and the constant
\begin{equation}
	\mathcal{N}_\mathfrak{s} = \frac{2^{-\mathfrak{s}} \Gamma(\tfrac{p - \mathfrak{s}}{2})}{\pi^{\tfrac{p}{2}} \Gamma(\tfrac{\mathfrak{s}}{2})}~,
\end{equation}
has been chosen in order to have a unit-normalized propagator in position space.
In complete analogy with the LRI, the long-range $O(N)$ model can be written as a defect for the free, massless $O(N)$ vector model \cite{Giombi:2019enr}.
Since the interaction is marginal when $\eps = 2\mathfrak{s} - p$ vanishes, loop diagrams can generate inverse powers of $\eps$. Even though we are interested in finite $\eps$, we will treat these as poles to be subtracted. This is a scheme that preserves the property that observables accurate to $O(\eps^n)$ require diagrams with up to $n$ loops. 
The two-loop fixed point in this scheme is \cite{Giombi:2019enr} (see \cite{Benedetti:2020rrq} for a three-loop result)
\begin{align}
	\lambda_* &= \frac{\Gamma(\tfrac{p}{2})}{3\pi^{p/2}} \eps + \frac{2\Gamma(\tfrac{p}{2})}{9\pi^{p/2}}\left[\psi \left ( \frac{p}{2} \right ) - 2\psi \left ( \frac{p}{4} \right ) + \psi(1)\right] \eps^2 + O(\eps^3)~, && N = 1~, \nonumber \\
	\lambda_* &= \frac{\Gamma(\tfrac{p}{2})}{2\pi^{p/2}(N + 8)} \eps + \frac{\Gamma(\tfrac{p}{2})(5N + 22)}{\pi^{p/2}(N + 8)^3}\left[\psi \left ( \frac{p}{2} \right ) - 2\psi \left ( \frac{p}{4} \right ) + \psi(1)\right] \eps^2 + O(\eps^3)~, && N > 1~. \label{lambda-fp}
\end{align}

Around the short-range end we will only consider the $N = 1$ action:
\begin{equation}
	S = S_{\text{SRI}} + \mathcal{N}_{\mathfrak{s}} \mathcal{N}_{-\mathfrak{s}} \int d^p\tau_1 d^p\tau_2 \frac{\hat{\chi}(\tau_1) \hat{\chi}(\tau_2)}{|\tau_{12}|^{p - \mathfrak{s}}} + \int d^p\tau g \sigma \hat{\chi}~, \label{full-flow2}
\end{equation}
which leads to results depending on the short-range CFT data quoted below \cite{Simmonsduffin:2017,Reehorst:2021hmp}.\footnote{See \cite{ElShowk:2013} for some bootstrap estimates of this data in fractional dimension.}
\begin{align}
	&\D^*_\sigma = \frac{1}{8}&&\D^*_\epsilon = 1 &&&c_T = 1 &&&&(p=2)\nonumber\\
	&\D^*_\sigma = 0.518157&&\D^*_\epsilon = 1.41265 &&&c_T = 1.419815 &&&&(p=3)\label{rigorous-data}
\end{align}
Taking $\delta = \frac{p - \mathfrak{s}}{2} - \D^*_\sigma$ to be small, the two-loop beta function for $g$ can be computed numerically with conformal perturbation theory \cite{Zamolodchikov:1987ti}. This was done in \cite{Behan:2017emf} leading to the fixed point
\begin{align}
	g_*^2 = \begin{cases}
		0.788392\delta + O(\delta^2)~, & p = 2 \\
		0.8155(3)\delta + O(\delta^2)~, & p = 3 \label{g-fp1}
	\end{cases}~.
\end{align}
It is clear that in principle, \eqref{g-fp1} can be generalized to $O(N)$ fixed points, although separate numerical calculations would be needed for each value of $N$.\footnote{The duality between the large-$N$ versions of \eqref{flow1} and \eqref{flow2} has been explored in \cite{Chai:2021arp}. References~\cite{DiPietro:2020fya,DiPietro:2023zqn} discuss a similar co-dimension one duality between large-$N$ free/critical vector modes coupled to a free massless bulk scalar field with Dirichlet/Neumann boundary conditions.}
We will also speculate later on about long-range fixed points based on minimal models. The most optimistic hypothesis one can derive from \cite{Behan:2017emf} is that \textit{any} Ginzburg-Landau model with a nonlocal kinetic term can be reached by coupling its short-range partner to a GFF.

\paragraph{Anomalous dimensions near the mean-field end\\}

Let us now discuss the anomalous dimensions of LRI operators.
We shall recall that the fundamental field $\hphi$ does not renormalize, its scaling dimension being fixed by the nonlocal equation of motion. For the leading scalar $\widehat{\phi^2} = \hphi\cdot \hphi / \sqrt{2N}$ among $O(N)$ singlets, the anomalous dimension is computed at three loops in \cite{Benedetti:2020rrq,bgh24} (two-loop results were originally obtained in \cite{Fisher:1972zz}) and reads:
\begin{align}\label{anphi2}
	\D_{\widehat{\phi^2}} &= \frac{p}{2} + \frac{\eps}{2} \frac{N - 4}{N + 8} + \eps^2 \frac{(N + 2)(7N + 20)}{(N + 8)^3} \left [\psi \left ( \frac{p}{2} \right ) - 2\psi \left ( \frac{p}{4} \right ) + \psi(1) \right] \\
	&- \eps^3 \frac{(N + 2)}{4(N + 8)^5} (19N^3 - 60N^2 - 432N - 256) \left [ \psi \left ( \frac{p}{2} \right ) - 2\psi \left ( \frac{p}{4} \right ) + \psi(1) \right ]^2 \nonumber \\
	&+ \eps^3 \frac{3(N + 2)}{4(N + 8)^4} (N^2 - 12N - 16) \left [ \psi'(1) - \psi' \left ( \frac{p}{2} \right ) \right ] - \eps^3 \frac{(N + 2)(5N + 22)}{(N + 8)^4} \alpha_{I_4} + O(\eps^4)~. \nonumber
\end{align}
The last coefficient is only known numerically and it was evaluated as
\begin{align}\label{addendum-i4}
	\alpha_{I_4} = \begin{cases}
		76.62828703(7)~, & p = 2~, \\
		30.1026152(6)~, & p = 3~.
	\end{cases}
\end{align}
in \cite{bgh24}.

The next quantity that is known, is the two-loop scaling dimension $\D_{[\hphi \hphi]_{n,\ell}}$ of the spin-$\ell$ operator $[\hphi \hphi]_{n,\ell}$, which was computed in \cite{Behan:2019lyd} (see also  \cite{Behan:2018hfx}) for the case of $n = 0$ and $N=1$.
As discussed there, we can compute this quantity by considering the three-point function $\langle \hphi \hphi [\hphi \hphi]_{n,\ell} \rangle$, and the first correction comes from the two-loop diagram:
\begin{align}
	\begin{tikzpicture}[baseline,valign,decoration={markings, 
			mark= at position 0.6 with {\arrow{stealth}}},scale=2]
		\draw[postaction=decorate] (0.5,0) -- (-0.5,0) arc(0:180:-0.5) --cycle;
		\draw[postaction=decorate] (-0.5,0) to (0.5,0);
		\draw[postaction=decorate] (-1.5,0) -- (-0.5,0);	
		\draw[postaction=decorate] (0.5,0) -- (1.5,0);
		\draw[postaction=decorate] (-0.5, 0) to[in=180,out=90] (-0.05, 0.5);
		\draw[postaction=decorate] (0.05,0.5) to[in=90,out=0] (0.5, 0);
		\node at (0,0.5) [wcirc] {};
		\node at (-0.5,0) [bcirc] {};
		\node at (0.5,0) [bcirc] {};
		\node at (-1,-0.2) [] {$k$};
		\node at (1,-0.2) [] {$k$};
		\node at (0,-0.7) [] {$k - k' - k''$};
		\node at (0,-0.2) [] {$k'$};
		\node at (-0.5,0.4) [] {$k''$};
		\node at (0.55,0.4) [] {$k''$};
	\end{tikzpicture}~.
\end{align}
Here, the solid lines represent (scalar) propagators, while the arrows indicate the flow of momentum. The white dot represents a composite operator, in this case $[\hphi \hphi]_{n,\ell}$. The black dots indicate the $\widehat{\phi^4}$ vertices proportional to $\lambda$.
Setting for simplicity the insertion of the spinning operator to zero external momentum and repeating the calculations of \cite{Behan:2019lyd}, we obtain the following result for the two-loop correction:
\begin{align}
	\frac{6(N + 2)(\lambda \mu^\eps)^2}{\mathcal{N}_\mathfrak{s}^4}& \int \frac{dk' dk''}{(2\pi)^{2p}} \frac{k''_{(a_1} \dots k''_{a_\ell)}}{|k'|^\mathfrak{s} |k - k' - k''|^\mathfrak{s} |k''|^{2(\mathfrak{s} - n)}} \nonumber\\
	&=  \frac{6(N + 2) \lambda^2}{(4\pi)^p \mathcal{N}_\mathfrak{s}^4} \frac{\Gamma(\ell + n + \tfrac{p}{2} - \mathfrak{s}) \Gamma(\tfrac{p - \mathfrak{s}}{2})^2 \Gamma(2\mathfrak{s} - p - n) \Gamma(n + p - \mathfrak{s}) \Gamma(2\mathfrak{s} - \tfrac{p}{2})}{\Gamma(\mathfrak{s}) \Gamma(\tfrac{\mathfrak{s}}{2})^2 \Gamma(\ell + 2n + \tfrac{3}{2}p - 2\mathfrak{s}) \Gamma(p - \mathfrak{s}) \Gamma(2\mathfrak{s} - n - \tfrac{p}{2})} \left | \frac{\mu}{k} \right |^\eps \nonumber \\
	&= 6(N + 2)\lambda^2 \frac{\pi^p (-1)^n \Gamma(n + \ell) \Gamma(\tfrac{p}{2} + n)}{\eps n! \Gamma(\tfrac{p}{2}) \Gamma(\tfrac{p}{2} + 2n + \ell) \Gamma(\tfrac{p}{2} - n)} + O(1) \label{only-momentum}~,
\end{align}
where we divided by the tree-level result.\footnote{As the spinning operator does not carry momentum, derivatives will produce the same effect no matter how they are distributed: they simply contribute the same factor of $k''_{a_1} \dots k''_{a_\ell} (k'' \cdot k'')^n$. The tree-level diagram is also proportional to the same factor. See \cite{Benedetti:2023} for some other calculations where the distribution of derivatives drops out.} Since there is no wave-function renormalization for $\hphi$, the $1/\eps$ pole in this diagram is removed by a wave-function renormalization for $[\hphi \hphi]_{n,\ell}$, which in turn leads to the following scaling dimension for $[\hphi \hphi]_{n,\ell}$ at the IR fixed point:
\begin{align}\label{andt}
	\D_{[\hphi \hphi]_{n,\ell}} = \frac{p + 4n + 2\ell}{2} - \frac{\eps}{2} - \eps^2 \frac{N + 2}{(N + 8)^2} \frac{(-1)^n}{n!} \frac{3\Gamma(\tfrac{p}{2}) \Gamma(n + \ell) \Gamma(\tfrac{p}{2} + n)}{\Gamma(\tfrac{p}{2} + 2n + \ell) \Gamma(\tfrac{p}{2} - n)} + O(\eps^3)~.
\end{align}
This generalizes the result in \cite{Behan:2018hfx,Behan:2019lyd} and also corrects a factor of 2 for the leading twist $N=1$ result.

\paragraph{Anomalous dimensions near the short-range end\\}
There are fewer perturbative results near the short-range end. 
The leading scalar's anomalous dimension was computed at two loops in the $\delta$ expansion in \cite{Behan:2017emf}, which found
\begin{align}
	\D_\epsilon = \begin{cases}
		1 + O(\delta^2)~, & p = 2~, \\
		\D^*_\epsilon + 0.27 \delta + O(\delta^2)~, & p = 3~.
	\end{cases} \label{sri-eps}
\end{align}

The anomalous dimension of the leading spin-two operator $T_{\mu\nu}$ can be obtained from multiplet recombination \cite{Rychkov:2015naa,Giombi:2016hkj} (see also \cite{Behan:2020nsf}), i.e. by exploiting the fact that conservation of the SRI's stress tensor $T_{\mu\nu}$ fails as soon as we turn on $g$. 
In turn, this strategy allows us to gain one order in perturbation theory. The resulting scaling dimension of $T$ depends on the central charge in \eqref{rigorous-data} and is \cite{Behan:2017emf,Behan:2017dwr}:
\begin{align}\label{recomb-spin2}
	\D_T = p + \frac{8 \pi^p}{c_T \Gamma(\tfrac{p}{2})^2} \frac{\D_\sigma^*(p - \D_\sigma^*)}{p^2 + p - 2} g_*^2 + O(g_*^4)~.
\end{align}

For the specific case of the $p=2$ LRI we can use Virasoro multiplet recombination methods to compute the anomalous dimensions of higher-spin Virasoro currents in the SRI.\footnote{The GFF $\hat{\chi}$ is what allows one to construct the operators which can serve as divergences for these broken currents. The SRI on its own does not have any. At present, this is one of the main obstacles to formulating the local Wilson-Fisher fixed point as an expansion around the 2d Ising model. See \cite{Zhou:2022} for recent comments on this problem.} At spin-4, the dimension of $\Lambda \equiv \left ( L_{-4} - \frac{5}{3} L_{-2}^2 \right ) | 0 \rangle$ has been computed in \cite{Behan:2018hfx} to be:
\begin{align}\label{recomb-spin4}
	\D_\Lambda = 4 + \frac{1335}{2048} \pi^2 g_*^2 + O(g_*^4)~.
\end{align}
Increasing the spin leads to a proliferation of currents and a computationally intensive problem, but for anomalous dimensions up to spin-10 it should become manageable using the techniques of \cite{Antunes:2022vtb}. We leave this problem for future work.

Results in the literature in both perturbative settings have mostly focused on anomalous dimensions and fewer results are available for the OPE coefficients. We will present some new results for OPE coefficients in both settings in the next sections.

\subsection{Using OPE relations near the short-range end}

Let us first discuss how to use the OPE relations near the short-range end, so that $g_*^2\sim \delta $ and $0 < \delta \ll 1$. 
Importantly, if we want to keep the co-dimension smaller than 2 (so we can compare to the predictions from the MFT end), the expansion does not start from the trivial defect $b_0^{\phi,-} = 0$, but rather from $b_0^{\phi,+} = 0$, see eq.~\eqref{bminusSRI}. For the same reason, the `$\pm$' modes are exchanged: they are $\sigma$ and $\hat{\chi}$ respectively, which means the OPE relations in \eqref{ssspinconstr2} read
\begin{align}\label{short-range-relations}
	\frac{\lambda_{\sigma\sigma\mathcal{O}}}{\lambda_{\sigma\hat{\chi}\mathcal{O}}} &= R(a_{\phi^2}) \frac{\Gamma(p/2 - \D_\sigma) \Gamma(\tfrac{\ell + \Delta}{2}) \Gamma(\tfrac{\ell + 2\D_\sigma - \D}{2})}{\Gamma(\D_\sigma - p/2) \Gamma(\tfrac{\ell + p - 2\D_\sigma + \D}{2}) \Gamma(\tfrac{\ell + p - \D}{2})}~, \nonumber \\
	\frac{\lambda_{\hat{\chi}\hat{\chi}\mathcal{O}}}{\lambda_{\sigma\hat{\chi}\mathcal{O}}} &= R(a_{\phi^2})^{-1} \frac{\Gamma(\D_\sigma - p/2) \Gamma(\tfrac{\ell + \Delta}{2}) \Gamma(\tfrac{\ell + 2p - 2\D_\sigma - \D}{2})}{\Gamma(p/2 - \D_\sigma) \Gamma(\tfrac{\ell - p + 2\D_\sigma + \D}{2}) \Gamma(\tfrac{\ell + p - \D}{2})}~. 
\end{align}
If $\mathcal{O}$ is a primary in the SRI which diagonalizes dilations, then $\lambda_{\hat{\chi}\hat{\chi}\mathcal{O}}$ starts at $O(g_*^2)$, while $\lambda_{\sigma\hat{\chi}\mathcal{O}}$ and $R(a_{\phi^2})^{-1}$ start at $O(g_*)$. The OPE relations can therefore give information about two-loop perturbation theory with one-loop data as input. In practice, the first non-vanishing terms in $\lambda_{\sigma\hat{\chi}\mathcal{O}}$ and $\lambda_{\hat{\chi}\hat{\chi}\mathcal{O}}$ can both be computed from the star-triangle relation in \eqref{3and4}, using it once for the former and twice for the latter. We will therefore compute the OPE coefficients in \eqref{short-range-relations} directly and treat $R(a_{\phi^2})$ as the unknown quantity. If we take $\mathcal{O}$ to be a scalar, the integrals arising in conformal perturbation theory can now be easily computed to get
\begin{align}\label{scalar-first}
	\lambda_{\sigma\hat{\chi}\mathcal{O}} &= \pi^{p/2} g_* \lambda^*_{\sigma\sigma\mathcal{O}} \frac{\Gamma(\D^*_\sigma - p/2) \Gamma(\tfrac{p + \D^*}{2} - \D^*_\sigma) \Gamma(\tfrac{p - \D^*}{2})}{\Gamma(p - \D^*_\sigma) \Gamma(\D^*/2) \Gamma(\D^*_\sigma - \D^*/2)} + O(g_*^3)~, \nonumber \\
	\lambda_{\hat{\chi}\hat{\chi}\mathcal{O}} &= \pi^p g_*^2 \lambda^*_{\sigma\sigma\mathcal{O}} \frac{\Gamma(\D^*_\sigma - p/2)^2 \Gamma(\tfrac{p + \D^*}{2} - \D^*_\sigma) \Gamma(p - \D^*_\sigma - \D^*/2)}{\Gamma(p - \D^*_\sigma)^2 \Gamma(\tfrac{\D^* - p}{2} + \D^*_\sigma) \Gamma(\D^*_\sigma - \D^*/2)} + O(g_*^4)~. 
\end{align}
Either one of these results is enough to compute $R(a_{\phi^2})$ upon going back to \eqref{short-range-relations} and the fact that they agree is a check. We find
\begin{align}\label{short-range-ratio}
	R(a_{\phi^2})^{-1} &= \pi^{p/2} g_* \frac{\Gamma(p/2 - \D^*_\sigma)}{\Gamma(p - \D^*_\sigma)} + O(g_*^3)~.
\end{align}
Upon plugging in $\D^*_\sigma$ and $g_*$ numerically and using \eqref{bpbmsol_d}, the result above gives the following $O(\delta)$ correction to \eqref{aexpanSRI}:
\begin{align}\label{short-range-1pt}
	a_{\phi^2} &= \begin{cases}
		-\frac{7}{8} + 9.896\delta + O(\delta^2)~, & p = 2~, \\
		-0.575408 + 37.13\delta + O(\delta^2)~, & p = 3~.
	\end{cases}
\end{align}
The power of this approach is that, with \eqref{short-range-ratio} in hand, the OPE relations can be used to quickly upgrade \eqref{scalar-first} to a spinning version of it. As a result,
\begin{align}\label{spinning-second}
	\lambda_{\sigma\hat{\chi}\mathcal{O}} &= \pi^{p/2} g_* \lambda^*_{\sigma\sigma\mathcal{O}} \frac{\Gamma(\D^*_\sigma - p/2) \Gamma(\tfrac{\ell + \Delta^*}{2}) \Gamma(\tfrac{\ell + 2\D^*_\sigma - \D^*}{2})}{\Gamma(p - \D^*_\sigma) \Gamma(\tfrac{\ell + p - 2\D^*_\sigma + \D^*}{2}) \Gamma(\tfrac{\ell + p - \D^*}{2})} + O(g_*^3)~, \nonumber \\
	\lambda_{\hat{\chi}\hat{\chi}\mathcal{O}} &= \pi^p g_*^2 \lambda^*_{\sigma\sigma\mathcal{O}} \frac{\Gamma(\D^*_\sigma - p/2)^2 \Gamma(\tfrac{\ell + p - 2\D^*_\sigma + \D^*}{2}) \Gamma(\tfrac{\ell + 2p - 2\D^*_\sigma - \D^*}{2} ) }{\Gamma(p - \D^*_\sigma)^2 \Gamma(\tfrac{\ell - p + 2\D^*_\sigma + \D^*}{2}) \Gamma(\tfrac{\ell + 2\D^*_\sigma - \D^*}{2})} + O(g_*^4)~, 
\end{align}
has been concluded using only the (scalar) star-triangle relation. Deriving it directly would have required the more complicated conformal integrals in \cite{SimmonsDuffin:2012uy}. We could now apply \eqref{spinning-second} for $p=3$, inputting the numerically determined CFT data from the SRI, some of which has been bounded rigorously in \cite{Reehorst:2021hmp}. For $p=2$, the caveat requiring $\mathcal{O}$ to diagonalize the generator of dilations becomes important. One way to satisfy this is to find the eigenstates at a given level of the Virasoro $\epsilon$ multiplet and choose $\mathcal{O}$ from this set. Conversely, operators in the Virasoro identity multiplet will mix with derivatives of $\sigma \hat{\chi}$ which precludes a diagonalization solely within the SRI.

\subsection{Using OPE relations near the mean-field end}\label{LRI2loops}

Let us now consider the mean-field end, where $g_*\sim \eps$, $0 < \eps \ll 1$ and the expansion starts from the trivial defect. This time the `$\pm$' modes are $\hat{\phi}$ and $\widehat{\phi^3}$ respectively, and the OPE relations in \eqref{ssspinconstr2} read:
\begin{align}\label{mean-field-relations}
	\frac{\hf_{\hphi \hphi \mathcal{O}}}{\hf_{\hphi \widehat{\phi^3} \mathcal{O}}} &= R(a_{\phi^2}) \frac{\Gamma(p/2 - \D_\phi) \Gamma(\tfrac{\ell + \Delta}{2}) \Gamma(\tfrac{\ell + 2\D_\phi - \D}{2})}{\Gamma(\D_\phi - p/2) \Gamma(\tfrac{\ell + p - 2\D_\phi + \D}{2}) \Gamma(\tfrac{\ell + p - \D}{2})}~, \nonumber \\
	\frac{\hf_{\widehat{\phi^3} \widehat{\phi^3} \mathcal{O}}}{\hf_{\hphi \widehat{\phi^3} \mathcal{O}}} &= R(a_{\phi^2})^{-1} \frac{\Gamma(\D_\phi - p/2) \Gamma(\tfrac{\ell + \Delta}{2}) \Gamma(\tfrac{\ell + 2p - 2\D_\phi - \D}{2})}{\Gamma(p/2 - \D_\phi) \Gamma(\tfrac{\ell - p + 2\D_\phi + \D}{2}) \Gamma(\tfrac{\ell + p - \D}{2})}~. 
\end{align}
All operators are taken to be unit-normalized.
Setting $\D_\phi = \frac{p - \eps}{4}$ shows that the OPE relations now constrain the CFT data in a different way.
By plugging the $\hf_{ijk}$ coefficients at $O(\eps^n)$ into \eqref{mean-field-relations}, we can learn about $R(\aphisq)$ and $\D$ at $O(\eps^{n+1})$.

We begin with the concrete example of $\D_{\widehat{\phi^2}}$ and $R(\aphisq)$ at $O(\eps^2)$.
The process starts by inputting the tree-level OPE coefficients
\begin{align}\label{OPEcoeffstree}
	\hf_{\hphi\hphi \widehat{\phi^2}}^{(0)} = \sqrt{2}~, \quad \hf_{\hphi \widehat{\phi^3} \widehat{\phi^2}}^{(0)} = \sqrt{3}~, \quad
	\hf_{\widehat{\phi^3} \widehat{\phi^3} \widehat{\phi^2}}^{(0)} = 3\sqrt{2}~,
\end{align}
into the first two OPE relations in eq.~\eqref{mean-field-relations}.
By further expanding them in $\eps$, it is a simple exercise to show that the OPE relations are satisfied if and only if
\begin{align}\label{easiest-part}
	\D^{(1)}_{\widehat{\phi^2}} = -1/6~,\quad R^{(0)}=0~,\quad R^{(1)}=\frac{\Gamma \left(1-\frac{p}{4}\right) \Gamma \left(\frac{p}{2}\right)}{3 \sqrt{6} \Gamma \left(\frac{p}{4}+1\right)}~.
\end{align}
In particular, we have reproduced the known anomalous dimension $\gamma^{(1)}_{\widehat{\phi^2}}$ and also $R^{(1)}$, which is computed directly in appendix \ref{app:FeynDiagON}. 

At the next-to-leading order, we shall need the OPE coefficients at $O(\eps)$. Computing $\hf^{(1)}_{ijk}$ in position space starts with integrating a GFF four-point function with respect to $\tau$ while keeping $\eps$ finite. 
In all cases, this produces a $1/\eps$ pole to be subtracted, a logarithmic term associated with anomalous dimensions, and the finite term we are after. As we will see, the latter is always proportional to
\begin{align}
	\mathcal{A}_p\equiv  \psi\left ( \frac{p}{2} \right ) - 2\psi\left ( \frac{p}{4} \right ) + \psi(1)~. \label{ap-def}
\end{align}
For $\hf^{(1)}_{\hphi\hphi\widehat{\phi^2}}$ the unique diagram is:
\begin{align}
	-\frac{\lambda}{\sqrt{2}} \;\; \begin{tikzpicture}[baseline,valign]
		\draw (-1,0.5) -- (0,0);
		\draw (0,0) -- (-1,-0.5);
		\draw (0,0) to[in=140,out=40] (1.5,0);
		\draw (0,0) to[in=-140,out=-40] (1.5,0);
		\node at (0,0) [bcirc] {};
		\node at (1.5,0) [wcirc] {};
	\end{tikzpicture} 
\end{align}
where the factor $1/\sqrt{2}$ comes from the insertion of $\widehat{\phi^2}$, and we are integrating over all possible insertions of the $\widehat{\phi^4}$ vertex proportional to $\lambda$. Here and below, solid lines indicate scalar propagators, while black dots indicate  $\widehat{\phi^4}$ interaction vertices. The white dots represent composite operators, e.g.  $\widehat{\phi^2}$,  $\widehat{\phi^3}$,  $\widehat{\phi^4}$.
The expression for the diagram follows straightforwardly from the master integrals given in appendix~\ref{app:FeynDiagON}, as is discussed there. Hence, we find
\begin{align} \label{mi1}
	I_{1,1,2} &= -\frac{\lambda \pi^{p/2}}{\sqrt{2}} \frac{\Gamma(\tfrac{p}{2})^{-1}}{|\tau_{13}|^{\tfrac{p - \eps}{2}} |\tau_{23}|^{\tfrac{p - \eps}{2}}} \left [ \frac{2}{\eps} + \mathcal{A}_p + \log \left | \frac{\tau_{13} \tau_{23}}{\tau_{12}} \right | + O(\eps) \right ]~.
\end{align}

For $\hf^{(1)}_{\hphi \widehat{\phi^3} \widehat{\phi^2}}$ we have three diagrams:
\begin{align}
	-\frac{3\lambda}{2\sqrt{3}}\;\; \begin{tikzpicture}[baseline,valign]
		\draw (-1,0.5) to[in=90,out=70] (0,0);
		\draw (-1,0.5) to[in=-110,out=-110] (0,0);
		\draw (-1,-0.5) to[in = -120,out = 0] (1.5,0);
		\draw (0,0) to[in=140,out=40] (1.5,0);
		\draw (0,0) -- (1.5,0);
		\node at (0,0) [bcirc] {};
		\node at (1.5,0) [wcirc] {};
		\node at (-1,0.5) [wcirc] {};
	\end{tikzpicture}  - \frac{\lambda}{\sqrt{3}} \;\; \begin{tikzpicture}[baseline,valign]
		\draw (-1,0.5) to[in=110,out=40] (0,0);
		\draw (-1,-0.5) to[in=-70,out=10] (-0.5,0);
		\draw (-0.5,0) to[in = -20,out = 110] (-1,0.5);
		\draw (0,0) to[in=140,out=40] (1.5,0);
		\draw (0,0) to[in=-140,out=-40] (1.5,0);
		\draw (0,0) -- (1.5,0);
		\node at (0,0) [bcirc] {};
		\node at (1.5,0) [wcirc] {};
		\node at (-1,0.5) [wcirc] {};
	\end{tikzpicture}   - \frac{3\lambda}{\sqrt{3}} \;\; \begin{tikzpicture}[baseline,valign]
		\draw (-1,0.5) to[in=140,out=30] (1.5,0);
		\draw (-1,0.5) -- (0,0);
		\draw (-1,-0.5) -- (0,0);
		\draw (0,0) to[in=-140,out=-40] (1.5,0);
		\draw (0,0) -- (1.5,0);
		\node at (0,0) [bcirc] {};
		\node at (1.5,0) [wcirc] {};
		\node at (-1,0.5) [wcirc] {};
	\end{tikzpicture}~.
\end{align}
Using the results of appendix~\ref{app:FeynDiagON} we see that the first diagram has no $O(1)$ term, while the second starts at $O(\eps)$. The third diagram has a finite part which is proportional to \eqref{mi1} leading to
\begin{align}\label{mi2}
	I_{1,3,2} = -\frac{3\lambda \pi^{p/2}}{\sqrt{3}} \frac{\Gamma(\tfrac{p}{2})^{-1}}{|\tau_{12}|^{\tfrac{p - \eps}{2}} |\tau_{23}|^{p - \eps}} \left [ \frac{4}{\eps} + \mathcal{A}_p + 2 \log \left | \frac{\tau_{12} \tau_{23}^2}{\tau_{13}} \right | + O(\eps) \right ]~.  
\end{align}

Finishing with $\hf^{(1)}_{\widehat{\phi^3} \widehat{\phi^3} \widehat{\phi^2}}$, we have five diagrams: 
\begin{align}
	-\frac{3\lambda}{\sqrt{2}}  \begin{tikzpicture}[baseline,valign]
		\draw (-1,0) -- (1,0); 
		\draw (-1,0) to[out=60,in=120] (0,0);
		\draw (0,0) to[out=60,in=120] (1,0);
		\draw (-1,0) to[out=-120,in=140] (0,-1.5);
		\draw (0,-1.5) to[out=40,in=-60] (1,0);
		\node at (0,0) [bcirc] {};
		\node at (1,0) [wcirc] {};
		\node at (-1,0) [wcirc] {};
		\node at (0,-1.5) [wcirc] {};
	\end{tikzpicture}  - \frac{\lambda}{\sqrt{2}} \left( \begin{tikzpicture}[baseline,valign]
		\draw (1,0) -- (0,0);
		\draw (0,0) to[out=60,in=120] (1,0);
		\draw (0,0) to[out=-60,in=-120] (1,0);
		\draw (-1,0) to[out=60,in=120] (0,0);
		\draw (-1,0) to[out=-70,in=160] (0,-1);
		\draw (-1,0) to[out = -10,in= 100] (0,-1);
		\node at (0,0) [bcirc] {};
		\node at (1,0) [wcirc] {};
		\node at (-1,0) [wcirc] {};
		\node at (0,-1) [wcirc] {};
	\end{tikzpicture} \; + \; \begin{tikzpicture}[baseline,valign]
		\draw (-1,0) -- (0,0);
		\draw (-1,0) to[out=60,in=120] (0,0);
		\draw (-1,0) to[out=-60,in=-120] (0,0);
		\draw (0,0) to[out=60,in=120] (1,0);
		\draw (0,-1) to[out=70,in=-160] (1,0);
		\draw (0,-1) to[out = 10,in= -100] (1,0);
		\node at (0,0) [bcirc] {};
		\node at (1,0) [wcirc] {};
		\node at (-1,0) [wcirc] {};
		\node at (0,-1) [wcirc] {};
	\end{tikzpicture} \right) \nonumber \\
	- \frac{6\lambda}{\sqrt{2}} \left(\begin{tikzpicture}[baseline,valign]
		\draw (-1,0) -- (0,0); 
		\draw (0,0) -- (0,-1);
		\draw (-1,0) to[out=60,in=120] (1,0);
		\draw (0,0) to[out=60,in=120] (1,0);
		\draw (0,0) to[out=-60,in=-120] (1,0);
		\draw (-1,0) to[out = -90,in= 180] (0,-1);
		\node at (0,0) [bcirc] {};
		\node at (1,0) [wcirc] {};
		\node at (-1,0) [wcirc] {};
		\node at (0,-1) [wcirc] {};
	\end{tikzpicture} \; + \; \begin{tikzpicture}[baseline,valign]
		\draw (1,0) -- (0,0); 
		\draw (0,0) -- (0,-1);
		\draw (-1,0) to[out=60,in=120] (1,0);
		\draw (-1,0) to[out=60,in=120] (0,0);
		\draw (-1,0) to[out=-60,in=-120] (0,0);
		\draw (0,-1) to[out = 0,in= -90] (1,0);
		\node at (0,0) [bcirc] {};
		\node at (1,0) [wcirc] {};
		\node at (-1,0) [wcirc] {};
		\node at (0,-1) [wcirc] {};
	\end{tikzpicture}\right) - \frac{3\lambda}{\sqrt{2}} \; \begin{tikzpicture}[baseline,valign]
		\draw (-1,0) -- (1,0);
		\draw (-1,0) to[out=40,in=140] (1,0);
		\draw (-1,0) to[out=90,in=90] (1,0);
		\draw (0,-1) to[out=60,in=-60] (0,0);
		\draw (0,-1) to[out = 120,in= -120] (0,0);
		\node at (0,0) [bcirc] {};
		\node at (1,0) [wcirc] {};
		\node at (-1,0) [wcirc] {};
		\node at (0,-1) [wcirc] {};
	\end{tikzpicture}~.
\end{align}
The first one is $72/\eps$ to the desired order, the second and third are zero to the desired order, and all $O(1)$ terms come from the last three diagrams. Using the master integrals in appendix~\ref{app:FeynDiagON} again, the result is:
\begin{align}\label{mi3}
	I_{3,3,2} = -\frac{3\lambda \pi^{p/2}}{\sqrt{2}} \frac{\Gamma(\tfrac{p}{2})^{-1}}{|\tau_{12}|^{p - \eps} |\tau_{13}|^{\tfrac{p - \eps}{2}} |\tau_{23}|^{\tfrac{p - \eps}{2}}} \left [ \frac{14}{\eps} + 5\mathcal{A}_p + 2 \log \left | \tau_{12}^5 \tau_{13} \tau_{23} \right | + O(\eps) \right ]~. 
\end{align}

The $1/\eps$ poles in \eqref{mi1}, \eqref{mi2} and \eqref{mi3} can be subtracted by a wave-function renormalization for $\widehat{\phi^2}$ and for $\widehat{\phi^3}$, which in turn give the correct anomalous dimensions at the IR fixed point, and from the finite part we get:
\begin{align}\label{coeff1loopchk}
	\hf^{(1)}_{\hphi\hphi\widehat{\phi^2}} = -\frac{\mathcal{A}_p}{3\sqrt{2}} ~,\quad \hf^{(1)}_{\hphi \widehat{\phi^3} \widehat{\phi^2}} = -\frac{\mathcal{A}_p}{\sqrt{3}} ~,\quad \hf^{(1)}_{\widehat{\phi^3} \widehat{\phi^3} \widehat{\phi^2}} =- \frac{5\mathcal{A}_p}{\sqrt{2}}~.
\end{align}

We can now plug these results into the OPE relations, and expanding in powers of $\eps$ we find that they can be satisfied if and only if
\begin{align}\label{finalOPE1l}
	\D^{(2)}_{\widehat{\phi^2}} = \frac{\mathcal{A}_p}{9}~,\quad R^{(2)}= -\frac{\Gamma \left(1-\frac{p}{4}\right) \Gamma \left(\frac{p}{2}\right)}{36 \sqrt{6} \Gamma \left(\frac{p}{4}+1\right)} \left [ 13\psi(\tfrac{p}{4}) + 3\psi(-\tfrac{p}{4}) - 8\psi(\tfrac{p}{2}) - 8\psi(1) \right ]~.
\end{align}
Again, we have reproduced the correct result for $\D_{\widehat{\phi^2}}$ -- see eq.~\eqref{anphi2} -- but this time we have also obtained the next-to-leading order prediction for $R(\aphisq)$.
Note that, via eq.~\eqref{bpbmsol_d}, the latter implies the following expansion for $\aphisq$:
\begin{align}
	\aphisq&=\eps^2 \aphisq^{(2)}+\eps^3\aphisq^{(3)}+O(\eps^4)~,\label{mean-field-1pt}\\
	\aphisq^{(2)} &= -\frac{\Gamma \left(\frac{p}{2}\right)^2}{3 \sqrt{6} \Gamma (p)} R^{(1)}~,\nonumber\\
	\aphisq^{(3)} &=-\frac{\Gamma(\tfrac{p}{2})^3 \Gamma(1 - \tfrac{p}{2}) \Gamma(1 - \tfrac{p}{4})}{216 \Gamma \left(\frac{p}{4} + 1\right) \Gamma (p+1)}  \left[4 - \frac{\Gamma(\tfrac{p}{4})^2\Gamma(1 - \tfrac{p}{4})^2}{\Gamma(\tfrac{p}{2})\Gamma(-\tfrac{p}{2})} - \frac{96 \sqrt{6} \Gamma \left(\frac{p}{4}+1\right)}{\Gamma \left(-\frac{p}{4}\right) \Gamma \left(\frac{p}{2}\right)}R^{(2)}\right]~.\nonumber
\end{align}
In particular, for $p=2$ and $p=3$ we get
\begin{align}
	\aphisq &= -\frac{1}{27}\eps^2+\left(\frac{1}{54}-\frac{8 \log 4}{81}\right)\eps^3+O(\eps^4) && (p = 2) \nonumber \\
	&= -0.037037\eps^2 - 0.118399\eps^3 + O(\eps^4)~, \nonumber \\
	\aphisq &= -\frac{\pi ^{5/2}}{576 \sqrt{2} \Gamma \left(\frac{7}{4}\right)^2}\eps^2+\frac{4 \sqrt{2 \pi }}{6075}(13 \pi -28-64 \log 2) \Gamma \left(\frac{9}{4}\right)^2\eps^3+O(\eps^4) && (p = 3) \nonumber \\
	&= -0.0254242 \eps^2 - 0.0667824 \eps ^3+O(\eps^4)~, \label{mean-field-1pt-num} 
\end{align}
which are analogous to \eqref{short-range-1pt}.

We can additionally use the OPE relations to compute anomalous dimensions of $\hphi^m$ operators.
We use eq.~\eqref{ssspinconstr} with $\hat{T}=\widehat{\phi^m}$, $\widehat{\mathcal{O}}=\widehat{\phi^{m-1}}$, as well as the $\eps$ expansion of $R(\aphisq)$ computed earlier. 
The relevant OPE coefficients all come from the same integrals which appeared in the derivation of \eqref{coeff1loopchk}. Working out the correct combinatorial factors, they read:
\begin{align}
	\hf_{\hphi \widehat{\phi^{m-1}} \widehat{\phi^m}}&=\sqrt{m}\left(1-\frac{\eps}{6} (m-1)  \mathcal{A}_p+O(\eps^2)\right)~,\nonumber\\
	\hf_{\widehat{\phi^3} \widehat{\phi^{m-1}} \widehat{\phi^m}}&={(m-1) \sqrt{\frac{3m}{2}}} \left(1-\frac{\eps}{6} (3m-4)  \mathcal{A}_p+O(\eps^2)\right)~. \label{phi-powers-1loop}
\end{align}
Plugging these into the OPE relations and expanding for small $\eps$, at one loop we find that
\begin{align}
	12 \left [ \D^{(1)}_{\widehat{\phi^{m-1}}}- \D^{(1)}_{\widehat{\phi^m}} \right ]+4 m-7 = 0~, \label{phi-power-recursion}
\end{align}
which allows $\D^{(1)}_{\widehat{\phi^0}}=0$ to be used as a boundary condition. Solving \eqref{phi-power-recursion} and also converting the result to an anomalous dimension yields:
\begin{align}
	\D^{(1)}_{\widehat{\phi^m}}&=\frac{m (2 m-5)}{12}~, \quad \gamma^{(1)}_{\widehat{\phi^m}} = \frac{m(m - 1)}{6}~.
\end{align}
Knowing \eqref{easiest-part}, this result can be written as $\gamma^{(1)}_{\widehat{\phi^m}} = \binom{m}{2} \gamma^{(1)}_{\widehat{\phi^2}}$ which is the same relation between one-loop anomalous dimensions that holds at the local Wilson-Fisher fixed point \cite{Rychkov:2015naa}. This is not surprising because Wick contractions in these theories have the same structure.
At two loops we find instead:
\begin{align}
	18 \left [ \D^{(2)}_{\widehat{\phi^{m - 1}}}- \D^{(2)}_{\widehat{\phi^m}} \right ]-(m-1) (3 m-8)\mathcal{A}_p =0~,
\end{align}
and so,
\begin{align}
	\gamma^{(2)}_{\widehat{\phi^m}} = \D^{(2)}_{\widehat{\phi^m}} &= -\frac{m(m-1)(m-3)}{18}\mathcal{A}_p~.\label{phi-power-2loop}
\end{align}
To our knowledge, for $m>4$, this quantity has not been computed before since the three-loop results of \cite{Benedetti:2020rrq} stop at $m=4$. Notice that \eqref{phi-power-2loop} vanishes for $m=1$ and $m=3$, as it should since the `$\pm$'
modes have protected scaling dimensions. The OPE relations allowed us to obtain a two-loop result using only CFT data at lower order.

\subsection{Generalization to long-range $O(N)$ models}
Long-range $O(N)$ models are on the same conceptual footing as the LRI model. It should therefore be no surprise that the calculations in the previous subsection can be generalized to the $O(N)$ case. We will demonstrate this only for the mean-field end, although we expect the same strategy to work for the short-range end.\footnote{To this end we would need extensive CFT data for the short-range $O(N)$ models. These are in principle available at large $N$. It would also be interesting to use the $O(2)$ model data from \cite{Liu:2020tpf} which was extracted using the techniques of \cite{Chester:2019ifh}.}

This time the defect modes of the free bulk scalar are in the fundamental of $O(N)$ and read
\begin{align}
	\psi_0^{(+,I)} = \hphi^I~, \quad \psi_0^{(-,I)} = \frac{1}{\sqrt{2(N + 2)}} (\hphi \cdot \hphi) \hphi^I~.\label{modesON}
\end{align}
The overall coefficients are chosen such that they are unit-normalized. For the long-range $O(N)$ models one can recover a set of exact OPE relations that is completely analogous to eq.~\eqref{ssspinconstr} with the same gamma functions.
We will exploit such OPE relations to compute anomalous dimensions of (unit-normalized) operators 
\begin{align}
	\Sigma_n = \frac{1}{2^n\sqrt{n!(N/2)_n}} (\hphi \cdot \hphi)^n~, \quad \W^I_n  = \frac{1}{2^n\sqrt{n!(N/2 + 1)_n}} (\hphi \cdot \hphi)^n \hphi^I~.\label{opsON}
\end{align}
The set of perturbative OPE coefficients we need are generalizations of \eqref{phi-powers-1loop}, which are computed in appendix \ref{app:FeynDiagON} up to $O(\eps^2)$ corrections. They read:\footnote{A three-point function with two $\W^I_n$-type operators is proportional to $\delta^{IJ}$, so we can drop the $I$ label in the corresponding OPE coefficient.}
\begin{align}\label{OPEcoeffstreenON1loop}
	\hf_{\W_0 \W_n \Sigma_{n}}&=\hf^{(0)}_{\W_0\W_n\Sigma_n}\left(1-\eps\frac{3n}{N+8}  \mathcal{A}_p+O(\eps^2)\right)~,\nonumber\\
	\hf_{\W_0 \W_n \Sigma_{n+1}}&=\hf^{(0)}_{\W_0\W_n\Sigma_{n+1}}\left(1-\eps\frac{6n+N+2}{2(N+8)}  \mathcal{A}_p+O(\eps^2)\right)~,\nonumber\\
	\hf_{\W_1 \W_n \Sigma_{n}}&=\hf^{(0)}_{\W_1\W_n\Sigma_{n}}\left(1-\eps\frac{  54 n+7 N-16 }{6 (N+8)}\mathcal{A}_p+O(\eps^2)\right)~,\nonumber\\
	\hf_{\W_1 \W_n \Sigma_{n+1}}&=\hf^{(0)}_{\W_1\W_n\Sigma_{n+1}}\left(1-\eps\frac{ n (54 n+17 N+28)+3 (N+2) }{ (N+8)(6 n+N+2)}\mathcal{A}_p+O(\eps^2)\right)~,
\end{align}
where the disconnected contributions are
\begin{align}
	\hf_{\W_0\W_n\Sigma_n}^{(0)} &=\frac{\sqrt{\left(\frac{N}{2}+1\right)_n}}{\sqrt{\left(\frac{N}{2}\right)_n}}~,\quad \hf_{\W_1 \W_n \Sigma_n}^{(0)} = \frac{3n \sqrt{2\left(\frac{N}{2}+1\right)_n}}{\sqrt{(N + 2)\left(\frac{N}{2}\right)_n}}~, \nonumber\\
	\hf_{\W_0\W_n\Sigma_{n+1}}^{(0)} &=\frac{\sqrt{(n + 1)\left(\frac{N}{2}+1\right)_n}}{\sqrt{\left(\frac{N}{2}\right)_{n+1}}}~, \quad \hf_{\W_1 \W_n \Sigma_{n+1}}^{(0)} = \frac{(6 n+N+2)\sqrt{2(n + 1)\left(\frac{N}{2}\right)_{n+1}}}{N \sqrt{(N + 2) \left(\frac{N}{2}+1\right)_n}}~.\label{OPEcoeffstreenON}
\end{align}

The calculation again starts with the minimal number of fields. Taking $n = 1$ and $\hat{T}=\Sigma_1$, eq.~\eqref{ssspinconstr2} gives expressions for the ratios $\hf_{\W_0 \W_0 \Sigma_1} / \hf_{\W_0 \W_1 \Sigma_1}$ and $\hf_{\W_1 \W_1 \Sigma_1} / \hf_{\W_0 \W_1 \Sigma_1}$. Demanding that they agree with the appropriate cases of \eqref{OPEcoeffstreenON} yields
\begin{align}
	R^{(0)}&=0~,\quad R^{(1)}=\frac{\sqrt{N+2}~,\Gamma \left(1-\frac{p}{4}\right) \Gamma \left(\frac{p}{2}\right)}{\sqrt{2} (N+8) \Gamma \left(\frac{p}{4}+1\right)}~, \nonumber \\
	R^{(2)} &=	\frac{\sqrt{N+2} \Gamma \left(1-\frac{p}{4}\right) \Gamma \left(\frac{p}{2}\right)}{4 \sqrt{2} (N+8)^3 \Gamma \left(\frac{p}{4}+1\right)}  \Bigg [ (N^2 - 64N - 288)\psi(\tfrac{p}{4}) - (N + 8)^2\psi(-\tfrac{p}{4}) \nonumber \\
	&+ 8(5N + 22)\psi(\tfrac{p}{2}) + 8(5N + 22)\psi(1) \Bigg ]~, \label{ratioON}
\end{align}
at $O(\eps^2)$, along with anomalous dimensions of $\Sigma_1$ which agree with \cite{Giombi:2019enr}. This results in the following one-point function coefficient of $\phi^2$:
\begin{align}
	a^{(2)}_{\phi^2} &= \frac{(N+2) \Gamma \left(1-\frac{p}{4}\right) \Gamma \left(\frac{p}{2}\right)^3}{2(N+8)^2 \Gamma(\tfrac{p}{4} + 1) \Gamma (p)}~, \nonumber \\
	a^{(3)}_{\phi^2} &=-\frac{(N+2) \Gamma(1 - \tfrac{p}{4}) \Gamma(\tfrac{p}{2})^3}{8(N+8)^4 \Gamma(\tfrac{p}{4} + 1) \Gamma (p)}  \Bigg[(N^2 - 144N - 640)\psi(\tfrac{p}{4}) - (N + 8)^2\psi(-\tfrac{p}{4}) \nonumber \\
	&+ 16(5N + 22)\psi(\tfrac{p}{2}) + 16(5N + 22)\psi(1)\Bigg]~. \label{hard-simp}
\end{align}

To go further and compute anomalous dimensions of \eqref{opsON}, we shall simply use eq.~\eqref{ssspinconstr}, once with $\hat{T}=\Sigma_n$ and $\widehat{\mathcal{O}}=\W_{n}$, and once with $\hat{T}=\Sigma_{n+1}$ and $\widehat{\mathcal{O}}=\W_{n}$. Using the known $R^{(1)}$, we obtain the coupled equations
\begin{align}
	& \D^{(1)}_{\W_n}= \D^{(1)}_{\Sigma_n}+\frac{6 n}{N+8}-\frac{1}{4}~, \nonumber \\
	& 0=2 (N+8)(\D^{(1)}_{\Sigma_n}-\D^{(1)}_{\Sigma_{n+1}})+24 n+N-4~,
\end{align}
at one loop. Their solution is:
\begin{align}\label{sol-1loop}
	\D^{(1)}_{\Sigma_n} = \frac{n   (12 n+N-16)}{2 (N+8)}~,\quad 		\D^{(1)}_{\W_n} = \frac{24 n^2+2 n (N-4)-N-8}{4 (N+8)}~.
\end{align}
Analogous equations can be found at two loops as $R^{(2)}$ is also known. They are solved by:
\begin{align}\label{sol-2loop}
	\D^{(2)}_{\Sigma_n} &=\frac{n \left[36 n^2 (N+8)+12 n (N^2-N-54)-N (19 N+58)+320\right]}{(N+8)^3}\mathcal{A}_p~,\nonumber\\
	\D^{(2)}_{\W_n} &=\frac{6 n (n-1) [6 n (N+8)+N (2 N+13)+12]}{(N+8)^3}\mathcal{A}_p~.
\end{align}
Once again, the virtue of the OPE relations is that they allow to gain one order in perturbation theory, as shown in these examples.

A few comments are in order. First, our results for $\D^{(1)}_{\Sigma_1}$ and $\D^{(2)}_{\Sigma_1}$ agree with equations (5.6) of \cite{Giombi:2019enr} and (3.20) of \cite{Benedetti:2020rrq}. The quantity $\D^{(2)}_{\Sigma_2}$ was also computed in \cite{Benedetti:2020rrq}, and we agree with their result. Second, both $\D^{(2)}_{\W_0}$ and $\D^{(2)}_{\W_1} $ vanish as they should, since the corresponding operators are protected defect modes. Finally, for $N=1$ and any $n$ we recover the results of subsection \ref{LRI2loops}.

\subsection{Taking stock}
Of the new results we have computed so far, the ones which will be used most immediately are \eqref{short-range-1pt} and \eqref{mean-field-1pt}. These express the one-point function $a_{\phi^2}$ near the short-range and mean-field end of the LRI respectively. Estimates of this quantity will prove to be an important guide for the numerical bootstrap. However, the conceptual novelty of this section is that it is based on new tools for streamlining perturbation theory in the long-range Ising and $O(N)$ models, namely the OPE relations in eq. \eqref{ssspinconstr}. These were powerful enough to provide new two-loop anomalous dimensions for $\Sigma_n$ and $\W_n^I$, defined in \eqref{opsON}.

When the special case $\gamma^{(2)}_{\Sigma_1}$ was first found in \cite{Giombi:2019enr}, more standard Feynman diagram calculations were used. It would have also been possible to find all $\gamma^{(2)}_{\Sigma_n}$ in this way because the required number of diagrams stabilizes for sufficiently large $n$. If desired, one can now turn the process around and predict the $O(\eps^{-1})$ parts of these new diagrams using our scaling dimensions \eqref{sol-2loop}. One might hope for a further generalization by including derivatives in \eqref{opsON}. The OPE relations explored here hold for operators with spin and they can certainly be used to compute their anomalous dimensions. The main obstacle one faces is that the linear combinations of derivatives which lead to a conformal primary are non-trivial. Finding a way around this was possible when we computed \eqref{andt} but only because the operator $[\hphi \hphi]_{n,\ell}$ was limited to two fields. With arbitrarily many fields, the presence of derivatives introduces a book-keeping exercise that seems hard to avoid with either the OPE relations or pre-bootstrap methods.

On the other hand, one statement of this form that we can make is $\lambda^{(1)}_{\hphi\hphi \hat{\mathcal{O}}} = 0$ when $\hat{\mathcal{O}}$ is an operator involving four fields $\hphi$ and potentially many derivatives.\footnote{This is in contrast to the local Wilson-Fisher fixed point which has $\lambda^{(1)}_{\hphi\hphi \hat{\mathcal{O}}} \neq 0$. The calculation leading to this simplification is similar to the one showing that $\gamma^{(2)}_{\hphi} = 0$ in the LRI but not the SRI.} This result will become important in the next section which discusses a method for extracting CFT data along a trajectory of operators in a four-point function. When computing $\langle \hphi(x_1) \hphi(x_2) \hat{\mathcal{O}}(x_3) \rangle$ to one loop, each distribution of derivatives is handled using the diagrams
\begin{align}\label{mirror-diagrams}
	\begin{tikzpicture}[baseline,valign]
		\draw (0,0) -- (1,0);
		\draw (-1,0) to[out=-90,in=180] (0,-1);
		\draw (0,-1) -- (0,0);
		\draw (0,-1) to[out=60,in=-60] (0,0);
		\draw (0,-1) to[out = 120,in= -120] (0,0);
		\node at (0,0) [bcirc] {};
		\node at (0,-1) [wcirc] {};
	\end{tikzpicture} \; + \; \begin{tikzpicture}[baseline,valign]
		\draw (0,0) -- (-1,0);
		\draw (1,0) to[out=-90,in=0] (0,-1);
		\draw (0,-1) -- (0,0);
		\draw (0,-1) to[out=60,in=-60] (0,0);
		\draw (0,-1) to[out = 120,in= -120] (0,0);
		\node at (0,0) [bcirc] {};
		\node at (0,-1) [wcirc] {};
	\end{tikzpicture}~.
\end{align}
Following the steps used for \eqref{andt}, these are computed using either
\begin{align}
	\int \frac{d^p\tau_0 \tau_{03}^{a_1} \dots \tau_{03}^{a_l}}{|\tau_{01}|^{\tfrac{1}{2}(p - \eps)} |\tau_{03}|^{\tfrac{3}{2}(p - \eps) + 2l}} &= \frac{\Gamma(p + l - \eps)}{\Gamma(\tfrac{1}{4}(p - \eps))\Gamma(\tfrac{3}{4}(p - \eps) + l)} \nonumber \\
	\times \int_0^1 dx \int d^p\tau_0 &\frac{x^{\tfrac{1}{4}(p - \eps) - 1} (1 - x)^{\tfrac{3}{4}(p - \eps) + l - 1}}{[\tau_0^2 + x(1 - x)\tau_{13}^2 ]^{p + l - \eps}} (\tau_0 + x\tau_{13})^{a_1} \dots (\tau_0 + x\tau_{13})^{a_l}~, \label{nice-vanishing}
\end{align}
or its image under $(1 \leftrightarrow 2)$. Without loss of generality, we have taken the factors of $\tau_{03}^a$ (if they are contracted at all) to be contracted with the factors of $\tau_{23}^a$ outside the integral rather than each other. While \eqref{nice-vanishing} clearly vanishes when $l$ is odd, a simple calculation shows that all $l/2$ terms of it are $O(\eps)$ when $l$ is even. This means that the three-point function, which includes $\lambda_*$, is $O(\eps^2)$.

\section{Inversion formula}
\label{sec:inversion}
One of our perturbative results in the previous section applies to an infinite family of operators labelled by spin -- namely $[\hat{\phi} \hat{\phi}]_{n, \ell}$ for $\ell > 0$. These are trajectories of double-twist operators built from the fundamental field. Many studies in recent years have computed analogous families of anomalous dimensions using an efficient technique that sidesteps most of the need for diagrammatic computations and implements analytic bootstrap for perturbative CFTs. This is the Lorentzian inversion formula \cite{Caron-Huot:2017vep}, designed to yield CFT data in a form that is analytic in spin down to a critical value determined by the growth of the correlator in the Regge limit. 

The main goal is as usual to extract CFT data, which is done by looking at the contribution of the operators in a four-point correlator through the crossing equation. We start from the non-trivial function $G(z, \bar{z})$ which appears in the four-point correlator of some (defect) scalar primary $\Psi$:
\begin{equation}
	\langle \Psi(\tau_1)\Psi(\tau_2)\Psi(\tau_3)\Psi(\tau_4) \rangle = \frac{G(z, \bar{z})}{|\tau_{12}|^{2\D_\Psi} |\tau_{34}|^{2\D_\Psi}} \, .
\end{equation}
Its arguments are related to the positions through
\begin{equation}\label{cross-ratios}
	U=z \bar{z} = \frac{\tau_{12}^2 \tau_{34}^2}{\tau_{13}^2 \tau_{24}^2} \,\quad \ V=(1-z)(1- \bar{z}) = \frac{\tau_{14}^2 \tau_{23}^2}{\tau_{13}^2 \tau_{24}^2} \ .
\end{equation}
By using the OPE twice, one can expand $G(z, \bar{z})$ in terms of the conformal blocks
\begin{equation}
	G(z, \bar{z}) =  \sum_{\Delta_{\mathcal{O}}, \ell_{\mathcal{O}}} a_{\Delta_{\mathcal{O}}, \ell_{\mathcal{O}}} \ g_{\Delta_{\mathcal{O}},\ell_{\mathcal{O}}}(z,\bar{z}) \ ,
\end{equation}
where $a_{\Delta_{\mathcal{O}}, \ell_{\mathcal{O}}} = \lambda^2_{\Psi \Psi \mathcal{O}}$ is the OPE coefficient squared, and the sum runs over all possible exchanged primaries, telling us which operators contribute to the correlator. The conformal blocks are theory-independent and arise as solutions of the quadratic Casimir equation. In $p=2$, the differential operator factorizes, $\mathcal{D}=\mathcal{D}_z + \mathcal{D}_{\bar{z}}$, and as a consequence, the conformal blocks factorize as well:
\begin{equation}
	g_{\Delta,\ell}(z,\bar{z}) = \frac{k_{\Delta-\ell}(z) k_{\Delta+\ell}(\bar{z}) + (z \leftrightarrow \bar{z})}{1+\delta_{\ell 0}} \ ,
\end{equation}
where
\begin{equation}
	k_\beta(x) = x^{\beta/2} {_2F}_1 \left (\tfrac{\beta}{2}, \tfrac{\beta}{2}, \beta;x \right ) \ .
\end{equation}
In the framework of the inversion formula, detailed knowledge of the correlator is optional for applying it as long as one has access to a simpler piece called the double discontinuity. It is defined as the expectation value of a squared commutator in real Minkowski spacetime. Equivalently, it can also be written as the difference between the Euclidean correlator and its two possible analytic continuations around $\bar{z}=1$:\footnote{In doing this, one treats $z$ and $\bar{z}$ as independent variables and rotate around the $\bar{z} = 1$ branch point.}
\begin{equation}
	\mathrm{dDisc}[G(z, \bar{z})] = G(z, \bar{z}) - \frac{1}{2} G^{\circlearrowright}(z, \bar{z}) - \frac{1}{2} G^{\circlearrowleft}(z, \bar{z})~. \label{ddisc}
\end{equation}
Specializing to identical external scalars in $p = 2$, the inversion integral is:
\begin{align}
	c(\Delta, \ell) &= \frac{1 + (-1)^\ell}{4} \frac{\Gamma(\tfrac{\Delta + \ell}{2})^4}{2\pi^2 \Gamma(\Delta + \ell) \Gamma(\Delta + \ell - 1)} \int_0^1 \frac{dz}{z^2} \int_0^1 \frac{d\bar{z}}{\bar{z}^2} g_{\Delta - 1, \ell + 1}(z, \bar{z}) \text{dDisc}[G(z, \bar{z})] \nonumber \\
	&= \frac{\Gamma(\tfrac{\Delta + \ell}{2})^4}{2\pi^2 \Gamma(\Delta + \ell) \Gamma(\Delta + \ell - 1)} \int_0^1 \frac{dz}{z^2} \int_0^1 \frac{d\bar{z}}{\bar{z}^2} k_{\ell - \Delta + 2}(z) k_{\Delta + \ell}(\bar{z}) \text{dDisc}[G(z, \bar{z})]~, \label{lif}
\end{align}
where we integrate the double discontinuity of the reduced correlator over a compact domain. The kernel is an unphysical conformal block and the prefactor comes from the fact that 
\begin{equation}
	c(\Delta,\ell) = c^t(\Delta,\ell)+(-1)^\ell c^u(\Delta,\ell) \ .
\end{equation} 
In the case of identical external scalars, the two contributions are the same and one extracts the direct channel CFT data by looking at the crossed channel. Moreover, we will set $(-1)^\ell$ to $1$ since all exchanged operators have even spin. 

The spectral density encodes the direct channel OPE data through its poles, so that
\begin{equation}
	c(\Delta, \ell) = -\sum_{\mathcal{O}} \frac{\lambda^2_{\Psi \Psi \mathcal{O}}}{\Delta - \Delta_\mathcal{O}}~, \label{spectral-density}
\end{equation}
where $\Delta_{\mathcal{O}}$ is the dimension of the exchanged operator. 
Our main goal in this section is to improve upon the $p = 2$ case of our results for $[\hat{\phi} \hat{\phi}]_{n, \ell}$ by extending them to three loops.

\subsection{Overview}

When applying the Lorentzian inversion formula, a useful fact is that double-twist operators produce integer powers of $1 - \bar{z}$ in the crossed channel which have a vanishing double discontinuity. This means that if $\Psi$ is a generalized free field, all contributions to \eqref{lif} come from $(1 - \bar{z})^{-\Delta_\Psi}$ associated with the identity operator. As we turn on deformations, however, other non-trivial terms (referred to as Casimir singular in \cite{Simmonsduffin:2017}) start to appear. These include other fractional powers of $1 - \bar{z}$ and logarithms appearing quadratically or higher. These satisfy
\begin{align}
	\text{dDisc}[(1 - \bar{z})^\xi] &= (1 - \bar{z})^\xi 2 \sin^2(\pi \xi)~, \nonumber \\
	\text{dDisc}[\log(1 - \bar{z})^2] &= 4\pi^2~, \nonumber \\
	\text{dDisc}[\log(1 - \bar{z})^3] &= 12\pi^2 \log(1 - \bar{z})~. \label{ddiscs}
\end{align}
We will take $\Psi = \hat{\phi}$, which allows the deformation to be parameterized by $\varepsilon$. CFT data will then be expanded as:
\begin{align}
	a_{\mathcal{O}} &= a_{\mathcal{O}}^{(0)} + \varepsilon a_{\mathcal{O}}^{(1)} + \varepsilon^2 a_{\mathcal{O}}^{(2)} + \varepsilon^3 a_{\mathcal{O}}^{(3)} + O(\varepsilon^4)~, \nonumber \\
	\Delta_{\mathcal{O}} &= \Delta_{\mathcal{O}}^{(0)} + \varepsilon \gamma_{\mathcal{O}}^{(1)} + \varepsilon^2 \gamma_{\mathcal{O}}^{(2)} + \varepsilon^3 \gamma_{\mathcal{O}}^{(3)} + O(\varepsilon^4)~, \label{ag-expansions}
\end{align}
where the OPE coefficients are fixed by the canonical normalization of the operators. We also remind the reader that $\Delta_{\mathcal{O}}^{(0)}$ and $a_{\mathcal{O}}^{(0)}$ in this expansion depend on $\varepsilon$ in our theory. Plugging \eqref{ag-expansions} into \eqref{spectral-density}, the behaviour of the spectral density as $\Delta \to \Delta_\mathcal{O}^{(0)}$ is:
\begin{align}
	-c(\Delta, \ell) &\sim \frac{\langle a_{\mathcal{O}}^{(0)} \rangle}{\Delta - \Delta_\mathcal{O}^{(0)}} + \varepsilon \left [ \frac{\langle a_{\mathcal{O}}^{(1)} \rangle}{\Delta - \Delta_\mathcal{O}^{(0)}} + \frac{\langle a_{\mathcal{O}}^{(0)} \gamma_{\mathcal{O}}^{(1)} \rangle}{(\Delta - \Delta_\mathcal{O}^{(0)})^2} \right ] \nonumber \\
	+ \varepsilon^2 &\left [ \frac{\langle a_{\mathcal{O}}^{(2)} \rangle}{\Delta - \Delta_\mathcal{O}^{(0)}} + \frac{\langle a_{\mathcal{O}}^{(0)} \gamma_{\mathcal{O}}^{(2)} + a_{\mathcal{O}}^{(1)} \gamma_{\mathcal{O}}^{(1)} \rangle}{(\Delta - \Delta_\mathcal{O}^{(0)})^2} + \frac{\langle a_{\mathcal{O}}^{(0)} \gamma_{\mathcal{O}}^{(1)2} \rangle}{(\Delta - \Delta_\mathcal{O}^{(0)})^3} \right ]  \label{expanded-density} \\
	+ \varepsilon^3 &\left [ \frac{\langle a_{\mathcal{O}}^{(3)} \rangle}{\Delta - \Delta_\mathcal{O}^{(0)}} + \frac{\langle a_{\mathcal{O}}^{(0)} \gamma_{\mathcal{O}}^{(3)} + a_{\mathcal{O}}^{(1)} \gamma_{\mathcal{O}}^{(2)} + a_{\mathcal{O}}^{(2)} \gamma_{\mathcal{O}}^{(1)} \rangle}{(\Delta - \Delta_\mathcal{O}^{(0)})^2} + \frac{\langle 2a_{\mathcal{O}}^{(0)} \gamma_{\mathcal{O}}^{(1)} \gamma_{\mathcal{O}}^{(2)} + a_{\mathcal{O}}^{(1)} \gamma_{\mathcal{O}}^{(1)2} \rangle}{(\Delta - \Delta_\mathcal{O}^{(0)})^3} + \frac{\langle a_{\mathcal{O}}^{(0)} \gamma_{\mathcal{O}}^{(1)3} \rangle}{(\Delta - \Delta_\mathcal{O}^{(0)})^4} \right ]~. \nonumber
\end{align}
Since there will be generically many operators with dimension $\Delta_\mathcal{O}^{(0)} + O(\varepsilon)$, we have used the $\langle \dots \rangle$ notation to refer to a sum over this degenerate eigenspace. In the following, we will refer to the coefficients of $\varepsilon^n$ in \eqref{expanded-density} as $c^{(n)}(\Delta,\ell)$.

One can now go through the same procedure and expand the four-point function to find
\begin{align}
	G(&z, \bar{z}) = \sum_{\Delta^{(0)}_{\mathcal{O}}, \ell_{\mathcal{O}}} \Bigg[ \langle a^{(0)}_\mathcal{O} \rangle + \varepsilon \langle a^{(1)}_\mathcal{O} + a^{(0)}_\mathcal{O} \gamma^{(1)}_\mathcal{O} \partial_\Delta \rangle \nonumber \\
	&+ \varepsilon^2 \langle a^{(2)}_\mathcal{O} + \left ( a^{(1)}_\mathcal{O} \gamma^{(1)}_\mathcal{O} + a^{(0)}_\mathcal{O} \gamma^{(2)}_\mathcal{O} \right ) \partial_\Delta + \frac{1}{2} a^{(0)}_\mathcal{O} \gamma^{(1)2}_\mathcal{O} \partial^2_\Delta \rangle  \nonumber \\
	&+ \varepsilon^3 \Bigg < a^{(3)}_\mathcal{O} + \left ( a^{(2)}_\mathcal{O} \gamma^{(1)}_\mathcal{O} + a^{(1)}_\mathcal{O} \gamma^{(2)}_\mathcal{O} + a^{(0)}_\mathcal{O} \gamma^{(3)}_\mathcal{O} \right ) \partial_\Delta + \left ( \frac{1}{2} a^{(1)}_\mathcal{O} \gamma^{(1)2}_\mathcal{O} + a^{(0)}_\mathcal{O} \gamma^{(1)}_\mathcal{O} \gamma^{(2)}_\mathcal{O} \right ) \partial^2_\Delta + \frac{1}{6} a^{(0)}_\mathcal{O} \gamma^{(1)3}_\mathcal{O} \partial^3_\Delta \Bigg > \nonumber\\
	&+  O(\varepsilon^4) \Bigg] \left | \frac{z}{1 - z} \right |^{2\Delta_\phi} g_{\Delta_{\mathcal{O}}^{(0)},l_{\mathcal{O}}} (1 - z, 1 - \bar{z})~, \label{expanded-blocks}
\end{align}
for the crossed channel conformal block expansion. Notice that in the usual convention, the contribution of the identity is singled out, while here we are including it in the $O(\varepsilon^0)$ term.

If we temporarily focus on double-twist operators, all contributions to the double discontinuity come from terms with two or more $\partial_\Delta$ derivatives. Their coefficients at $O(\varepsilon^n)$ only involve CFT data up to $O(\varepsilon^{n - 1})$. A naive reading of \eqref{expanded-blocks} therefore suggests an iterative procedure wherein the results of the inversion formula at one order are fed back into it at the next. At high enough orders, this logic breaks down for three reasons:
\begin{enumerate}
	\item Multi-twist operators will eventually be exchanged, which can produce a double discontinuity both with and without two derivatives. 
	\item Even for double-twist operators, those of low spin are not captured by the inversion formula and need to be put in by hand.
	\item Averages like $\langle a^{(0)}_\mathcal{O} \gamma^{(1)2}_\mathcal{O} \rangle$ are only known in terms of $\langle a^{(0)}_\mathcal{O} \rangle$ and $\langle a^{(0)}_\mathcal{O} \gamma^{(1)}_\mathcal{O} \rangle$ if one works with a large system of correlators to resolve the degeneracy.
\end{enumerate}
For us, new types of operators will appear in a controlled way and degeneracy will not be an obstacle until after three loops. This differs from the situation in holographic theories which are more severely affected by degeneracy because they involve infinitely many generalized free fields \cite{AldayBissi,adhp1,adhp2,Alday:2017}.

\subsection{Results at two loops}

We will now make the above statements more precise. To start, the $\varepsilon = 0$ OPE (apart from the identity) contains only double-twist operators $[\hat{\phi} \hat{\phi}]_{n,\ell}$ labelled by $n \geq 0$ and even $\ell \geq 0$. These have conformal dimensions $\Delta^{(0)}_{n, \ell} = 2\Delta_\phi + 2n + \ell$ and squared OPE coefficients
\begin{align}
	a^{(0)}_{n, \ell} = \frac{1 + (-1)^\ell}{n! (n + \ell)!} \frac{\Gamma(\Delta_\phi + n)^2 \Gamma(\Delta_\phi + n + \ell)^2}{\Gamma(\Delta_\phi)^4} \frac{\Gamma(2\Delta_\phi + n - 1) \Gamma(2\Delta_\phi + n + \ell - 1)}{\Gamma(2\Delta_\phi + 2n - 1) \Gamma(2\Delta_\phi + 2n + 2\ell - 1)}~. \label{gff-ope}
\end{align}
We can derive \eqref{gff-ope} by inverting the identity or by using the original brute-force methods of \cite{Heemskerk:2009pn}. It is now possible to say that $a^{(0)}_\mathcal{O}$ and $a^{(1)}_\mathcal{O}$ are only non-zero for double-twist operators but we can actually make a stronger statement. After inserting $\lambda \hat{\phi}^4$ once, the naive expectation is that a quadruple-twist operator $\mathcal{O}$ will have $\lambda_{\hat{\phi} \hat{\phi} \mathcal{O}} = O(\varepsilon)$. In fact, our calculation in the last section showed that $\lambda_{\hat{\phi} \hat{\phi} \mathcal{O}} = O(\varepsilon^2)$, which means that also $a^{(2)}_\mathcal{O}$ and $a^{(3)}_\mathcal{O}$ are only non-zero for double-twist operators. The second major simplification is that $\gamma^{(1)}_{n, \ell}$ is only non-zero for $n = \ell = 0$. The necessity of $\ell = 0$ is already clear from the fact that there is no double discontinuity at $O(\varepsilon)$. To prove the more general claim, consider
\begin{align}
	\langle [\hat{\phi} \hat{\phi}]_{n,\ell}(\tau_1) [\hat{\phi} \hat{\phi}]_{n,\ell}(\tau_2) \hphi^4(\tau_3) \rangle = \sum_{i, j} \langle D_A^{(i)} \phi D_B^{(i)} \phi(\tau_1) D_A^{(j)} \phi D_B^{(j)} \phi(\tau_2) \hphi^4(\tau_3) \rangle~, \label{3pt-regularity}
\end{align}
where $D_A$ and $D_B$ are differential operators. They ultimately act on $|\tau_{13}|^{-2\Delta_\phi}$ and $|\tau_{23}|^{-2\Delta_\phi}$ after we apply Wick's theorem to the right hand side of \eqref{3pt-regularity}. This three-point function is therefore regular as $\tau_1 \to \tau_2$. This property is only consistent with the Polyakov form
\begin{align}
	\langle [\hat{\phi} \hat{\phi}]_{n,\ell}(\tau_1) [\hat{\phi} \hat{\phi}]_{n,\ell}(\tau_2) \hphi^4(\tau_3) \rangle = \frac{\lambda_{[\hat{\phi} \hat{\phi}]_{n,\ell} [\hat{\phi} \hat{\phi}]_{n,\ell} \hphi^4}}{|\tau_{12}|^{4n + 2\ell} |\tau_{13}|^{4\Delta_\phi} |\tau_{23}|^{4\Delta_\phi}}~, \label{3pt-polyakov}
\end{align}
if $\lambda_{[\hat{\phi} \hat{\phi}]_{n,\ell} [\hat{\phi} \hat{\phi}]_{n,\ell} \hphi^4}$ vanishes whenever $n$ and $\ell$ are not both zero. First-order perturbation theory then implies that the same holds for $\gamma^{(1)}_{n, \ell}$. An alternative way to prove that $\gamma^{(1)}_{n,\ell} \propto \delta_{n, 0} \delta_{\ell, 0}$ is discussed in appendix~\ref{ss:bulk-2pt}.

Returning to \eqref{expanded-blocks}, we now only need to input two quantities to solve for the double discontinuity at $O(\varepsilon^2)$. These are:
\begin{equation}
	a^{(0)}_{\hat{\phi}^2} = 2~, \quad \gamma^{(1)}_{\hat{\phi}^2} = \frac{1}{3}~. \label{data-start}
\end{equation}
If we now let $\partial_\Delta^2$ act on the overall power law in $\left | \frac{z}{1 - z} \right |^{2\Delta_\phi} g_{\Delta, 0}(1 - z, 1 - \bar{z}) \bigl |_{\Delta = 2\Delta_\phi}$ (as opposed to the hypergeometric functions), it follows that the desired term in the spectral density is:
\begin{align}
	c^{(2)}(\Delta, \ell) &= \frac{\Gamma(\tfrac{\Delta + \ell}{2})^4}{18 \Gamma(\Delta + \ell) \Gamma(\Delta + \ell - 1)} \nonumber \\
	&\times \int_0^1 \frac{dz}{z^2} \int_0^1 \frac{d\bar{z}}{\bar{z}^2} k_{\ell - \Delta + 2}(z) k_{\Delta + \ell}(\bar{z}) \left | \frac{z}{1 - z} \right |^{1 - \frac{\varepsilon}{2}} |k_{1 - \frac{\varepsilon}{2}}(1 - z)|^2~. \label{density-we}
\end{align}
If we are content with extracting CFT data to an accuracy of $O(\varepsilon^2)$, it is consistent to set $\varepsilon = 0$ in the integrand and evaluate
\begin{align}
	c^{(2)}(\Delta, \ell) \bigl |_{\varepsilon = 0} &= \frac{\Gamma(\tfrac{\Delta + \ell}{2})^4}{18 \Gamma(\Delta + \ell) \Gamma(\Delta + \ell - 1)} \int_0^1 \frac{d\bar{z}}{\bar{z}^2} \bar{z}^{\frac{\Delta + \ell}{2}} {}_2F_1(\tfrac{\Delta + \ell}{2}, \tfrac{\Delta + \ell}{2}; \Delta + \ell; \bar{z}) \bar{z}^{\frac{1}{2}} {}_2F_1(\tfrac{1}{2}, \tfrac{1}{2}; 1; 1 - \bar{z}) \nonumber \\
	&\times \int_0^1 \frac{dz}{z^2} z^{\frac{\ell - \Delta + 2}{2}} {}_2F_1(\tfrac{\ell - \Delta + 2}{2}, \tfrac{\ell - \Delta + 2}{2}; \ell - \Delta + 2; z) z^{\frac{1}{2}} {}_2F_1(\tfrac{1}{2}, \tfrac{1}{2}; 1; 1 - z)~. \label{density-woe}
\end{align}
The factored integrals can be done in at least three ways:
\begin{enumerate}
	\item By using the Mellin-Barnes representation of the hypergeometric function ${_2F_1}$ and picking the proper contour, it is possible to evaluate the related double sum and find $\frac{\Gamma(\Delta + \ell)}{\Gamma \left ( \tfrac{1}{2}(\Delta + \ell) + 1 \right )^2} \left ( \frac{\Delta + \ell}{\Delta + \ell - 1} \right )^2$ for the first integral. The second is simply the shadow with $\Delta \mapsto 2 - \Delta$. 
	\item One can get the same result by exploiting the inner product which makes the quadratic Casimir $\mathcal{D}$ of $\mathfrak{sl}(2)$ self-adjoint \cite{Hogervorst:2017sfd}. Since each integral is an inner product of two eigenfunctions, one with eigenvalue $-\frac{1}{4}$ and the other with eigenvalue $h(h - 1)$ for $h \in \{ \tfrac{\Delta + \ell}{2}, \tfrac{\ell - \Delta + 2}{2} \}$, we can use $\langle f, g \rangle = \frac{1}{h(h - 1) + \tfrac{1}{4}} \left [ \langle \mathcal{D}f, g \rangle - \langle f, \mathcal{D}g \right \rangle ]$ which localizes to a boundary term.
	\item Finally, \eqref{density-we} is nothing but $\frac{\Gamma \left ( \tfrac{\Delta + \ell}{2} \right )^4}{18 \Gamma(\Delta + \ell) \Gamma(\Delta + \ell - 1)} \mathcal{I}(\ell - \Delta + 2, \tfrac{2 - \varepsilon}{2}, \tfrac{2 - \varepsilon}{4}) \mathcal{I}(\Delta + \ell, \tfrac{2 - \varepsilon}{2}, \tfrac{2 - \varepsilon}{4})$, where $\mathcal{I}(\Delta_s, \Delta_t, \Delta_\phi)$ is the crossing kernel in one dimension \cite{Hogervorst:2017sfd,Liu:2020tpf,Gopakumar:2018xqi}.
\end{enumerate}
It is worth reviewing the crossing kernel since this is the approach that worked for evaluating \eqref{density-we} even with $\varepsilon \neq 0$. This will be crucial for the three-loop computations in the next subsection.

Conformal blocks for $\mathfrak{sl}(2)$ are simply $k_{\Delta_s}(z)$ for the $s$-channel and $\left ( \frac{z}{1 - z} \right )^{\Delta_\phi} k_{\Delta_t}(1 - z)$ for the $t$-channel. The crossing kernel projects one onto the other. Using the techniques of \cite{Gopakumar:2018xqi}, it becomes
\begin{align}
	\mathcal{I}(\Delta_s, \Delta_t, \Delta_\phi) &= \int_0^1 \frac{dz}{z^2} k_{\Delta_s}(z) \left ( \frac{z}{1 - z} \right )^{\Delta_\phi} k_{\Delta_t}(1 - z) \nonumber  \\
	&= \frac{\Gamma(\Delta_s) \Gamma(\Delta_t)}{\Gamma(\tfrac{\Delta_s}{2})^2 \Gamma(\tfrac{\Delta_t}{2})^2} \int_{-i\infty}^{i\infty} \frac{ds dt}{2\pi i}K(s,t) \delta(\Delta_\phi + \tfrac{\Delta_s - \Delta_t}{2} - 1 + s - t)~, \label{kernel-def}
\end{align}
where
\begin{align}
	K(s,t)\equiv \Gamma(-s) \Gamma(-t) \frac{\Gamma(\tfrac{\Delta_s}{2} + s)^2}{\Gamma(\Delta_s + s)} \frac{\Gamma(\tfrac{\Delta_t}{2} + t)^2}{\Gamma(\Delta_t + t)}~,
\end{align}
which leaves a single Mellin-Barnes integral. The formula
\begin{align}
	& {}_7F_6 \left [ \begin{tabular}{ccccccc} $a$ & $1 + \frac{1}{2} a$ & $b$ & $c$ & $d$ & $e$ & $f$ \\ $\frac{1}{2} a$ & $1 + a - b$ & $1 + a - c$ & $1 + a - d$ & $1 + a - e$ & $1 + a - f;$ & $1$ \end{tabular} \right ] \label{7f6-form} \\
	&= \frac{\Gamma(1 + a - b) \Gamma(1 + a - c)\Gamma(1 + a - d) \Gamma(1 + a - e) \Gamma(1 + a - f)}{\Gamma(1 + a) \Gamma(b) \Gamma(c) \Gamma(d) \Gamma(1 + a - c - d) \Gamma(1 + a - b - d) \Gamma(1 + a - b - c) \Gamma(1 + a - e - f)} \nonumber \\
	& \times\int_{-i\infty}^{i\infty} \frac{du}{2\pi i} \frac{\Gamma(-u) \Gamma(1 + a - b - c - d - u) \Gamma(b + u) \Gamma(c + u) \Gamma(d + u) \Gamma(1 + a - e - f + u)}{\Gamma(1 + a - e + u) \Gamma(1 + a - f + u)}~, \nonumber
\end{align}
is then available for turning \eqref{kernel-def} into a single, very well-poised hypergeometric function evaluated at $1$. This can be done in four different ways. One can choose $1 + a - e - f + u$ to be either $\frac{\Delta_s}{2} + s$ or $\frac{\Delta_t}{2} + t$ and also eliminate either $s$ or $t$ in \eqref{kernel-def}. We will make $\frac{\Delta_s}{2} + s$ privileged and eliminate $t$ since this is the only choice whose hypergeometric function approaches $1$ as $\varepsilon \to 0$. The other choices would significantly disguise the simplicity of the final result. Accordingly, we will write:
{\footnotesize
	\begin{align}
		& \mathcal{I}(\Delta_s, \Delta_t, \Delta_\phi) = \frac{\Gamma(\Delta_t) \Gamma(\tfrac{\Delta_s}{2} + \Delta_\phi - 1)^2 \Gamma(\tfrac{\Delta_t}{2} - \Delta_\phi + 1) \Gamma(\Delta_s + \tfrac{\Delta_t}{2} + \Delta_\phi - 1)}{\Gamma(\tfrac{\Delta_s + \Delta_t}{2})^2 \Gamma(\tfrac{\Delta_s + \Delta_t}{2} + \Delta_\phi - 1)^2} \times \label{kernel-pref} \\
		& {}_7F_6 \left [ \begin{tabular}{ccccccc} $\frac{2\Delta_s + \Delta_t + 2\Delta_\phi - 4}{2}$ & $\frac{2\Delta_s + \Delta_t + 2\Delta_\phi}{4}$ & $\frac{\Delta_s + 2\Delta_\phi - 2}{2}$ & $\frac{\Delta_s + 2\Delta_\phi - 2}{2}$ & $\frac{\Delta_s}{2}$ & $\frac{\Delta_s}{2}$ & $\frac{\Delta_t + 2\Delta_\phi - 2}{2}$ \\ $\frac{2\Delta_s + \Delta_t + 2\Delta_\phi - 4}{4}$ & $\Delta_s$ & $\frac{\Delta_s + \Delta_t}{2}$ & $\frac{\Delta_s + \Delta_t}{2}$ & $\frac{\Delta_s + \Delta_t + 2\Delta_\phi - 2}{2}$ & $\frac{\Delta_s + \Delta_t + 2\Delta_\phi - 2}{2};$ & $1$ \end{tabular} \right ]~. \nonumber
	\end{align}
	\normalsize}
Mixed correlators have more than a four-fold ambiguity and sometimes it is even best to forgo \eqref{7f6-form} altogether and instead simplify the crossing kernel by manipulating the Mellin-Barnes integral directly \cite{Behan:2021pzk}.

The purely two-loop result 
\begin{align}
	\varepsilon^2 c^{(2)}(\Delta, \ell) \bigl |_{\varepsilon = 0} &= \frac{\varepsilon^2}{9} \frac{\Gamma(\tfrac{\Delta + \ell}{2})^4}{2 \Gamma(\Delta + \ell) \Gamma(\Delta + \ell - 1)} \mathcal{I}(\ell - \Delta + 2, 1, \tfrac{1}{2}) \mathcal{I}(\Delta + \ell, 1, \tfrac{1}{2}) \nonumber \\
	&= \frac{\varepsilon^2}{9} \frac{\Gamma(\tfrac{\Delta + \ell}{2})^4}{2 \Gamma(\Delta + \ell - 1)} \frac{ \Gamma(\ell - \Delta + 2)}{\Gamma(\tfrac{\Delta + \ell}{2} + 1)^2 \Gamma(\tfrac{\ell - \Delta + 2}{2} + 1)^2} \frac{(\Delta + \ell)^2 (\ell - \Delta + 2)^2}{(\Delta + \ell - 1)^2 (\ell - \Delta + 1)^2}~, \label{final-2loop}
\end{align}
has a double pole as $\Delta \to \ell + 1$. Expanding around this pole and comparing the result to \eqref{expanded-density} gives:
\begin{equation}
	a^{(2)}_{0, \ell} = -\frac{2}{9} \frac{\Gamma(\ell + \tfrac{1}{2})^2}{\pi \ell^2 \Gamma(2\ell)} \left [ \psi(2\ell) - \psi(\ell) - 4\log 2 - \ell^{-1} \right ], \quad \gamma^{(2)}_{0, \ell} = -\frac{1}{9\ell}~. \label{dim-ope-2loop}
\end{equation}
The anomalous dimension (which required us to divide by \eqref{gff-ope}) agrees with the diagrammatic result. Interestingly, \eqref{final-2loop} only has single poles at $\ell + 2n + 1$ for $n > 0$. We therefore see that the subleading double twists $[\hat{\phi} \hat{\phi}]_{n,\ell}$, despite being independent operators not subject to any equation of motion, have vanishing anomalous dimensions to two loops. On the other hand, their OPE coefficients receive the correction
\begin{equation}
	a^{(2)}_{n, \ell} = -\frac{1}{18} \frac{n^{-2} (n + \ell)^{-2} \Gamma(n + \ell + \tfrac{1}{2})^2}{\Gamma(2n) \Gamma(2n + 2\ell) \Gamma(\tfrac{1}{2} - n)^2}~, \label{ope-2loop}
\end{equation}
which can be seen from the coefficient of the single pole.

\subsection{Results at three loops}

By looking at the expansion of the four-point correlator in \eqref{expanded-blocks}, the most obvious terms that contribute to the double discontinuity at $O(\varepsilon^3)$ are\footnote{Notice that the third derivative of the conformal block will lead to two terms entering the double discontinuity.} 
\begin{equation}
	\sum_{\Delta^{(0)}_{\mathcal{O}}, \ell_{\mathcal{O}}} \varepsilon^3 \langle \left ( \frac{1}{2} a^{(1)}_\mathcal{O} \gamma^{(1)2}_\mathcal{O} + a^{(0)}_\mathcal{O} \gamma^{(1)}_\mathcal{O} \gamma^{(2)}_\mathcal{O} \right ) \partial^2_\Delta + \frac{1}{6} a^{(0)}_\mathcal{O} \gamma^{(1)3}_\mathcal{O} \partial^3_\Delta \rangle \left |\frac{z}{1 - z} \right |^{2\Delta_\phi} g_{\Delta_{\mathcal{O}}^{(0)},l_{\mathcal{O}}} (1 - z, 1 - \bar{z}) \ . \label{3loop}
\end{equation}
As explained before, we have kept terms containing at least two derivatives of the conformal blocks, since each differentiation produces one logarithm.
In \eqref{3loop}, we set $\varepsilon = 0$ in $\Delta_{\mathcal{O}}^{(0)}$ to not spoil the overall order in $\varepsilon$. Given that $a_{\mathcal{O}}^{(0)}$ and $a_{\mathcal{O}}^{(1)}$ are non-zero only for double-twist operators, and all terms of \eqref{3loop} include $\gamma^{(1)}_{n,\ell} \propto \delta_{n, 0} \delta_{\ell, 0}$, we can still compute the double discontinuity at this order from the CFT data of $\hat{\phi}^2$. These are:
\begin{equation}
	a^{(0)}_{\hat{\phi}^2} = 2, \quad a^{(1)}_{\hat{\phi}^2}=\frac{4}{3} \left [\psi\left(\frac{1}{2}\right)-\psi(1)\right ], \quad \gamma^{(1)}_{\hat{\phi}^2} = \frac{1}{3} , \quad \gamma^{(2)}_{\hat{\phi}^2} = \frac{2}{9} \left [\psi(1)-\psi\left(\frac{1}{2}\right) \right ]~. \label{data-start2}
\end{equation}
In addition to \eqref{3loop}, we have to consider the $O(\varepsilon^2)$ parts of \eqref{expanded-blocks} again because the scaling dimensions $\Delta_{\mathcal{O}}^{(0)}$ in the blocks being differentiated have $O(\varepsilon)$ terms. The result is a contribution to the double discontinuity at this order which is the same as \eqref{density-we}, but now keeping the $O(\varepsilon)$ correction. All together we find:
\begin{align}
	\varepsilon^3 &\bigg [c^{(3)}(\Delta, \ell)+\frac{\partial}{\partial \varepsilon} c^{(2)}(\Delta,\ell)  \bigg] \biggl |_{\varepsilon = 0} =  \frac{\varepsilon^3}{54} \frac{\Gamma(\tfrac{\Delta + \ell}{2})^4}{\Gamma(\Delta + \ell) \Gamma(\Delta + \ell - 1)} \nonumber\\
	&\times\int_0^1 \frac{d\bar{z}}{\bar{z}^2} \int_0^1 \frac{dz}{z^2} \bar{z}^{\frac{\Delta + \ell+1}{2}} {}_2F_1(\tfrac{\Delta + \ell}{2}, \tfrac{\Delta + \ell}{2}; \Delta + \ell; \bar{z})z^{\frac{\ell - \Delta + 3}{2}} {}_2F_1(\tfrac{\ell - \Delta + 2}{2}, \tfrac{\ell - \Delta + 2}{2}; \ell - \Delta + 2; z)  \nonumber \\
	&\times  \bigg \{ \big ( \mathcal{A}_2+\log |1-z| -\tfrac{3}{2} \log|z| \big ) {}_2F_1(\tfrac{1}{2}, \tfrac{1}{2}; 1; 1 - \bar{z}) {}_2F_1(\tfrac{1}{2}, \tfrac{1}{2}; 1; 1 - z)  \nonumber \\
	&-\frac{1}{2} \eval{\partial_{\Delta_{\phi^2}} \left [{}_2F_1(\tfrac{\Delta_{\phi^2}}{2}, \tfrac{\Delta_{\phi^2}}{2}; \Delta_{\phi^2}; 1 - \bar{z}) {}_2F_1(\tfrac{\Delta_{\phi^2}}{2}, \tfrac{\Delta_{\phi^2}}{2}; \Delta_{\phi^2}; 1 - z)\right ]}_{\varepsilon=0} \bigg \}  ~,\label{3loopspec}
\end{align}
where $\Delta_{\phi^2}=1-\varepsilon/2$.

Due to the complexity of the various terms, we will follow the approach that led to the greatest flexibility in the two-loop case.\footnote{Some of the terms can be computed directly by applying the techniques of \cite{Hogervorst:2017kbj}. By rewriting the $\log$s in a convenient way and expanding the integrand, we can integrate termwise and then resum. The result is nicely written in terms of hypergeometric functions at $1$. We refer to the Mathematica notebook for the details.}
Specifically, we will write the inversion integral at three loops as a differential operator acting on the crossing kernel. In particular, we want to generate $\log |z|$, $\log|1-z|$, and the derivative of the ${_2F}_1$ with respect to its parameter, as shown in \eqref{3loopspec}. Notice that in our case $\Delta_s$ in $\mathcal{I}(\Delta_s,\Delta_t,\Delta_\phi)$ is always fixed by the structure of the inversion integral, but we can play with combinations of derivatives with respect to the other two parameters to reproduce our result. This leads to:
\begin{align}\label{diff-op-kernel}
	\varepsilon^3 \bigg [c^{(3)}(\Delta, \ell)+\frac{\partial}{\partial \varepsilon} c^{(2)}(\Delta,\ell)  \bigg] \biggl |_{\varepsilon = 0} = &\frac{\varepsilon^3}{54} \frac{\Gamma(\frac{\Delta+\ell}{2})^4}{ \Gamma(\Delta+\ell) \Gamma(\Delta+\ell-1)} \biggl [ \bigg (\mathcal{A}_2-\frac{3}{4} \frac{\partial}{\partial \Delta_\phi} - \frac{1}{2}  \frac{\partial}{\partial \Delta_t} \bigg ) \nonumber \\ 
	&\times \mathcal{I}(\ell - \Delta + 2,\Delta_t,\Delta_\phi) \mathcal{I}(\Delta + \ell,\Delta_t,\Delta_\phi) \biggr ] \bigg|_{\substack{\Delta_t = 1 \\ \hspace{2.5mm} \Delta_\phi = 1/2}}~. 
\end{align}
The existence of an expression like \eqref{diff-op-kernel} is no accident. For the part of the inversion integral that came from inputting \eqref{data-start2}, this is a consequence of the identity
\begin{align}
	\text{dDisc} \, \partial_\Delta^3 \left [ (1 - \bar{z})^\Delta f(\Delta, \bar{z}) \right ] &= \text{dDisc} (1 - \bar{z})^\Delta \left [ \log(1 - \bar{z})^3 f(\Delta, \bar{z}) + 3\log(1 - \bar{z})^2 \partial_\Delta f(\Delta, \bar{z}) \right ] \nonumber \\
	&= 12\pi^2 \partial_\Delta \left [ (1 - \bar{z})^\Delta f(\Delta, \bar{z}) \right ]~.
\end{align}
The rest of the inversion integral results from re-expanding a term that was already shown to be a product of crossing kernels. We will refer to these two parts of \eqref{diff-op-kernel} as $O(\lambda^3)$ and $O(\lambda^2)$ respectively. It will be instructive to consider them separately before setting the crossing kernel to \eqref{7f6-form} and acting on it with our differential operator. 

We are interested in reading off the poles and corresponding residues in order to extract the CFT data at three loops. 
A subtle point is that for the product of the two crossing kernels, the so-called double pole at $\Delta=\ell+1$ exhibited in the previous section is really a double pole at $\Delta = \ell + 2\Delta_\phi$. Therefore, when we naively hit it with the differential operator, it becomes a triple pole. By comparing with \eqref{expanded-density}, we do not expect this to be present since the leading spin-$\ell$ operator has no one-loop anomalous dimension. What it does have is a one-loop dimension shift $\Delta^{(1)}_{0,\ell}$ as defined in section \ref{sec:perturb}.
To account for this, instead of re-expanding the $O(\lambda^2)$ part right away by writing
\begin{align}
	\varepsilon^3 \left [ \frac{\partial}{\partial \varepsilon} c^{(2)}(\Delta,\ell) \right ] \biggl |_{\varepsilon = 0} = &-\frac{\varepsilon^3 \Gamma(\frac{\Delta+\ell}{2})^4}{ 18 \Gamma(\Delta+\ell) \Gamma(\Delta+\ell-1)} \biggl [ \bigg (\frac{1}{4} \frac{\partial}{\partial \Delta_\phi} + \frac{1}{2}  \frac{\partial}{\partial \Delta_t} \bigg ) \nonumber \\
	&\times \mathcal{I}(\ell - \Delta + 2,\Delta_t,\Delta_\phi) \mathcal{I}(\Delta + \ell,\Delta_t,\Delta_\phi)\biggr ] \bigg|_{\substack{\Delta_t = 1 \\ \hspace{2.5mm} \Delta_\phi = 1/2}}~, \label{diff-op-l2}
\end{align}
we will extract its $\Delta = \ell + 2\Delta_\phi$ pole for finite $\varepsilon$ and expand only its coefficient. Expanding around the pole
$\Delta=\ell+1-\tfrac{\varepsilon}{2}$, the crossing kernels become:
\begin{equation}\label{expand-kernels}
	\mathcal{I}(\Delta+\ell,\tfrac{2-\varepsilon}{2},\tfrac{2-\varepsilon}{4}) \mathcal{I}(\ell-\Delta+2,\tfrac{2-\varepsilon}{2},\tfrac{2-\varepsilon}{4}) \to 4 \ \frac{ \mathcal{I}(2\ell + 1 - \tfrac{\varepsilon}{2},\tfrac{2-\varepsilon}{2},\tfrac{2-\varepsilon}{4}) \Gamma(\tfrac{2-\varepsilon}{2})}{\Gamma(\tfrac{2-\varepsilon}{4})^2 (\Delta-1-\ell+\tfrac{\varepsilon}{2})^2} \ .
\end{equation}
After including prefactors, the derivative with respect to $\varepsilon$ should act on just the numerator of the previous expression, so that:
\begin{align}
	\varepsilon^3 \left [ \frac{\partial}{\partial \varepsilon} c^{(2)}(\Delta,\ell) \right ] \biggl |_{\varepsilon = 0} = &\frac{\varepsilon^3}{9\pi} \frac{\Gamma(\ell + \tfrac{1}{2})^4}{\Gamma(2\ell) \Gamma(2\ell + 1) (\Delta - \ell - 1 + \tfrac{\varepsilon}{2})^2} \biggl [ \bigg ( 2\frac{\partial}{\partial \varepsilon} - \frac{1}{2} \mathcal{A}_2 + \psi(2\ell) + \psi(2\ell + 1)  \nonumber \\
	&- 2\psi(\ell + \tfrac{1}{2}) \bigg ) \mathcal{I}(2\ell + 1 - \tfrac{\varepsilon}{2}, \tfrac{2 - \varepsilon}{2}, \tfrac{2 - \varepsilon}{4}) \biggr ] \biggl |_{\varepsilon = 0} \ .\label{3loop-l2}
\end{align}
Because of the $\varepsilon$ derivative which remains in \eqref{3loop-l2}, this expression is cumbersome but it can easily be evaluated numerically. One does not need to repeat \eqref{expand-kernels} for subleading trajectories. Even for finite $\varepsilon$, a simple exercise shows that $\Delta = 2\Delta_\phi + 2n + \ell$ is at most a single pole for $n > 0$. It will only be a double pole in the $O(\lambda^3)$ calculation to which we now turn.

Comparing \eqref{diff-op-kernel} to \eqref{diff-op-l2}, the differential operator acting on the crossing kernels and returning the $O(\lambda^3)$ part, is $\mathcal{A}_2+\partial_{\Delta_t}$. Since this does not include $\partial_{\Delta_\phi}$, no triple pole can be created, which is an important check. We also remarked below \eqref{3loop} that it is consistent at this order to set $\Delta_\phi = \frac{1}{2}$. Even though we expect the coefficient of $(\Delta - \ell - 1 + \tfrac{\varepsilon}{2})^{-2}$ to be cumbersome, it is now clear how to proceed. One simply needs to expand around the pole as before and then act on this expression with $\mathcal{A}_2+\partial_{\Delta_t}$. We refer readers to the attached Mathematica notebook for the result. However, this is not the end of the story, because double poles at $\Delta=\ell+2n+1$, for $n > 0$, tell us about the anomalous dimension of the subleading double-twist operators.
The coefficients of these double poles turn out to be particularly nice and compact. By looking carefully at the structure of our results, the ${_7F_6}$ functions that are not acted on by the derivative will become just $1$. Indeed, all physical poles of the original crossing kernel come from the gamma functions that have double poles at $\Delta=\ell+1$ and single poles at $\Delta=\ell+1+2n$, for $n > 0$. To get these subleading double poles, we then need our differential operator to hit the gamma functions instead of the ${_7F_6}$. We can finally extract the three-loop anomalous dimensions of these Regge trajectories by extracting the coefficients of these poles and dividing by $(-a_{n,l}^{(0)})$:
\begin{equation}
	\gamma^{(3)}_{n,\ell} = \frac{1}{36n(\ell+n)} \ .
\end{equation}

Since the anomalous dimension of $\hat{\phi}^2$ is known at $O(\varepsilon^3)$, it is tempting to look ahead to $c^{(4)}(\Delta, \ell)$. This is part of the four-loop spectral density along with re-expansion terms like $\frac{\partial}{\partial \eps} c^{(3)}(\Delta, \ell)$ and $\frac{\partial^2}{\partial \eps^2} c^{(2)}(\Delta, \ell)$. It is known that the four-loop double discontinuity in the SRI is affected by both double-twist and quadruple-twist operators. The latter contribution is much more complicated because the degeneracy among quadruple-twist operators can only be resolved with a large correlator system that involves external spin. Instead, \cite{Alday:2017zzv} was able to avoid the mixing problem and conjecture a formula for the double discontinuity by appealing to a transcendentality principle. As seen in \eqref{mirror-diagrams}, such a quadruple-twist contribution at four loops is completely absent in the LRI. This reduces the problem to evaluating an infinite sum of double-twist conformal blocks which are weighted by $a^{(0)}_{0, \ell} \gamma^{(2)2}_{0, \ell}$. While it is common to use transcendentality methods for this sum as well, an algorithm developed in \cite{Behan:2022uqr} makes it possible to extract large-spin CFT data from it without any extra assumptions.

\subsection{Further comments}

The analytic bootstrap is most powerful when applied to theories that have a weakly broken degeneracy in their spectrum of twists \cite{Alday:2016}. Generalized free theories provide the simplest examples. In this section, we have applied the Lorentzian inversion formula to the four-point function of $\hat{\phi}$ which is the fundamental field of the LRI near the mean-field end. What about going to the short-range end and considering the four-point function of $\hat{\chi}$?

The main difference in this case is that no double-twist operator has an anomalous dimension at $O(\delta)$. It is easy to compute a few renormalized integrals such as:
\begin{equation}
	\int d^p\tau_2 d^p\tau_3 \left \langle \hat{\chi}^2(0) \sigma\hat{\chi}(\tau_2) \sigma\hat{\chi}(\tau_3) \hat{\chi}^2(\infty) \right \rangle~, \label{try-integral}
\end{equation}
and see why -- the integrand completely vanishes after power law divergences are subtracted. This is a straightforward repetition of the argument in \cite{Behan:2017emf} for why $\hat{\chi}$ itself does not renormalize. This can be physically understood from the fact that, at $O(\delta)$, the Ising model coupled to a generalized free field by $\sigma \hat{\chi}$ is indistinguishable from \textit{two} generalized free fields coupled by $\sigma \hat{\chi}$. The statements made in \cite{Witten:2001ua}, about certain operators having no anomalous dimension in this theory, can be extended to all operators, because quadratic actions cannot produce loop diagrams that are logarithmically divergent. The $O(\delta)$ correction to the four-point function as a whole (which indeed has no logarithmic terms) can be computed as:
\begin{align}\label{chi-corr}
	\langle \hat{\chi}(\tau_1)\hat{\chi}(\tau_2)\hat{\chi}(\tau_3)\hat{\chi}(\tau_4) \rangle =& \left [ (|\tau_{12}||\tau_{34}|)^{-2\Delta_\chi} + (|\tau_{13}||\tau_{24}|)^{-2\Delta_\chi} + (|\tau_{14}||\tau_{23}|)^{-2\Delta_\chi} \right ] \nonumber \\
	& \times\left [ 1 + 2g_*^2 \pi^p \frac{\Gamma(\tfrac{p}{2} - \Delta_\sigma)}{\Gamma(\Delta_\sigma)} \frac{\Gamma(\tfrac{p}{2} - \Delta_\chi)}{\Gamma(\Delta_\chi)} + O(g_*^4) \right ]~,
\end{align}
using only the chain integral. This shows that all OPE coefficients for $[\hat{\chi} \hat{\chi}]_{n,\ell}$ undergo the same shadow-symmetric rescaling.

\section{Numerical results}
\label{sec:numerics}

Up until this point, we have been focused on perturbative results. Although these are only strictly valid close to
$\mathfrak{s} = \frac{p}{2}$ or $\mathfrak{s} = p - 2\D_\sigma^*$, they improve our expectations for where non-trivial LRI models should be found in the space of scaling dimensions. When this space is cut down to a manageable size, it can be searched with the numerical bootstrap \cite{Rattazzi:2008pe}. In the most favourable cases, one can force the theory of interest to live in an ``island'' by rigorously excluding the possibility that a consistent CFT lives at any of the surrounding points \cite{Kos:2014bka,Kos:2015mba,Rong:2018okz,Atanasov:2018kqw,Kousvos:2018,Erramilli:2022kgp}. More commonly, a precursor to this type of result called a ``kink'' is what provides evidence that a theory has been found. The goal of this section is to produce kinks for general $\mathfrak{s}$
which agree with perturbation theory near the two endpoints and therefore give non-perturbative predictions for LRI critical exponents in between.\footnote{Non-perturbative checks using Monte Carlo results are also possible in between but we only know a few values of $\mathfrak{s}$ where enough data has been taken to make a non-trivial comparison. While Monte Carlo simulations of the LRI in $p > 1$ have been done since \cite{earlysim2,earlysim1}, most of the early work focused on checking that the spin field is indeed protected and did not extract predictions for unprotected operators.} The main analysis will be done for $p=2$ systems and presented in figures~\ref{fig:spin2gap_vs_aphi2_wide} through \ref{fig:isolated}. Afterwards, we will move onto $p = 3$ with figure \ref{fig:kinks_3d} which improves the results of \cite{Behan:2018hfx}.

Our numerical bootstrap setup is similar to that of \cite{Behan:2020nsf,Behan:2021tcn}, except that now we have an extra parameter $\ps$ which enters the crossing equations via the scaling dimensions of the external operators.
As explained in appendix~\ref{ss:crossingeq}, here we combine the system of 5 crossing equations for the four-point correlation functions involving $\psi_0^{(\pm)}$, with all the exact OPE relations \eqref{ssspinconstr}. It is worth explaining how this approach differs from that of \cite{Behan:2018hfx}, which ignored all of the OPE relations except:
\begin{align}
	\lambda_{11 \mathcal{O}} \lambda_{22 \mathcal{O}} = \frac{\Gamma(\frac{\ell + \Delta_\mathcal{O}}{2})^2 \Gamma(\frac{\ell + p + q - 2 - \Delta_\mathcal{O}}{2}) \Gamma(\frac{\ell + p - q + 2 - \Delta_\mathcal{O}}{2})}{\Gamma(\frac{\ell + p - \Delta_\mathcal{O}}{2})^2 \Gamma(\frac{\ell + 2 - q + \Delta_\mathcal{O}}{2}) \Gamma(\frac{\ell - 2 + q + \Delta_\mathcal{O}}{2})} \lambda^2_{12 \mathcal{O}}~. \label{cancel-ratio}
\end{align}
In \cite{Behan:2018hfx} this choice was made in order to remove the dependence on $R(a_{\phi^2}) = R(a_{\chi^2})^{-1}$ in the crossing equations, but the studies \cite{Behan:2020nsf,Behan:2021tcn} later showed that it is better to keep this ratio in the problem and vary it manually within the region depicted in figure~\ref{Unitarityregion}. Besides allowing us to impose more OPE relations, scanning over a parameter other than a scaling dimension is necessary for making progress on the two-dimensional LRI. Indeed, it is known that single-correlator exclusion plots for the dimensions of relevant operators (which have no LRI kink) are not improved by extra correlators unless one imposes model-dependent gaps \cite{Fuente:2019}. As we will see, this new type of scan is also sufficient for finding the LRI in $p = 2$.\footnote{See \cite{Agmon:2019imm} for a different bootstrap study which benefited from scanning over more than just scaling dimensions.}

\subsection{Warm-up}

For a given external dimension $\D_\phi$, we can bound the scaling dimensions of exchanged operators. As discussed above, we will bound them as functions of the unknown ratio $R(a_{\phi^2})$, We will use $a_{\phi^2}$ as a proxy for this ratio and label our bounds by $\D_\phi$. If one prefers $a_{\chi^2}$, which vanishes at $\mathfrak{s} = p - 2\D_\sigma^*$ instead of $\mathfrak{s} = \frac{p}{2}$, it is easy to convert between them using eq.~\eqref{a-relation}. 
\begin{figure}
	\centering
	\includegraphics[scale=0.8]{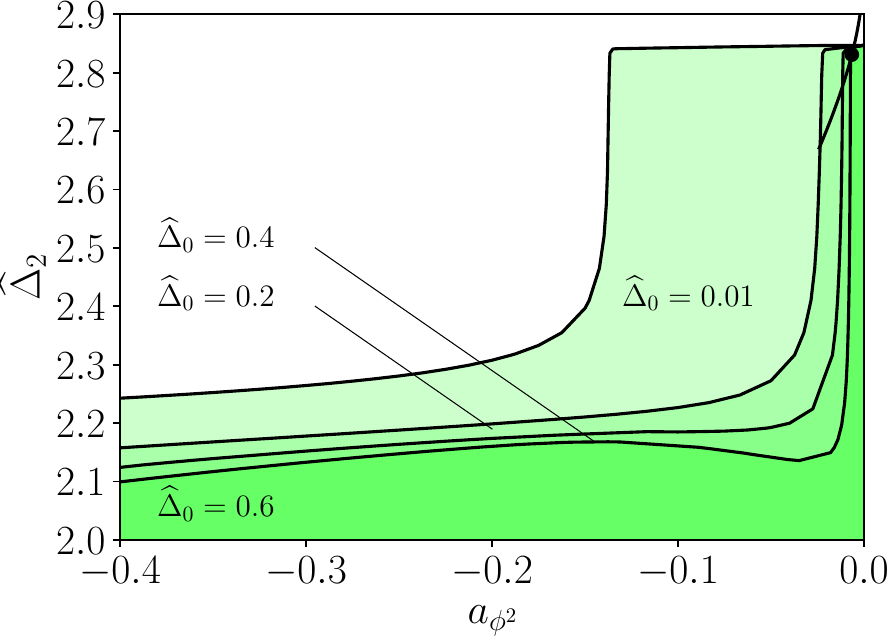}
	\caption{Bounds on the dimension of the leading spin-2 operator for $\D_\phi = 0.425$ in two dimensions. This corresponds to a value of $\eps$ which singles out the dot on the perturbative line in the upper right corner. The lightest region shows a nearly universal bound with only a spin-0 gap of $0.01$ imposed for stability. The darker regions impose more restrictive spin-0 gaps and are needed to see evidence of the LRI model for this value of $\D_\phi$.}
	\label{fig:spin2gap_vs_aphi2_wide}
\end{figure}
The crossing equations used throughout, given explicitly in appendix \ref{ss:crossingeq}, are the ones appropriate for a three-correlator system involving $\hat{\phi}\equiv \psi_0^{(+)}$ and its shadow $\psi_0^{(-)}$.
Imposing the OPE relations \eqref{ssspinconstr} on odd-spin operators forces them to have scaling dimensions in the discrete set $p + \ell + 2n$ for $n \geq 0$. We will treat $\ell = 1$ as an exception and demand that there is no spin-1 operator of dimension $p + 1$.
A primary with these quantum numbers cannot exist in the LRI near either endpoint. At the mean-field end, the only candidate is a descendant of $\hat{\phi}^4$ while at the short-range end the only candidate is a descendant of $T_{\mu\nu}$, see e.g. the discussion in ref.~\cite{Behan:2017emf}.\footnote{In the case of unitary and local BCFTs, \cite{Behan:2020nsf} proved that a vector of dimension $d$ is absent from the spectrum of boundary primaries. One can find counter-examples when unitarity is given up \cite{Chalabi:2022qit}. Via the same reasoning, one can prove that for unitary and local $p$-dimensional conformal defects a vector of dimension $p+1$ should be absent from the spectrum of defect primaries.} Finally, we will impose \eqref{ssspinconstr} for all operators with $\ell = 0, 2$. This is a slight simplification compared to the BCFT setup in \cite{Behan:2020nsf,Behan:2021tcn} which required there to be a discrete set of even-spin operators (like the displacement) which evade the OPE relations.

The type of plot made in \cite{Behan:2020nsf,Behan:2021tcn}, which maximizes the gap $\hD_2$ for spin-2 operators, will be a useful starting point. Whenever a non-trivial upper bound on $\hD_2$ exists, a theory saturating it is guaranteed to be nonlocal.
The plot is shown in figure \ref{fig:spin2gap_vs_aphi2_wide} for an external dimension of $\D_\phi = 0.425$.
The most permissive region in this plot has a kink at around $a_{\phi^2} = -0.15$.
This is an order of magnitude off from the horizontal position of the dot which estimates $a_{\phi^2}$ in the $\D_\phi = 0.425$ LRI by plugging $\varepsilon = 0.3$ into the perturbative results of eq.~\eqref{mean-field-1pt}. The vertical position of this dot is the leading spin-2 anomalous dimension from Lorentzian inversion.
\begin{figure}
	\centering
	\includegraphics[scale=0.8]{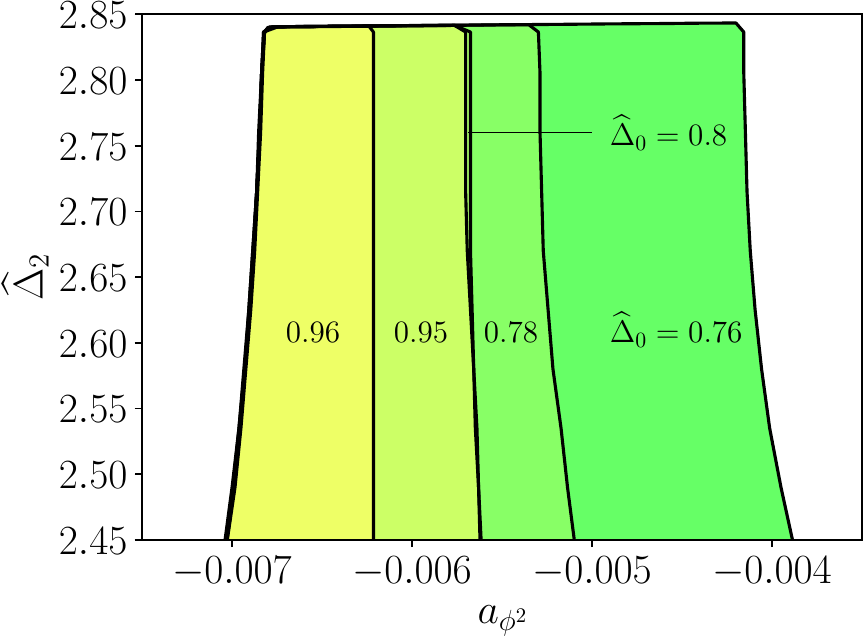}
	\caption{Bounds on the spin-2 gap for larger values of the imposed spin-0 gap which would be hard to see on the scale of figure \ref{fig:spin2gap_vs_aphi2_wide}. Yellow regions correspond to more stringent gaps and the smallest one, which fixes $a_{\phi^2}$ to $0.1\%$, is a good candidate for where the $\D_\phi = 0.425$ LRI model should live.}
	\label{fig:spin2gap_vs_aphi2_zoom}
\end{figure}
As noted in \cite{Behan:2021tcn} however, plots of this type can show more interesting features once we restrict our search to CFTs obeying a certain gap $\hD_0$ on internal operators of spin-0. As we raise this gap in figure \ref{fig:spin2gap_vs_aphi2_wide}, the vertical wall moves rapidly at first but then slows down. By $\hD_0 = 0.6$, which is well below the mean-field value of $\hD_0 = 1$, it has already stabilized at a position close to the right hand side. As we continue to raise $\hD_0$, something even more interesting happens. A second vertical wall starts moving to the \textit{left}. The zoomed-in plot figure \ref{fig:spin2gap_vs_aphi2_zoom} shows the spike which is obtained in this way.
The leftward motion stops as well at around $\hD_0 = 0.8$ making the spike temporarily stable. At about $\hD = 0.95$, the leftward motion starts again resulting in complete disappearance once we get to $\hD_0 = 0.97$. This is strikingly close to the scalar gap for the LRI with $\D_\phi = 0.425$ as predicted by perturbation theory.

\subsection{Kinks in two dimensions}

\begin{figure}
	\centering
	\includegraphics[scale=0.8]{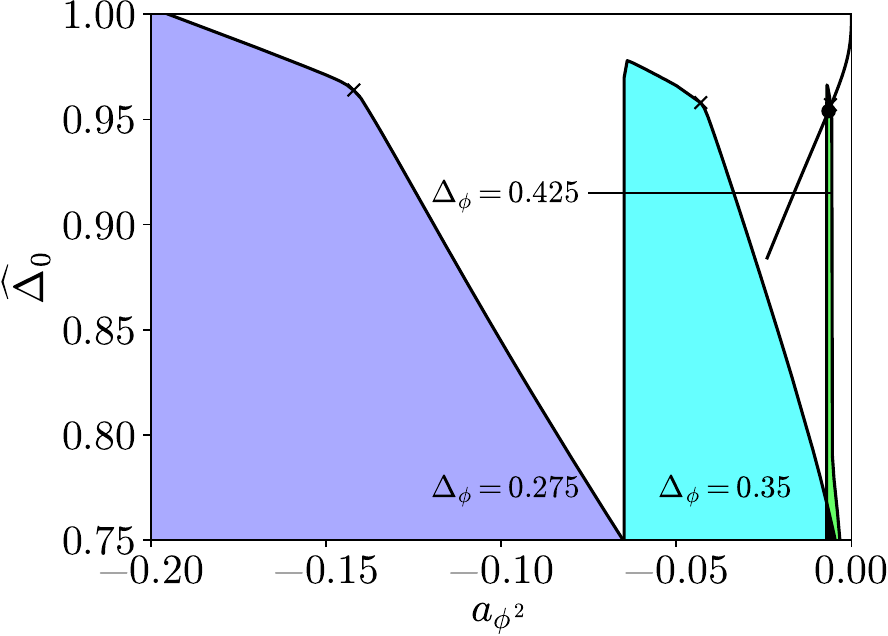}
	\caption{Bounds on the dimension of the leading exchanged scalar for $\D_\phi = 0.425$ (narrowest), $\D_\phi = 0.35$ and $\D_\phi = 0.275$ (widest), in two dimensions. For aesthetic reasons, these bounds were made subject to the constraints $\hD_2 \geq 2.4$, $\hD_2 \geq 2.3$, and $\hD_2 \geq 2.2$ respectively. The rightmost kinks indicated on the plot with x's are stable with respect to the spin-2 gap and serve as candidates for the 2d LRI.}
	\label{fig:spin0gap_vs_aphi2_low}
\end{figure}
\begin{figure}
	\centering
	\includegraphics[scale=0.8]{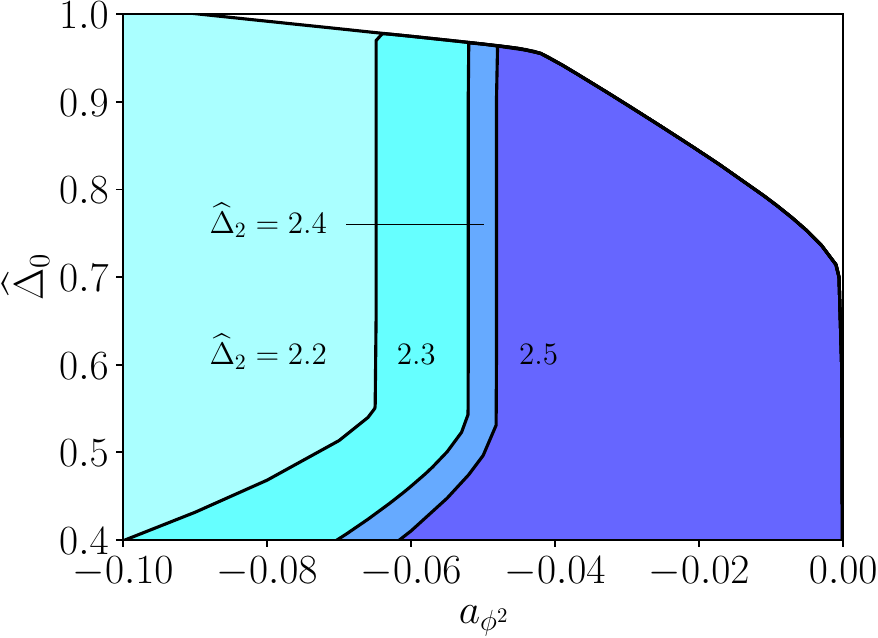}
	\caption{Allowed regions in $(a_{\phi^2}, \hD_0)$ space for $\D_\phi = 0.35$ in two dimensions. These include the region from figure \ref{fig:spin0gap_vs_aphi2_low} with $\hD_2 \geq 2.3$ but also three other versions of it with different spin-2 gaps. A darker region corresponds to a larger spin-2 gap. The prevailing trend is that the left edge moves while the right edge (which contains the LRI kink) does not. This happens for all external dimensions so we have chosen $\D_\phi = 0.35$ as a ``generic'' example not close to $\frac{p}{4}$.}
	\label{fig:spin0gap_many_spin2gap}
\end{figure}

As the spin-0 gap is varied, the bounds above sometimes change quickly and sometimes change slowly.
Plots bounding $\hD_0$ should therefore be just as interesting as those bounding $\hD_2$. Our main conjecture regarding numerics is that this exercise allows one to find a kink corresponding to the LRI model for arbitrary $\D_\phi$ without any assumption on the spin-2 gap. Figure \ref{fig:spin0gap_vs_aphi2_low} shows almost this result except with a modest spin-2 gap for readability.
Specifically, we have imposed
\begin{equation}
	\hD_2 \geq \frac{4}{3} \D_\phi - \frac{1}{6}~, \label{toy-gap}
\end{equation}
to prevent the allowed regions from overlapping heavily.
This arbitrary gap is easily obeyed by the LRI model and the features we will now discuss would still be visible without it. Two of the allowed regions in Figure \ref{fig:spin0gap_vs_aphi2_low} show one kink on the left and another on the right. The third region would as well if the horizontal axis were extended. It turns out that the solution to crossing at the leftmost kink has a spin-2 operator saturating \eqref{toy-gap} whereas the leading spin-2 operator dimension at the rightmost kink is much higher. This can be seen in figure \ref{fig:spin0gap_many_spin2gap} which compares four different allowed regions for $\D_\phi = 0.35$ with each one having a different constraint on $\hD_2$.

Evidently, there is some critical spin-2 gap which makes the allowed region have only one kink. For $\D_\phi = 0.425$, this is about $\hD_2 = 2.84$ in agreement with figure \ref{fig:spin2gap_vs_aphi2_zoom}. As the spin-2 constraint is relaxed, making the region wider, the single kink splits into two with one staying still and the other moving to the left. Based on this pattern, the stationary kink on the right should be the one associated with the LRI model. Figure \ref{fig:spin0gap_vs_aphi2_low} shows that it is about $0.3\%$ above the perturbative prediction around the mean-field end despite having a much less noticeable error in the horizontal position.
\begin{figure}[h]
	\centering
	\includegraphics[scale=0.8]{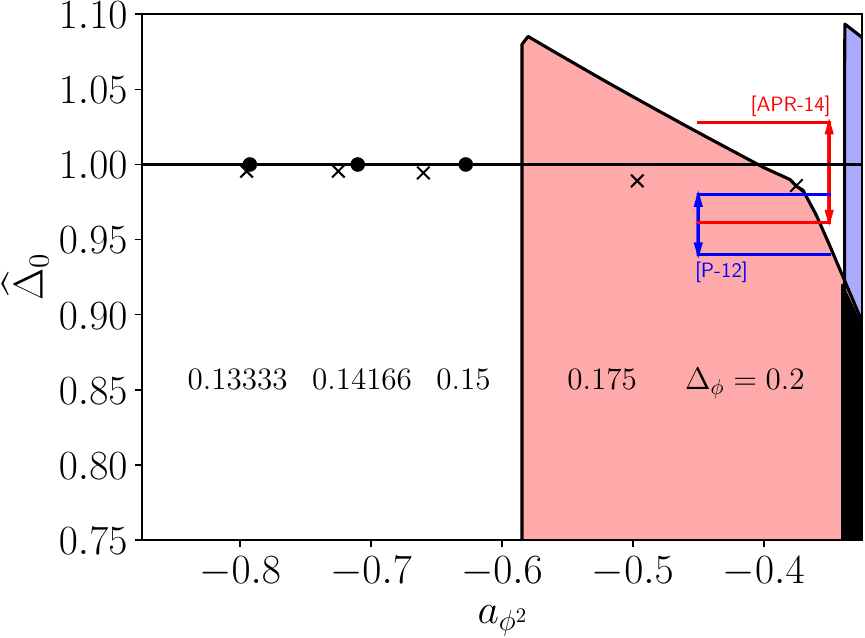}
	\caption{A plot of the kinks (denoted by x's) attributed to the 2d LRI model for five more values of $\D_\phi$. For $\D_\phi = 0.2$, the full allowed region subject to $\hD_2 \geq 2.1$ is shown as well. The error bars close to this kink denote the findings for the $\nu$ exponent from the Monte Carlo studies \cite{p12,apr14}. Closer to the left, dots are given representing perturbative results so that the kinks at $\D_\phi = 0.15$, $\D_\phi = 0.14166$ and $\D_\phi = 0.13333$ may be compared to $O(\delta)$ perturbation theory.}
	\label{fig:spin0gap_vs_aphi2_high}
\end{figure}
Reasons for this will be discussed in the next subsection which deals with resummations. The moving kink bears some resemblance to the one in \cite{Behan:2021tcn} which was associated with minimal models. To check whether the same might be true here, we can follow the kink to the leftmost edge of $a_{\phi^2} = -\frac{7}{8}$ where the theory becomes local. Interestingly, this gives it an approximate spin-0 gap of
\begin{equation}
	\hD_0 = \frac{8}{3} \D_\phi + \frac{2}{3}~. \label{toy-evidence}
\end{equation}
This is precisely the relation between $\D_{(1,2)}$ and $\D_{(1,3)}$ for the $m$'th Virasoro minimal model if we eliminate $m$ to express one scaling dimension in terms of the other. As remarked in \cite{Behan:2017emf}, there is no reason why these other models cannot be coupled to a generalized free field in the same vein as the Ising model to generate many nonlocal CFTs. The match with \eqref{toy-evidence} can then be taken as evidence that these ``long-range minimal models'' are being singled out by the numerical bootstrap. To put it another way, the kink for $\D_\phi$ which saturates the imposed spin-2 gap might be constructed by taking the $m$'th minimal model for some $m$ and coupling its $\phi_{(1,2)}$ operator to a generalized free field of dimension $2 - \D_\phi$. The precise value of $m$ will be difficult to determine in general, but we believe that many checks of this proposal can be done in perturbation theory. As usual, it will be important to consider non-integer values of $m$ and construct these RG flows using generalized minimal models. Generalized minimal models are non-unitary but it is known that their low-lying correlators manifest this in a very limited way \cite{Behan:2018}.

As shown in eq.~\eqref{unitaryintervalaphisq}, we have that $-\frac{7}{8} \leq a_{\phi^2} \leq 0$. Our conclusions so far have been based on values of $a_{\phi^2}$ in the upper half of this range. We have also computed a bound on $\hD_0$ for $\D_\phi = 0.2$ and still found a kink in the upper half of this range. The basic structure of perturbation theory makes it clear that four evenly spaced values of $\D_\phi$ in the range $\frac{1}{8} \leq \D_\phi \leq \frac{1}{2}$ will be far from evenly spaced with respect to $a_{\phi^2}$. After all, the expansion of the latter starts at $O(1)$ around the short-range end but $O(\varepsilon^2)$ around the mean-field end. We have therefore picked another four values of $\D_\phi$ between $\frac{1}{8}$ and $0.2$ for testing how well the putative LRI kinks agree with the $\delta$ expansion. The kinks themselves are shown in figure \ref{fig:spin0gap_vs_aphi2_high}. Their positions were computed from the second derivative of a bound on $\hD_0$ but the rest of the bound has been omitted to avoid clutter. The perturbative line (which always has $\hD = 1$ since $\D_{\hat{\epsilon}} = 1 + O(\delta^2)$) shows the desired convergence between the numerical and analytic values of $a_{\phi^2}$ on the left hand side. The right hand side of figure \ref{fig:spin0gap_vs_aphi2_high} shows partially overlapping error bars obtained by Monte Carlo simulations which both considered $\mathfrak{s} = 1.6$ or $\D_\phi = 0.2$. After using the standard relation
\begin{equation}
	\D_{\hat{\phi}^2} = p - \nu^{-1}~, \label{nu-exp}
\end{equation}
it becomes clear that our kink is compatible with $\nu^{-1} = 0.996(33)$ found by \cite{apr14} but not $\nu = 0.96(2)$ found by \cite{p12}.

\subsection{Checks of the critical exponents}
\label{ss:betternumerics}

The most important next step is to verify that the numerical bootstrap, beyond simply ``knowing'' about the LRI model, can actually be used to derive its critical exponents to high precision. To do this, we should address the fact that the data in figures \ref{fig:spin0gap_vs_aphi2_low} and \ref{fig:spin0gap_vs_aphi2_high} only agree with the $\varepsilon$ and $\delta$ expansions respectively within a narrow window. It is also clear that the kinks demonstrating the best agreement still lie noticeably below the perturbative line. These discrepancies can be diminished with simple improvements to both the numerics and perturbation theory.

On the numerical side, we should refine the discrete grid of points discussed in appendix \ref{ss:crossingeq}. In all previous plots, the OPE relations have been imposed for 3000 closely spaced values of $\D$. Increasing this number appears to make all of the kinks move monotonically upwards. This is expected in the numerical bootstrap since constraints at the new points make it harder to find functionals for ruling out CFTs. We have gone to 12000 which appears to be large enough that further errors are dominated by the resolution in $a_{\phi^2}$.
\begin{figure}[h]
	\centering
	\includegraphics[scale=0.8]{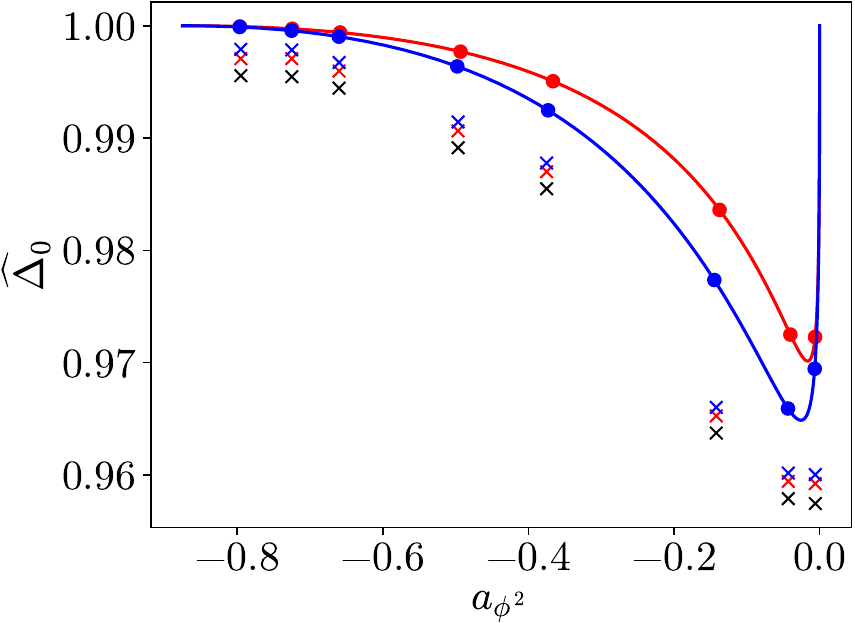}
	\caption{Comparison between perturbative and non-perturbative estimates of the leading internal scalar dimension in the 2d LRI model. The blue line is a Pad\'{e} approximant based on all known perturbative data while the red one omits $O(\varepsilon^3)$. The dots on them correspond to the eight external dimensions considered throughout this section ranging from $\D_\phi = 0.13333$ on the left to $\D_\phi = 0.425$ on the right. The x's denote kinks from the numerical runs which are sensitive to the number of grid points. We show results for 3000, 6000 and 12000 points in black, red and blue respectively.}
	\label{fig:pade_comparison}
\end{figure}
On the perturbative side, we will use the approximants $\text{Pade}_{[m, n]}$ which are rational functions with $m + n + 1$ coefficients designed to resum an asymptotic series. While this can be done for either the $\varepsilon$ or $\delta$ expansions individually, it is much better to use a two-sided approximant.\footnote{See ref.~\cite{Rong:2024vxo} for alternative strategies of resumming mean-field end perturbative data using Borel-Pad\'{e} and conformal mapping.} Picking the same one for both axes, we will write
\begin{equation}
	\hD_0(\mathfrak{s}) = \frac{\sum_{j = 0}^m a_j \mathfrak{s}^j}{1 + \sum_{j = 1}^n b_j \mathfrak{s}^j}~, \quad a_{\phi^2}(\mathfrak{s}) = \frac{\sum_{j = 0}^m a'_j \mathfrak{s}^j}{1 + \sum_{j = 1}^n b'_j \mathfrak{s}^j}~, \label{pade-form}
\end{equation}
and fix the coefficients by demanding that $\mathfrak{s} = 1 + \frac{\varepsilon}{2}$ and $\mathfrak{s} = 2(1 - \D^*_\sigma - \delta)$ reproduce the expected behaviour around the mean-field and short-range ends respectively. Solutions for the coefficients (if they exist) are admissible if they lead to real-valued functions that have no poles between $\varepsilon = 0$ and $\delta = 0$. If we perturb up to $O(\delta)$ and $O(\varepsilon^{m - 1})$, the $\text{Pade}_{[m,1]}$ approximant has admissible solutions for $m = 3$ and $m = 4$. As shown in figure \ref{fig:pade_comparison}, the latter gets us significantly closer to where the kinks are. By bounding the spin-2 gap at each kink, it is also possible to repeat this check for spin-2 operators. Adding $\hD_2$ as a third axis, we obtain the plot in Figure \ref{fig:pade_3d}.

\begin{figure}
	\centering
	\includegraphics[scale=0.8]{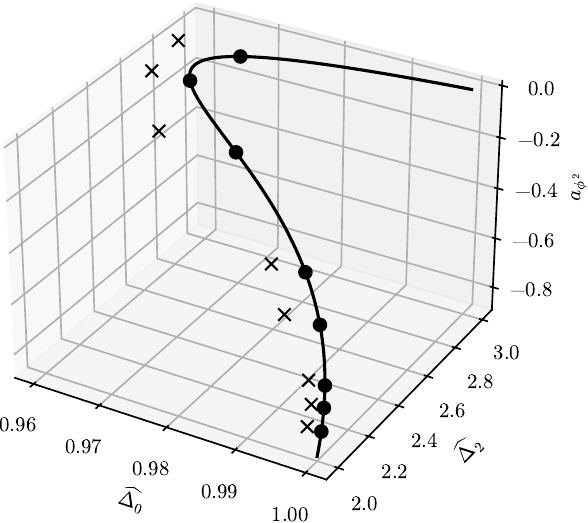}
	\caption{The comparison from figure \ref{fig:pade_comparison} extended to include spin-2 operators as well. For clarity, we only show data with the highest number of sample points (blue points from figure \ref{fig:pade_comparison}). The perturbative $[\hphi \hphi]_{0,2}$ dimensions, based on results up to $O(\eps^3)$ from the inversion formula, hardly show any deviation from their numerical bootstrap values.}
	\label{fig:pade_3d}
\end{figure}

\begin{figure}[h]
	\centering
	\includegraphics[scale=0.8]{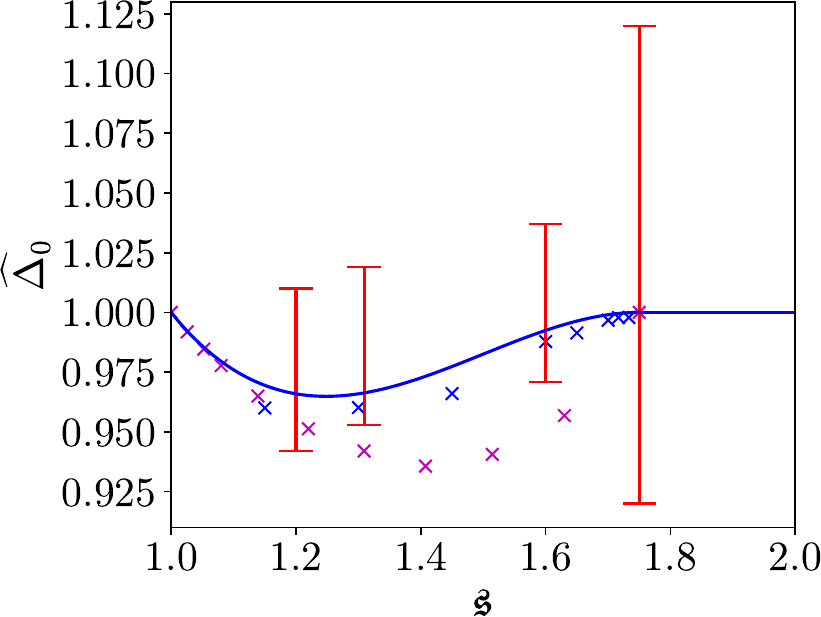}
	\caption{The dimension of the leading unprotected scalar $\hD_0$ as obtained by four methods in recent years. All of these methods have access to the parameter $\mathfrak{s}$ but so far only perturbation theory and the bootstrap have produced results for $a_{\phi^2}$. As in figure \ref{fig:pade_comparison}, the blue line and blue x's represent the best known perturbative and bootstrap results respectively. The red error bars are Monte Carlo predictions from \cite{apr14}. The purple x's were derived from an approximation to the effective action in \cite{Defenu:2015}. They agree well with the other data close to the mean-field and short-range ends but there appears to be room to push the accuracy further in between.}
	\label{fig:mc_frg}
\end{figure}

While Pad\'{e} approximants indicate that the kinks are trustworthy, it is also important to see how well the bootstrap agrees with other non-perturbative methods. The Monte Carlo estimates of $\nu$ from \cite{p12,apr14}, mentioned briefly in the last subsection, are ideal for this purpose. Figure \ref{fig:spin0gap_vs_aphi2_high} shows that our kink for $\D_\phi = 0.2$ lies slightly outside the error bars predicted for this LRI in \cite{p12}. This becomes even more prominent after increasing the number of sample points to 12000. Conversely, \cite{apr14} found a larger value of $\nu$ and explained the discrepancy by noting that the LRI has significant finite size effects which can be mitigated by simulating a different Hamiltonian in the same universality class. Three other values of $\nu$ were extracted in \cite{apr14} and may be compared to both bootstrap results and estimates obtained with the functional renormalization group technique \cite{Defenu:2015,Defenu:2023}. Figure \ref{fig:mc_frg} plots this data as a function of $\mathfrak{s}$ and shows that bootstrap results are all within the Monte Carlo error bars. The functional renormalization group results are not but it is possible that this can be addressed by refining the approximation made in \cite{Defenu:2015}.

\begin{figure}[h]
	\centering
	\includegraphics[scale=0.8]{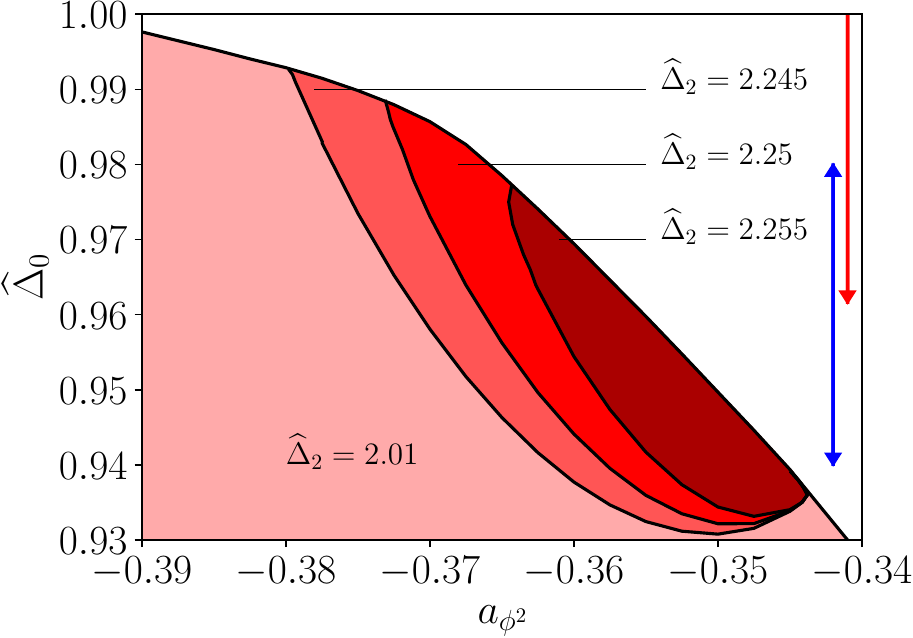}
	\caption{The vicinity of the $\D_\phi = 0.2$ kink in $(a_{\phi^2}, \hD_0)$ space attributed to 2d LRI, when we scan over the dimension of a single exchanged relevant scalar instead of maximizing the scalar gap. The lightest region is essentially the one seen in figure \ref{fig:spin0gap_vs_aphi2_high} while the darker islands start to form after a significant spin-2 gap $\hD_2$ is imposed. The blue and red error bars follow from the Monte Carlo results of \cite{p12} and \cite{apr14} respectively.}
	\label{fig:isolated}
\end{figure}

An intriguing possibility is using the numerical bootstrap to derive, not just upper bounds on $\hD_0$, but lower bounds as well since these could conceivably exclude the earlier Monte Carlo result. We have not been able to find lower bounds that are strong enough but it is instructive to see how close they come. The key is that the LRI (and the SRI as well) has one relevant operator that is $\mathbb{Z}_2$-even. This is of course $\hphi^2$ in one description and $\epsilon$ in the other. A bound on the spin-0 gap $\hD_0$ says nothing about how many scalars there are with $\hD_0 < \D < 2$. A better strategy is to demand that all scalars exchanged in our crossing equations have $\D > 2$ except for one with dimension $\hD_0$. Tuning this value and testing for feasibility of the crossing equations every time is what produces upper and lower bounds.
The lower bound we find with no extra assumption is well below the bottom of the blue error bar in figure \ref{fig:spin0gap_vs_aphi2_high}, but more interesting things happen when we raise the spin-2 gap $\hD_2$. When doing simple gap maximization, it has already been seen in figure \ref{fig:spin0gap_many_spin2gap} that the upper bound in the vicinity of the kink is insensitive to $\hD_2$. When we take $\hD_0$ to be isolated, this continues to hold for the upper bound but not the lower bound. The lower bound becomes increasingly restrictive and turns the allowed region into an island. Figure \ref{fig:isolated} shows three of these islands starting with $\hD_2 = 2.245$, which is almost large enough to rule out the kink. By the time we reach $\hD_2 = 2.255$, the kink has become disallowed but the island is still large enough to be compatible with the error bars of both \cite{p12} and \cite{apr14}. Unfortunately, the lower bounds do not improve significantly when we add the $\mathbb{Z}_2$-even scalar as a third external operator using the larger system of crossing equations in \eqref{bigger-system}.

\subsection{Kinks in three dimensions}

It remains to be seen that finding the LRI model numerically works equally well in three dimensions. This time, it is harder to get a point of comparison perturbatively.
In figure \ref{fig:pade_comparison}, the red and blue curves had similar values of $a_{\phi^2}$ because the $\varepsilon$ and $\delta$ expansions approximated each other fairly well. For example, $\delta = 0$ corresponds to $\varepsilon = \frac{3}{2}$ in two dimensions and this underestimates $a_{\phi^2}$ by about a factor of 2. In three dimensions, this is instead a factor of 7, leading to predictions which fluctuate wildly upon using different Pad\'{e} approximants.
Improving this with higher-loop perturbation theory around the short-range end will lead to integrals with more than four insertions of $\sigma$ in the 3d Ising model. Approximating one of these correlators will require higher-point CFT data and techniques for extracting them are currently in their infancy \cite{Poland:2023vpn}.

\begin{figure}
	\centering
	\includegraphics[scale=0.8]{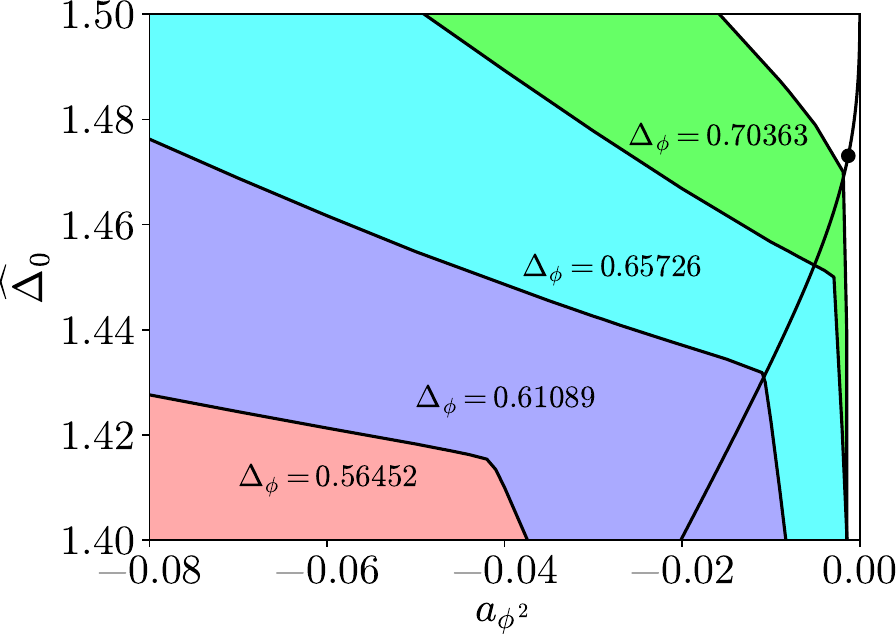}
	\caption{Allowed regions in $(a_{\phi^2}, \hD_0)$ space for three dimensions and four equally spaced external dimensions between $\D_\phi = \frac{3}{4}$ and $\D_\phi = \D_\sigma^*$. Again, we conjecture that each kink is a 3d LRI model. The black line shows perturbative results up to $O(\eps^3)$ with the dot corresponding to $\D_\phi = 0.70363$.}
	\label{fig:kinks_3d}
\end{figure}

Although the checks we can do are limited, it is plausible that the kinks plotted in figure \ref{fig:kinks_3d} describe the LRI model once again. For one thing, their vertical coordinates interpolate between $\hD_0 = \frac{3}{2}$ and $\hD_0 = \D_\epsilon^*$ as expected. Their horizontal coordinates all appear to be an order of magnitude smaller than the short-range value $a_{\phi^2} = -0.575408$, but we know that the $O(\delta)$ deformation changes this rapidly. Numerical runs for $\D_\phi$ very close to $\D_\sigma^*$ indeed show that the motion of the kink becomes almost completely horizontal so that the correct limiting behaviour is recovered.

A final point is that the regions in figure \ref{fig:kinks_3d} do not change after imposing a spin-2 gap. The moving kinks in figures \ref{fig:spin0gap_vs_aphi2_low} and \ref{fig:spin0gap_vs_aphi2_high}, which depend on this gap, are therefore unique to two dimensions. This supports the interpretation that they arise as a deformation of Virasoro minimal models which almost all have an upper critical dimension below $p = 3$.

\section{Conclusions and outlook}
\label{sec:outlook}

In this work, we have investigated the long-range Ising model, a particular and well-known example of a nonlocal CFT. Being a line of fixed points, it brings about a departure from the standard universality paradigm.
We have used different techniques, both perturbative and non-perturbative, to determine various CFT data. A central role is played by the use of exact relations between OPE coefficients, which can be implemented in the numerical bootstrap as well as employed in perturbative computations to gain an extra loop order in the results. These OPE relations emerge from the setup chosen in this paper, where we consider the $p$-dimensional LRI as a defect in a free bulk CFT. We explored both weakly coupled descriptions of the LRI, which have been dubbed the mean-field and short-range ends, 
although most of our new perturbative results were for the mean-field end. 

Besides combining the OPE relations with standard Feynman diagram techniques to obtain anomalous dimensions and OPE coefficients, as done in section~\ref{sec:perturb}, one can also use Lorentzian inversion of the OPE to obtain CFT data perturbatively. This was initiated in section~\ref{sec:inversion} yielding double-twist anomalous dimensions at three loops from the mean-field end. In addition to the original inversion formula \cite{Caron-Huot:2017vep}, which we have applied to a defect four-point function, there are also analogous formulae for bulk two-point functions \cite{Lemos:2017vnx,Liendo:2019jpu} which could lead to complementary results. In section~\ref{sec:numerics} the LRI was studied non-perturbatively using the numerical bootstrap. The numerics showed a number of kinks corresponding to both perturbative and non-perturbative (Monte Carlo) predictions for the LRI in two and three dimensions. While these showed that the LRI, and possibly other long-range models, can be bootstrapped in the sense of \cite{Rattazzi:2008pe}, a more detailed analysis will be possible once numerical spectra are extracted using the Extremal Functional Method (EFM) \cite{ElShowk:2013}.

One promising application of a precise spectrum would be passing it as input to the Lorentzian inversion formula away from the perturbative regime. This was explored before in \cite{Simmonsduffin:2017,Liu:2020tpf,Atanasov:2022} for local CFTs. It would also be interesting to investigate long-range minimal models since there appeared to be hints of them in the two-dimensional results of section~\ref{sec:numerics}. This suspicion could be tested by using conformal perturbation theory and comparing it to a numerical spectrum extracted using the EFM. Additionally, \cite{levelcross} recently found numerical evidence of level repulsion for the CFTs of the Wilson-Fisher fixed point. A similar feature can be expected to appear when studying the LRI. By varying $\mathfrak{s}$ between its two endpoints and applying the EFM many times, this approach could shed light on how the $\sigma \chi$-type operators map onto the $\phi^4$-type operators beyond the known lowest-lying cases.

Although the extremal functional method works anywhere on the boundary of an allowed region, it becomes more powerful when it is used inside an island. In this work, our preliminary attempt to isolate the LRI was motivated by a tension in the Monte Carlo literature. In particular, the results of \cite{apr14} are consistent with the position of our kink, while those of \cite{p12} are not. Unfortunately, all of the islands we could obtain were compatible with the error bars of both studies even after making significant spin-2 gap assumptions. Making progress in this area will likely require more correlators. These numerical bootstrap studies will involve many parameters and are expected to be very time-consuming, especially considering the complications already present with three correlators. To overcome this, one good strategy would be running \texttt{SDPB} within the framework of \cite{navigator,skydiving}. Another would be using the newer software in \cite{outerapp} which could be a better fit for optimization problems involving the OPE relations.

The numerical bootstrap and the perturbative methods discussed here are available for studying CFTs in the continuum limit. If one works with the lattice Hamiltonian instead, it is possible that an understanding of the long-range Ising model can be achieved along different lines. For computing the spectrum of the short-range Ising model, a recent breakthrough was made in \cite{zhhhh22} which used the fuzzy sphere as a regulator. It will be important to check whether this approach still performs well when there are long-range interactions. There is also a recently developed lattice bootstrap method from \cite{cglrsy22} which gave good results for the non-critical SRI. Applying this to the LRI is another possible future direction considering the interesting phase structure that can exist for nonlocal QFTs \cite{hjjy22}. Some of these have been studied with the functional renormalization group in \cite{Defenu:2017,Defenu:2023}.

Lastly, this work can be viewed as part of a broader effort to classify defect-localized interactions. The OPE relations hold in all such theories, not just the long-range Ising/$O(N)$ models and the BCFTs studied in \cite{Behan:2020nsf,Behan:2021tcn}. For this study of the LRI we have focused on integer-dimensional defects in a free bulk, with possible fractional co-dimension. Instead, we can also consider defects with fixed co-dimension which obey the unitarity constraints of \cite{Lauria:2020emq}. Monodromy defects provide interesting targets and some useful results about them including Ward identities have been computed in \cite{Bianchi:2021snj}. The works \cite{Soderberg:2017oaa,Giombi:2021uae,Gimenez-Grau:2021,Gimenez-Grau:2022}, although focused on interacting bulk theories, also helped clarify the kinematics of monodromy defects in general. It would be a pity not to combine these results with the techniques developed here.

\section*{Acknowledgements }

We would like to thank A. Gimenez-Grau, A. Tilloy, B. van Rees, D. Maz\'{a}\v{c}, G. Gori, J. Henriksson, K. Tiwana, L. F. Alday, M. Meineri, M. Paulos, N. Defenu, P. Liendo and S. Rychkov for discussions. Numerical calculations in this paper were done on the University of Oxford Advanced Research Computing (ARC) facility \cite{cluster}. CB is supported by the S\~{a}o Paulo Research Foundation (FAPESP) grants 2023/03825-2 and 2019/24277-8, and also received funding from the European Union (ERC, Analytic Conformal Bootstrap project, Grant Agreement no. 787185, held by L. F. Alday). EL is funded by the European Union (ERC, QFT.zip project, Grant Agreement no. 101040260, held by A. Tilloy). The work
of MN is supported by funding from the Mathematical Institute, University of Oxford. PvV is funded by the European Union (ERC, FUNBOOTS project, project number 101043588, held by M. Paulos) and acknowledges support from the DFG through the Emmy Noether research group ‘The Conformal Bootstrap Program’ project number 400570283, and through the German-Israeli Project Cooperation (DIP) grant ‘Holography and the Swampland’. Views and opinions expressed are however those of the authors only and do not necessarily reflect those of the European Union or the European Research Council Executive Agency. Neither the European Union nor the granting authority can be held responsible for them.

\appendix

\section{Details on the numerics}\label{ss:crossingeq}

\subsection{Crossing equations}

We want to bootstrap the system of four-point functions in $p$ spatial dimensions, involving two scalar operators $\psi_{1,2}$, with scaling dimensions $\frac{p \pm \mathfrak{s}}{2}$ with $\mathfrak{s} > 0$. 
From the 7 inequivalent crossing equations satisfied by $\psi_{1,2}$, we will truncate them to the 5 equations
\begin{align}
	0=&\sum_{\mathcal{O}}\hf^2_{11{\mathcal{O}}} F_{-,\Delta,l}^{11,11}(u,v)~, \nonumber \\
	0=&\sum_{\mathcal{O}}\hf^2_{22{\mathcal{O}}} F_{-,\Delta,l}^{22,22}(u,v)~,\nonumber \\				
	0=&\sum_{\mathcal{O}}\hf^2_{12{\mathcal{O}}} F_{-,\Delta,l}^{12,12}(u,v)~, \nonumber \\
	0=&\sum_{\mathcal{O}}(-1)^\ell \hf^2_{12{\mathcal{O}}} F_{\mp,\Delta,l}^{12,21}(u,v)\pm{}\hf_{11\mathcal{O}}  {}\hf_{22{\mathcal{O}}}  F_{\mp,\Delta,l}^{11,22}(u,v)~, \label{cross_eq_red}
\end{align}
where $(u,v)$ are the standard cross-ratios defined in \eqref{cross-ratios}.
We now input the exact OPE relations~\eqref{ssspinconstr} for all spins $\ell$. We take $\psi_1$ to be the operator with dimension $\frac{p - \mathfrak{s}}{2}$ (called $\hphi$ or $\sigma$ in the LRI) and $\psi_2$ the operator with dimension $\frac{p + \mathfrak{s}}{2}$ (called $\widehat{\phi^3}$ or $\hchi$ in the LRI). Since we wish to express results as functions of $a_{\phi^2}$, we must take $q = 2 - \mathfrak{s}$ and use OPE relations with $\psi_0^{(+)}$ as operator 1 and $\psi_0^{(-)}$ as operator 2.\footnote{If we wish instead to express results as functions of $a_{\chi^2}$, we must take $q = 2 + \mathfrak{s}$ and use OPE relations with $\psi_0^{(-)}$ as operator 1 and $\psi_0^{(+)}$ as operator 2.}
The crossing equations can be rewritten as follows
\begin{align}
	0=&\quad\vec{V}_{\id} + \sum_{\ell = \text{even}} \lambda^2_{12\mathcal{O}} \vec{V}_{+,\Delta,\ell} + \sum_{\ell = \text{odd}} \sum_{n = 0}^\infty \lambda^2_{12\mathcal{O}} \vec{V}_{-,p + 2n + \ell,\ell}~.
\end{align}
One can also add a third external scalar to the setup and find
\begin{align}
	0=&\quad\vec{V}_{\id}+\sum_{\ell= \text{even}}{}\begin{pmatrix}
		{}\hf_{12\mathcal{O}} &
		{}\hf_{13\mathcal{O}}
	\end{pmatrix} \vec{V}_{+,\Delta,\ell}\begin{pmatrix}
		{}\hf_{12\mathcal{O}}  \\
		{}\hf_{13\mathcal{O}} \\
	\end{pmatrix} \nonumber \\
	&+\sum_{\ell= \text{odd}}{}\hf^2_{13\mathcal{O}} \vec{V}_{0,\Delta,\ell}+\sum_{\ell = \text{odd}} \sum_{n = 0}^\infty \lambda^2_{12\mathcal{O}} \vec{V}_{-,p + 2n + \ell,\ell}~. \label{bigger-system}
\end{align}
In practice, we keep all spins in the range $0 \leq \ell \leq \ell_\text{max} = 30$ but only impose the relations for $\ell = 0$, $\ell = 2$ and the odd spins. To transform \eqref{cross_eq_red} into a finite problem, we have replaced each functional equation with 45 component equations obtained by evaluating its derivatives at the crossing symmetric point $u = v = \frac{1}{4}$. In the conventions of \cite{ElShowk:2013}, these derivatives are the ones selected by $n_\text{max} = 8$.

\subsection{Differences compared to integer co-dimension}
We use the semidefinite program solver \texttt{SDPB} \cite{SimmonsDuffin:2015,Landry:2019} which is designed for crossing equations that depend on $\Delta$ through rational functions. The algorithm in \cite{Kos:2014bka} can produce rational approximations of conformal blocks to any desired order but the OPE relations pose more of a challenge. The prefactors in \eqref{ssspinconstr} contain ratios of the form
\begin{align}
	\Gamma(\Delta + \delta_1) / \Gamma(\Delta + \delta_2)~. \label{example-ratio}
\end{align}
When $\delta_1 - \delta_2$ is an integer, \eqref{example-ratio} is manifestly rational. When $\delta_1 - \delta_2$ is a half-integer, \eqref{example-ratio} is not a rational function but its \textit{square} is to a high degree of precision \cite{Behan:2020nsf}. These are the only two cases which arise for integer co-dimension $q$. By contrast, this work considers continuously varying $q$ with each one corresponding to a different long-range Ising model between the non-trivial endpoints.

Without a rational function at our disposal, we have opted to approximate crossing equations by a large number of constant functions. As such, we do not demand that the OPE relations hold for all $\hD_0 < \Delta < \infty$ and $\hD_2 < \Delta < \infty$ when imposing the spin-0 gap $\hD_0$ and the spin-2 gap $\hD_2$. Instead, we demand that they hold for many closely spaced scaling dimensions in $(\hD_0, d + 18)$ for spin-0 operators and $(\hD_2, d + 20)$ for spin-2 operators. Most results in section \ref{sec:numerics} were obtained with 3000 points but for the high precision results of subsection \eqref{ss:betternumerics} we have sampled 12000. As the gaps are varied, the number of points stays constant rather than the grid spacing. This allows us to take advantage of the checkpoint feature of \texttt{SDPB}. We also recommend passing \texttt{-\--writeSolution=y} since files from old runs can take a non-trivial amount of time to delete without this.

\section{Perturbative computations}\label{app:FeynDiagON}

\subsection{Master integrals}\label{ss:masterint}

The massless integrals we will use can all be derived from Symanzik's formula for an integral with exponents $\alpha_i$ summing to $p$ \cite{Symanzik:1972}. This reads
\begin{align}
	\pi^{-p/2} \int d^p\tau_0 \prod_{i = 1}^n \frac{\Gamma(\alpha_i)}{|\tau_{i0}|^{2\alpha_i}} = \prod_{i < j} \int_{-i\infty}^{i\infty} \frac{d \delta_{ij}}{2\pi i} \Gamma(\delta_{ij}) |\tau_{ij}|^{-2\delta_{ij}}~, \label{symanzik}
\end{align}
where the variables on the right hand side obey the constraint $\sum_{i \neq j} \delta_{ij} = \alpha_i$. For $n = 3$ and $n = 4$, eq.~\eqref{symanzik} gives
\begin{align}
	\int \frac{d^p \tau_0}{|\tau_{01}|^{2\alpha_1} |\tau_{02}|^{2\alpha_2} |\tau_{03}|^{2\alpha_3}} &= \prod_{i = 1}^3 \frac{\Gamma(\tfrac{p}{2} - \alpha_i)}{\Gamma(\alpha_i)} \frac{\pi^{p/2}}{|\tau_{12}|^{p - 2\alpha_3} |\tau_{13}|^{p - 2\alpha_2} |\tau_{23}|^{p - 2\alpha_1}}~, \nonumber \\
	\int \frac{d^p \tau_0}{|\tau_{01}|^{2\alpha_1} |\tau_{02}|^{2\alpha_2} |\tau_{03}|^{2\alpha_3} |\tau_{04}|^{2\alpha_4}} &= \left | \frac{\tau_{24}}{\tau_{14}} \right |^{\alpha_{12}} \left | \frac{\tau_{14}}{\tau_{13}} \right |^{\alpha_{34}} \frac{\pi^{p/2}}{|\tau_{12}|^{\alpha_1 + \alpha_2} |\tau_{34}|^{\alpha_3 + \alpha_4}} \int_{-i\infty}^{i\infty} \frac{ds dt}{(2\pi i)^2} U^s V^{t - \tfrac{1}{2}(\alpha_2 + \alpha_3)} \nonumber \\
	\times \Gamma(\tfrac{\alpha_1 + \alpha_2 - 2s}{2}) \Gamma(\tfrac{\alpha_3 + \alpha_4 - 2s}{2}) &\Gamma(\tfrac{\alpha_2 + \alpha_3 - 2t}{2}) \Gamma(\tfrac{\alpha_1 + \alpha_4 - 2t}{2}) \Gamma(\tfrac{\alpha_1 + \alpha_3 - 2u}{2}) \Gamma(\tfrac{\alpha_2 + \alpha_4 - 2u}{2}) ~, \label{3and4}
\end{align}
where $s + t + u = p/2$. We will refer to the first line as the star-triangle relation. If we multiply these expressions by $|\tau_n|^{2\alpha_n}$ and take $\tau_n \to \infty$, we find related lower-point integrals where the exponents no longer sum to $p$.
\begin{align}
	\int \frac{d^p \tau_0}{|\tau_{01}|^{2\alpha_1} |\tau_{02}|^{2\alpha_2}} &= \frac{\Gamma(\alpha_1 + \alpha_2 - \frac{p}{2})}{\Gamma(p - \alpha_1 - \alpha_2)} \prod_{i = 1}^2 \frac{\Gamma(\tfrac{p}{2} - \alpha_i)}{\Gamma(\alpha_i)} \frac{\pi^{p/2}}{|\tau_{12}|^{2\alpha_1 + 2\alpha_2 - p}}~,\nonumber \\
	\int \frac{d^p \tau_0}{|\tau_{01}|^{2\alpha_1} |\tau_{02}|^{2\alpha_2} |\tau_{03}|^{2\alpha_3}} &= \pi^{p/2} |\tau_{13}|^{p - 2\alpha_1 - 2\alpha_2 - 2\alpha_3} \int_{-i\infty}^{i\infty} \frac{ds dt}{(2\pi i)^2} \left | \frac{\tau_{12}}{\tau_{13}} \right |^{2s - \alpha_1 - \alpha_2} \left | \frac{\tau_{23}}{\tau_{13}} \right |^{2t - \alpha_2 - \alpha_3} \nonumber \\
	\times \Gamma(\tfrac{\alpha_1 + \alpha_2 - 2s}{2}) &\Gamma(\tfrac{\alpha_3 + \alpha_4 - 2s}{2}) \Gamma(\tfrac{\alpha_2 + \alpha_3 - 2t}{2}) \Gamma(\tfrac{\alpha_1 + \alpha_4 - 2t}{2}) \Gamma(\tfrac{\alpha_1 + \alpha_3 - 2u}{2}) \Gamma(\tfrac{\alpha_2 + \alpha_4 - 2u}{2}) ~.\label{2and3}
\end{align}
We will refer to the first line as the chain integral. Note that the second lines of \eqref{3and4} and \eqref{2and3} can be expressed in terms of the so-called $\bar{D}$ functions \cite{Dolan:2000ut}.

When looking at bulk correlators at the end of this appendix, we will encounter ``massive'' integrals with transverse distances playing the role of mass. For these, it is helpful to use the identity
\begin{align}\label{mass-id}
	(\tau^2 + y^2)^{-\alpha} = |y|^{-2\alpha} \int_{-i\infty}^{i\infty} \frac{du}{2\pi i} \frac{\Gamma(u) \Gamma(\alpha - u)}{\Gamma(\alpha)} \left | \frac{y}{\tau} \right |^{2u}.
\end{align}

\subsection{Conventions}

We recall here our conventions. We denote bulk coordinates with $x^\mu = (\tau^a, y^i)$, where $a$ labels the parallel directions with respect to the flat $p$-dimensional defect. The bulk dimension is $d=p +q$. 

For the mean-field end description, we use
\begin{align}\label{modelON}
	S &= \mathcal{N}_{\mathfrak{s}} \mathcal{N}_{-\mathfrak{s}} \int d^p\tau_1 d^p\tau_2 \frac{\hat{\phi}(\tau_1) \cdot \hat{\phi}(\tau_2)}{|\tau_{12}|^{p + \mathfrak{s}}} + \int d^p\tau \frac{\lambda}{4} (\hphi \cdot \hphi)^2~,
\end{align}
with $\phi^I$ transforming as a vector of $O(N)$ and
\begin{equation}\label{Nconst}
	\mathcal{N}_\mathfrak{s} = \frac{2^{-\mathfrak{s}} \Gamma(\tfrac{p - \mathfrak{s}}{2})}{\pi^{\tfrac{p}{2}} \Gamma(\tfrac{\mathfrak{s}}{2})}~.
\end{equation}
The small parameter is $\eps \equiv 2\ps-p$, and the IR fixed point is
\begin{equation}\label{fixedptON}
	\lambda_* = \frac{\Gamma(\tfrac{p}{2})}{2\pi^{p/2}} \frac{\eps}{N + 8}  + O(\eps^2)\:.
\end{equation}
For the short-range end description, we use
\begin{equation}
	S = S_{\text{SRI}} + \mathcal{N}_{\mathfrak{s}} \mathcal{N}_{-\mathfrak{s}} \int d^p\tau_1 d^p\tau_2 \frac{\hat{\chi}(\tau_1) \hat{\chi}(\tau_2)}{|\tau_{12}|^{p - \mathfrak{s}}} + \int d^p\tau g \sigma \hat{\chi}~,
\end{equation}
where $\mathcal{N}_{\mathfrak{s}}$ is given in \eqref{Nconst}, the small parameter is $\delta \equiv \frac{p - \mathfrak{s}}{2} - \D^*_\sigma$ and the IR fixed point is
\begin{align}
	g_*^2 = \begin{cases}
		0.788392\delta + O(\delta^2), & p = 2 \\
		0.8155(3)\delta + O(\delta^2), & p = 3 \label{g-fp}
	\end{cases}~.
\end{align}

To realize these actions on a $p$-dimensional defect, we will take the co-dimension to be $q = 2 - \mathfrak{s}$ so that the bulk propagator is given by:
\begin{equation}\label{bulkpropON}
	\langle \phi^I(x_1)\phi^J(x_2)\rangle =  \frac{\delta^{IJ}}{(x_{12}^2)^{\D_\phi}}~,\quad \D_{\phi} = \frac{d}{2}-1\equiv \frac{p - \ps}{2}~.
\end{equation}
When $\eps \ll 1$, making the long-range $O(N)$ model a perturbation of the trivial defect, $\hphi^I$ will be the `$+$' mode of the bulk field i.e. $\hphi^I(\tau) =\phi^I(\tau, y^i = 0)$. When $\delta \ll 1$ (and $N = 1$), we will keep the co-dimension the same rather than changing the sign of $\mathfrak{s}$, which would make the starting defect trivial again. In this case, $\sigma$ becomes the `$+$' mode of $\phi$.

It is worth emphasizing that kinetic terms with $\mathcal{N}_{\mathfrak{s}} \mathcal{N}_{-\mathfrak{s}}$ have been chosen to give $\hphi$ and $\hat{\chi}$ unit-normalized two-point functions in position space.
As we will see in examples, a nice property of our convention is that operator norms stay the same at next-to-leading order for arbitrary powers of both $\hphi$ and $\hat{\chi}$. This can be seen by using the chain integral from \eqref{2and3} once and twice respectively. In the case of $\hphi$, we observe that all $O(\lambda_*)$ corrections to $\langle \widehat{\phi^m} \widehat{\phi^m}\rangle$ involve the subdiagram 
\begin{align}
	\begin{tikzpicture}[baseline,valign]
		\draw (-0.5,0) to[in=-90,out=-90] (0,0);
		\draw (-0.5, 0) to[in=90,out=90] (0, 0);
		\draw (0.5,0) to[in=90,out=90] (0, 0);
		\draw (0.5,0) to[in=-90,out=-90] (0, 0);
		\node at (0,0) [bcirc] {};
	\end{tikzpicture}
	=  \int \frac{d^p\tau_0}{|\tau_{01}|^{p - \eps} |\tau_{02}|^{p - \eps}} &= \frac{\pi^{p/2}}{|\tau_{12}|^{p - 2\eps}} \frac{\Gamma(\tfrac{\eps}{2})^2 \Gamma(\tfrac{p}{2} - \eps)}{\Gamma(\eps) \Gamma(\tfrac{p - \eps}{2})^2} = \frac{\pi^{p/2}}{|\tau_{12}|^{p - 2\eps}} \Gamma \left ( \frac{p}{2} \right )^{-1} \left [ \frac{4}{\eps} + O(\eps) \right ]~, \label{nice-int1}
\end{align}
where the black dot represents the four-scalar interaction vertex inserted at $\tau_0$, and the solid lines are propagators of the field $\hat{\phi}$.
This diagram has an $O(\eps^{-1})$ term which is compatible with the anomalous dimensions of these operators, but there is no $O(1)$ term which would lead to a correction of the normalization.
This also applies to powers of $\hat{\chi}$, for which the universal subdiagram at $O(g_*^2)$ is given by:
\begin{align}
	\begin{tikzpicture}[baseline,valign]
		\draw (-0.3,0) to (0,0);
		\draw[dashed] (0, 0) to (0.5, 0);
		\draw (0.5,0) to (0.8, 0);
		\node at (0,0) [dcirc] {};
		\node at (0.5,0) [dcirc] {};
	\end{tikzpicture}
	&= \int \frac{d^p \tau_2 d^p \tau_3}{|\tau_{12}|^{2\D_\chi} |\tau_{23}|^{2\D_\sigma^*} |\tau_{34}|^{2\D_\chi}} \nonumber \\
	&= \frac{\Gamma(\tfrac{p}{2} + \delta) \Gamma(\tfrac{p}{2} - \delta) \Gamma(\tfrac{p}{2} - \D_\sigma^*) \Gamma(\tfrac{p}{2} - \D_\sigma^* - 2\delta) \Gamma(\D_\sigma^* - \tfrac{p}{2} + \delta)^2}{|\tau_{14}|^{2\D_\chi - 2\delta} \pi^{-p} \Gamma(\delta)\Gamma(-\delta)\Gamma(\D_\sigma^*)\Gamma(\D_\sigma^* + \delta)\Gamma(p - \D_\sigma^* - \delta)^2}~. \label{nice-int2}
\end{align}
Here, the blue dots represent the $\sigma \hat{\chi}$ interaction vertex, the solid lines the propagator for $\hat{\chi}$, and the dotted line the propagator for $\sigma$. This diagram lacks not only an $O(1)$ term, but an $O(\delta^{-1})$ pole as well. This reflects the very simple behaviour of $\hat{\chi}$ correlators under $O(\delta)$ perturbations seen in \eqref{chi-corr}.

\subsection{Defect three-point functions}

In this section we compute the correlation functions
\begin{align}
	&\langle W_0 (\tau_1) W_n (\tau_2) S_n (\tau_3)\rangle~,\quad \langle W_0 (\tau_1) W_n (\tau_2) S_{n+1} (\tau_3)\rangle~,\nonumber\\
	&\langle W_1 (\tau_1) W_n (\tau_2) S_n (\tau_3)\rangle~,\quad \langle W_1 (\tau_1) W_n (\tau_2) S_{n+1} (\tau_3)\rangle~,
\end{align}
for the operators
\begin{align}
	S_n \equiv (\hphi \cdot \hphi)^n~, \quad W_n  \equiv (\hphi \cdot \hphi)^n \hphi^I~,
\end{align}
where we suppressed $O(N)$ indices for simplicity. The renormalized two-point functions of $S_n$ and $W_n$ are found to be
\begin{align}\label{resulpertSW}
	\langle S_n (\tau_1) S_n (\tau_2)\rangle &= \frac{4^{ n} n! \left({N}/{2}\right)_n {\left(1+O(\eps^2)\right)}}{(\tau^2_{12})^{2n\D_\phi + \frac{n (6 n+N-4)}{N+8}\eps}}~, \nonumber \\
	\langle W_n (\tau_1) W_n (\tau_2)\rangle &= \frac{4^{ n} n! \left({N}/{2}+1\right)_n {\left(1+O(\eps^2)\right)}}{(\tau^2_{12})^{(2n+1)\D_\phi + \frac{n (6 n+N+2)}{N+8}\eps}}~, 
\end{align}
where we removed the $1/\eps$ poles in the one-loop integrals via the the wave-function counterterms $Z_{\mathcal{O}} \equiv 1 + \delta Z_{\mathcal{O}}$ given by
\begin{align}\label{RGconsts}
	\delta Z_{S_n}=\frac{ 2 n  \pi ^{p/2} (6 n+N-4)}{  \Gamma \left(\frac{p}{2}\right)}\frac{\lambda}{\eps}+O(\lambda^2), \quad \delta Z_{W_n}=\frac{ 2 n \pi ^{p/2} (6 n+N+2)}{  \Gamma \left(\frac{p}{2}\right)}\frac{\lambda}{\eps}+O(\lambda^2)~.
\end{align}
These expressions are consistent with \eqref{sol-1loop}, as they should be. Next, we consider the three-point function with $W_0$ and $S_n$, which at tree level reads
\begin{align}
	\langle W_0 (\tau_1) W_n (\tau_2) S_n (\tau_3) \rangle^{(0)}=2^{2 n} n!  \left({N}/{2}+1\right)_n G_{12}G_{23}^{2 n} ~,
\end{align}
where we have defined $G_{ij} \equiv |\tau_{ij}|^{-2\Delta_\phi}$ for brevity.
The one-loop correction reads
\begin{align}\label{1loopcontrphinm1phinON}
	\langle W_0 (\tau_1) W_n (\tau_2) S_n (\tau_3) \rangle^{(1)} &=(\delta Z_{W_n}+ \delta Z_{S_n})\langle  W_0 (\tau_1) W_n (\tau_2) S_n (\tau_3) \rangle^{(0)} -\frac{\lambda}{4}~I_{0,n,n}~,
\end{align}
where the integral $I_{0,n,n}$ has the following form
\begin{align}
	I_{0,n,n}=\int d^p\tau_4~\left(a^{(1)}_n G_{14} G_{23}^{2n-1}G_{24}^2G_{34}+a^{(2)}_nG_{13} G_{24}^3G_{34}G_{23}^{2n-2}+a^{(3)}_n G_{12}G_{24}^2G_{34}^2G_{23}^{2n-2}\right)~,
\end{align}
for some combinatorial coefficients which we will fix shortly.
Using the master formulae given in section \ref{ss:masterint} it is not difficult to verify that
\begin{align}\label{int1}
	\int d^p&\tau_4~ G_{14} G_{24}^2G_{34} =\frac{\pi ^{p/2} G_{12}G_{23}}{\Gamma \left(\frac{p}{2}\right)}\left(\frac{2}{\eps}+ \mathcal{A}_p -2 \log \left(\frac{\tau_{13}}{\tau_{12} \tau_{23}}\right) +O(\eps)\right)~,
\end{align}
with $\mathcal{A}_p\equiv  \psi\left ( \frac{p}{2} \right ) - 2\psi\left ( \frac{p}{4} \right ) + \psi(1)$. The second addend in the integrand above is $O(\eps)$, and is therefore subleading at the IR fixed point. The last addend was computed in eq.~\eqref{nice-int1}. %
We also have that
\begin{align}
	a^{(1)}_n = 3\times 2^{2 n+3} n! n\left({N}/{2}+1\right)_n~,\quad a^{(3)}_n = 4^{n+1}  n! n(6 n+N-4) \left({N}/{2}+1\right)_n~.
\end{align}
When plugging these results into eq.~\eqref{1loopcontrphinm1phinON} and using \eqref{RGconsts}, %
we find that the $1/\eps$ poles cancel out, as they should. At the IR fixed point, the renormalized result becomes a conformal three-point correlation function with an overall coefficient given by
\begin{align}\label{res1}
	&C_{W_0 W_n S_{n}}=4^{ n} n!   \left({N}/{2}+1\right)_n \left(1-\eps\frac{3n}{N+8}  \mathcal{A}_p+O(\eps^2)\right)~.
\end{align}
Dividing this by the normalizations of the operators we find the first line of \eqref{OPEcoeffstreenON1loop}.

For the three-point function with $S_{n+1}$, at tree-level we have
\begin{align}
	\langle W_0 (\tau_1) W_n (\tau_2) S_{n+1} (\tau_3) \rangle^{(0)} =2^{2 n+1}(n+1)! \left({N}/{2}+1\right)_n G_{13}G_{23}^{2 n+1} ~.
\end{align}
The one-loop correction reads:
\begin{align}\label{1loopcontrphinm1phinp1ON}
	\langle W_0 (\tau_1) W_n (\tau_2) S_{n+1} (\tau_3) \rangle^{(1)} &=(\delta Z_{W_n}+ \delta Z_{S_{n+1}})\langle  W_0 (\tau_1) W_n (\tau_2) S_{n+1} (\tau_3) \rangle^{(0)}  -\frac{\lambda}{4}~I_{0,n,n+1}~,
\end{align}
where the integral $I_{0,n,n+1}$ has the following form
\begin{align}
	I_{0,n,n+1}=\int d^p\tau_4~\left(a^{(1)}_n G_{14} G_{23}^{2n}G_{24}G_{34}^2+a^{(2)}_n G_{13} G_{24}^2 G_{34}^2G_{23}^{2n-1}+a^{(3)}_n G_{12}G_{24}G_{34}^3G_{23}^{2n-1}\right)~.
\end{align}
The integrals are the same as what we have computed in \eqref{int1}, in particular the third addend above is again $O(\eps)$. The combinatorial coefficients are found to be:
\begin{align}
	a^{(1)}_n &= 2^{2 n+3} (n+1)! (6 n+N+2) \left({N}/{2}+1\right)_n~,\nonumber\\
	a^{(2)}_n &= 2^{2 n+3} (n+1)! n(6 n+N+2) \left({N}/{2}+1\right)_n~.
\end{align}
When plugging these results into eq.~\eqref{1loopcontrphinm1phinp1ON} and using \eqref{RGconsts}, %
we find that the $1/\eps$ poles cancel out, as they should. At the IR fixed point, the renormalized result becomes a conformal three-point correlation function with an overall coefficient given by
\begin{align}\label{res2}
	&C_{W_0 W_n S_{n+1}}=2^{2 n+1} (n+1)! \left({N}/{2}+1\right)_n \left(1-\eps\frac{ 2+6n+N}{ 2(N+8)}\mathcal{A}_p+O(\eps^2)\right)~.
\end{align}
Dividing this by the normalizations of the operators we find the second line of \eqref{OPEcoeffstreenON1loop}. 

Next, we compute the three-point function with $W_1$ and $S_n$.
At tree-level, we have:
\begin{align}
	\langle W_1 (\tau_1) W_n (\tau_2) S_n (\tau_3) \rangle^{(0)}  =3\times  2^{2 n+1} n!n \left({N}/{2}+1\right)_n G_{12}^2 G_{12}G_{23}^{2 n-1} ~,
\end{align}
while the one-loop correction reads:
\begin{align}\label{1loopcontrW1nm1phinON}
	\langle W_1 (\tau_1) W_n (\tau_2) S_n (\tau_3) \rangle^{(1)} &=(\delta Z_{W_1}+\delta Z_{W_n}+ \delta Z_{S_n})\langle  W_1 (\tau_1) W_n (\tau_2) S_n (\tau_3) \rangle^{(0)}  -\frac{\lambda}{4} ~I_{1,n,n}~.
\end{align}
The quantity $I_{1,n,n}$ is (we disregarded a tadpole diagram which vanishes in dimreg):
\begin{align}
	I_{1,n,n}=\int d^p\tau_4~&\left(a^{(1)}_n G_{12}G_{14}^2 G_{23}^{2n-1}G_{24}G_{34}+a^{(2)}_n G_{12} G_{13}G_{14}G_{24}^2G_{23}^{2n-2}G_{34}+a^{(3)}_n G_{12}G_{13}^2 G_{23}^{2n-3}G_{24}^3 G_{34}\right.~\nonumber\\
	&\left.+a^{(4)}_n G_{12}^2G_{14} G_{23}^{2n-4}G_{24}G_{34}^2+a^{(5)}_n G_{12}^2 G_{13}G_{23}^{2n-3}G_{24}^2G_{34}^2 +a^{(6)}_n G_{12}^3G_{23}^{2n-3}G_{24}G_{34}^3\right.\nonumber\\
	&\left.+a^{(7)}_n G_{13} G_{23}^{2n-1}G_{14}^2G_{24}^2+a^{(8)}_nG_{13}^2 G_{23}^{2n-2}G_{14}G_{24}^3+a^{(9)}_n G_{24}G_{23}^{2n}G_{14}^3\right),
\end{align}
for some combinatorial coefficients $a^{(k)}_n$. As it turns out, upon using the master formulae of section \ref{ss:masterint}, in the expression above the terms proportional to $a^{(3)}_n$, $a^{(6)}_n$, $a^{(8)}_n$ and $a^{(9)}_n$ are all $O(\eps)$. The relevant integrals were computed earlier, and for the combinatorial coefficients we find
\begin{align}
	&a^{(1)}_n =4^{2+n}n! n(8+N)\left(N/2+1\right)_n~,\nonumber\\
	& a^{(2)}_n =4^{n+2} n! n(18 n+N-10)\left(N/2+1\right)_n~,\nonumber\\
	& a^{(4)}_n=3\times 2^{2 n+3} n!n (6 n+N-4)\left(N/2+1\right)_n\nonumber\\
	&a^{(5)}_n =3\times 2^{2 n+3} n! (n-1) n (6 n+N-4)\left(N/2+1\right)_n~,\nonumber\\
	& a^{(7)}_n =2^{2 n+3} n!n (N+8)\left(N/2+1\right)_n~.
\end{align}
When plugging these results into eq.~\eqref{1loopcontrW1nm1phinON} and using \eqref{RGconsts}, %
the $1/\eps$ poles cancel. At the IR fixed point, the renormalized result becomes a conformal three-point correlation function with coefficient
\begin{align}\label{res3}
	&C_{W_1 W_n S_{n+1}}=\frac{3 \times 2^{2 n+1} n \Gamma (n+1) \Gamma \left(n+1+\frac{N}{2}\right)}{\Gamma \left(\frac{N}{2}+1\right)}\left(1-\eps\frac{  54 n+7 N-16 }{6 (N+8)}\mathcal{A}_p+O(\eps^2)\right)~.
\end{align}
Dividing this by the normalizations of the operators we find the third line of \eqref{OPEcoeffstreenON1loop}. 

With no extra complications, we can also compute the one-loop correction to the three-point function with $S_{n+1}$.
The final result reads:
\begin{align}\label{res4}
	C_{W_1 W_n S_{n+1}}&=\frac{1}{N}{4^{1+n} (6 n+N+2) \Gamma (n+2) \left(\frac{N}{2}\right)_{n+1}}{}\nonumber\\
	&\times\left(1-\eps\frac{ n (54 n+17 N+28)+3 (N+2) }{ (N+8)(6 n+N+2)}\mathcal{A}_p+O(\eps^2)\right)~.
\end{align}
Dividing this by the normalizations of the operators we find the fourth line of \eqref{OPEcoeffstreenON1loop}.

\subsection{Bulk two-point functions}\label{ss:bulk-2pt}

In this section, we collect some perturbative results for the bulk two-point function of $\Phi$ and its composites $(\Phi\cdot \Phi)^n$.
It will be convenient to work with the connected two-point function, defined for a generic operator $O$ as 
\begin{align}
	\langle O(x_1) O(x_2) \rangle_c = \langle O(x_1) O(x_2) \rangle - |x_{12}|^{-2\D_O}~.
\end{align}

\subsubsection{Two-point function of $\Phi$ near the short-range end}
We start from the bulk two-point function of $\Phi=\phi$ near the short-range end. At tree-level, this is the bulk-defect crossing symmetric solution with $b_0^{\phi,+} = 0$ -- see eq.~\eqref{bminusSRI} --, and the connected correlator is therefore completely specified. In order to extract $\aphisq$ we can take $x_1 \to x_2$ (meaning $\hat{r} \to 1$) to get
\begin{align}
	|y|^{2\D_\phi} \langle \phi(x) \phi(x) \rangle_c^{(0)} &= (b_0^{\phi,-})^2 {}_2F_1(p - \D_\phi, p/2; 1 - \D_\phi + p/2; 1) - {}_2F_1(\D_\phi, p/2; 1 + \D_\phi - p/2; 1) \nonumber \\
	&= \begin{cases}
		-\frac{7}{8} + \delta, & p = 2 \\
		-0.575 + 0.749 \delta, & p = 3
	\end{cases}, \label{1pt-preint}
\end{align}
matching \eqref{aexpanSRI}. 
Turning on interactions, the first correction to the result above comes from two $\sigma\hat{\chi}$ insertions and gives
\begin{align}
	\langle  \phi(x_1) \phi(x_2)  \rangle^{(2)}_c &=\frac{g^2 (b_0^{\phi,-})^2}{2} \int \frac{d^p \tau_3 d^p \tau_4}{|x_{13}|^{2\D^*_\sigma} |\tau_{34}|^{2\D_\chi} |x_{24}|^{2\D^*_\sigma}} \nonumber\\
	&= \frac{g^2 (b_0^{\phi,-})^2}{2} \frac{\pi ^p \Gamma \left(\frac{p}{2}-\D^*_\sigma\right)}{|\xper_1|^{\D^*_\sigma + \D_\chi} |\xper_2|^{\D^*_\sigma + \D_\chi}|\xpar_{12}|^{2\D^*_\sigma-2p}\Gamma \left(\D_\chi\right)^2 \Gamma \left(\D^*_\sigma\right)}  \\
	&\times \int_{-i \infty}^{+i \infty} \frac{dt du}{(2\pi i)^2} \frac{ \Gamma \left(\frac{p}{2}-t\right) \Gamma \left(\frac{p}{2}-u\right) \Gamma (\D_\chi-t) \Gamma (\D_\chi-u) \Gamma (-p+t+u+\D^*_\sigma)}{ \Gamma \left(- \D^*_\sigma+\frac{3 p}{2}-t-u\right)|\xpar_{12}|^{2 (t+u)}|\xper_1|^{-2 t}|\xper_2|^{-2 u}}~. \nonumber
\end{align}
To get to the second line we have used \eqref{mass-id}.
To compute this integral we deform the contours such that poles for positive $t,u$ and negative $t,u$ are separated, and then close to the left, picking up the poles with negative $t,u$. 
We can now take the residue at $u = p - \D_\sigma^* - t$ and use the first Barnes' lemma to conclude:
\begin{align}
	|y|^{2\D_\phi} \langle \phi(x) \phi(x) \rangle_c^{(2)} = \frac{g^2 (b_0^{\phi,-})^2}{2} \frac{\Gamma(\tfrac{p}{2} - \D^*_\sigma) \Gamma(\D^*_\sigma + 2\D_\chi - p) \Gamma(\D^*_\sigma + \D_\chi - \tfrac{p}{2})^2}{\Gamma(\tfrac{p}{2}) \Gamma(2\D^*_\sigma + 2\D_\chi - p) \Gamma(\D_\chi)^2}~.
\end{align}
It is now a simple matter to plug in the fixed point and the numerical value for $\D_\sigma^*$. Doing so and adding \eqref{1pt-preint} recovers the result \eqref{short-range-1pt} originally obtained from the OPE relations.

\subsubsection{Two-point function of $\Phi$ near the mean-field end}
We perform a similar computation near the mean-field end. Here the tree-level contribution is just the correlator in the presence of a trivial defect of co-dimension $q = 2 - \mathfrak{s}$, which has $b_0^{\phi,+} = 1$ and $b_0^{\phi,-} = 0$.
The first non-trivial correction to the connected correlator comes at $O(\lambda^2)$, and reads:
\begin{align}
	\langle  \phi^I(x_1) \phi^J(x_2)  \rangle^{(2)}_c &=2(N + 2)\delta^{IJ} \lambda^2 \int \frac{d^p \tau_3 d^p \tau_4}{|x_{13}|^{2\D_\phi} |\tau_{34}|^{6\D_\phi} |x_{24}|^{2\D_\phi}} \\
	&= 2(N + 2)\delta^{IJ} \lambda^2 \frac{\pi ^p \Gamma \left(\frac{p}{2}-3 \D_\phi\right)}{|\xper_1|^{2\D_\phi} |\xper_2|^{2\D_\phi}|\xpar_{12}|^{6\D_\phi-2p}\Gamma \left(\D_\phi\right)^2 \Gamma \left(3\D_\phi\right)} \nonumber \\
	&\times \int_{-i \infty}^{+i \infty} \frac{dt du}{(2\pi i)^2} \frac{ \Gamma \left(\frac{p}{2}-t\right) \Gamma \left(\frac{p}{2}-u\right) \Gamma (\D_\phi-t) \Gamma (\D_\phi-u) \Gamma (-p+t+u+3 \D_\phi)}{ \Gamma \left(-3 \D_\phi+\frac{3 p}{2}-t-u\right)|\xpar_{12}|^{2 (t+u)}|\xper_1|^{-2 t}|\xper_2|^{-2 u}}~. \nonumber
\end{align}
The computation is very similar to what we did before except now the defect will have three $\hphi$ propagators instead of one $\hat{\chi}$ propagator. We deform the contours again such that poles for positive $t,u$ and negative $t,u$ are separated, and then close to the left, picking up the poles with negative $t,u$.
This yields a result that can be decomposed into only the $s = 0$ blocks of eq.~\eqref{two_points_def} because there is no dependence on the transverse angle as expected. Adding back the tree-level contribution, this function is:
\begin{align}
	\langle  \phi^I(x_1) \phi^J(x_2)  \rangle = \frac{\delta^{IJ}}{(|\xper_1||\xper_2|)^{p/2}}\left((b_0^{\phi,+})^2 {\hat g}^{(+)}_{0}(\hr,\heta)+(b_0^{\phi,-})^2 {\hat g}^{(-)}_{0}(\hr,\heta)\right)+ O(\eps^3)~,
\end{align}
where
\begin{align}
	(b_0^{\phi,+})^2&=1+\frac{(N+2) \Gamma \left(-\frac{p}{4}\right) \Gamma \left(\frac{p}{2}\right)^2}{2(N+8)^2 \Gamma \left(\frac{3 p}{4}\right)}\eps ^2,\quad
	(b_0^{\phi,-})^2=\frac{(N+2) \Gamma \left(1-\frac{p}{4}\right)^2 \Gamma \left(\frac{p}{2}\right)^2}{2(N+8)^2 \Gamma(\tfrac{p}{4} + 1)^2}\eps^2~.
\end{align}
The ratio
\begin{align}
	(b_0^{\phi,-})^2/(b_0^{\phi,+})^2 = \frac{(N+2) \Gamma \left(1-\frac{p}{4}\right)^2 \Gamma \left(\frac{p}{2}\right)^2}{2 (N+8)^2 \Gamma \left(\frac{p}{4}+1\right)^2}\eps^2+ O(\eps^3)~,
\end{align}
can then be used to see that we get precisely \eqref{mean-field-1pt} to $O(\eps^2)$. Another way to extract $a_{\phi^2}$ quickly is to use the fact that it is the coefficient of $|y|^{-2\D_\phi}$ in the connected correlator when we set $y_1 = y_2 = y$ and take $\tau_{12} \to 0$. In this limit, only the residues with $t + u = p - 3\D_\phi$ contribute.

\subsubsection{Composite operators}

Finally, we will show how anomalous dimensions for defect operators can also be obtained by perturbing two-point functions in the bulk. The idea is to use an expansion very similar to \eqref{expanded-blocks} for a generic bulk operator $O$, namely:
\begin{align}
	(|y_1| |y_2|)^{\Delta_O} \langle O(x_1) O(x_2) \rangle = \sum_{\Delta_{\mathcal{O}}, s_{\mathcal{O}}} \left [ \omega^{(0)}_{\mathcal{O}} + \eps \left ( \omega^{(1)}_{\mathcal{O}} + \omega^{(0)}_{\mathcal{O}} \hgamma^{(1)}_{\mathcal{O}} \partial_\Delta \right ) + O(\eps^2)  \right ] \hat{g}_{\Delta_{\mathcal{O}}^{(0)}, s_{\mathcal{O}}}(\hat{r}, \hat{\eta})~, \label{expanded-defect-blocks}
\end{align}
where the squared bulk-defect OPE coefficient has been written as
\begin{align}
	(b^O_{\mathcal{O}})^2 \equiv \omega_{\mathcal{O}} = \omega_{\mathcal{O}}^{(0)} + \eps \omega_{\mathcal{O}}^{(1)} + O(\eps^2)~.
\end{align}
Since the bulk is free, $O$ must be a composite operator for unprotected defect operators to appear in \eqref{expanded-defect-blocks}. We will consider the important example of $\mathcal{S}_{n}(x)= (\phi \cdot \phi)^n(x) / \sqrt{4^n n! (N/2)_n}$ which is the bulk version of $\mathcal{S}_n(\tau)$ considered in section \ref{sec:perturb}. This will provide an alternative proof of the fact that a double-twist operator $[\hphi \hphi]_{n,\ell}$ only has a one-loop anomalous dimension when $n = 0$ (the need for $\ell = 0$ was more clear).

As appropriate for the mean-field end of a long-range $O(N)$ model, our starting point will be the two-point function in the presence of a trivial defect
\begin{align}
	\langle \mathcal{S}_n (x_1) \mathcal{S}_n (x_2)\rangle^{(0)} = \frac{1}{(x_{12}^2)^{2n\D_\phi}} \equiv \frac{1}{(x_{12}^2)^{\D_O}}~.
\end{align}
This has a decomposition into defect primaries of the form
\begin{align}
	\psi_{m,s} &\sim {(\partial^2_\perp)}^m \partial^{i_1}_\perp \dots \partial^{i_s}_\perp S_{n}-\text{traces}~, \\
	\Delta^{(0)}_{m,s} &= \D_O + 2m + s~, \nonumber
\end{align}
whose coefficients were found in \cite{Lemos:2017vnx}
\begin{align}\label{gfftrivialcoeffs}
	\omega^{(0)}_{m,s} &= \frac{2^s \Gamma \left(\frac{q}{2}+s\right) \Gamma \left(-\frac{d}{2}+m+\D_O +1\right) \Gamma (2 m+s+\D_O ) \Gamma \left(m+s+\D_O -\frac{p}{2}\right)}{m! s! \Gamma (\D_O ) \Gamma \left(-\frac{d}{2}+\D_O +1\right) \Gamma \left(m+\frac{q}{2}+s\right) \Gamma \left(2 m+s+\D_O -\frac{p}{2}\right)}~.
\end{align}
These coefficients are positive for $\D_O$ above the unitarity bound and integer $s$, as they should be.
As a check, when $\D_O=\D_\phi = d/2-1$, all but $\omega^{(0)}_{0,s}$ vanish, and we recover \eqref{two_points_def_freeLRI}. Note that for $\D_O\neq d/2-1$ there are infinitely many $s=0$ defect blocks that contribute.

The first correction appears at $O(\lambda)$ and is given by:
\begin{align}\label{Sigmabulktwopt}
	\langle \mathcal{S}_n (x_1) \mathcal{S}_n (x_2)\rangle &= \langle \mathcal{S}_n (x_1) \mathcal{S}_n (x_2)\rangle^{(0)} - \lambda n (6n + N - 4) \frac{1}{|x_{12}|^{4(n - 1)\D_\phi}} \int \frac{d^p \tau_3}{|x_{13}|^{4\D_\phi} |x_{23}|^{4\D_\phi}}~.
\end{align}
Note that there is no wave-function renormalization for $\mathcal{S}_n$. From the integral alone we get:
\begin{align}
	&\frac{1}{|x_{12}|^{4(n - 1)\D_\phi}}\sum_{i,j =0}^\infty \left[\frac{A_{ij}(\D_\phi)}{\left(\tau_{12}^2\right)^{2 \D_\phi +i+j}}(|\xper_1|^{2 j}  |\xper_2|^{-4 \D_\phi +2 i+p}+(1\leftrightarrow 2))+\frac{B_{ij}(\D_\phi) |\xper_2|^{2 i} |\xper_1|^{2 j}}{\left(\tau_{12}^2\right)^{ (4 \D_\phi + i+ j-p/2)}} \right]~,
\end{align}
with
\begin{align}
	A_{ij}(\D_\phi)& = \frac{\pi ^{p/2} (-1)^{i+j}  \Gamma (i+j+2 \D_\phi ) \Gamma \left(-i+2 \D_\phi -\frac{p}{2}\right) \Gamma \left(-j+\frac{p}{2}-2 \D_\phi \right)}{i! j! \Gamma (2 \D_\phi )^2 \Gamma \left(-i-j+\frac{p}{2}-2 \D_\phi \right)}~,\nonumber\\
	B_{ij}(\D_\phi)&=\frac{\pi ^{p/2} (-1)^{i+j} \Gamma \left(-i+\frac{p}{2}-2 \D_\phi \right) \Gamma \left(-j+\frac{p}{2}-2 \D_\phi \right) \Gamma \left(i+j+4 \D_\phi -\frac{p}{2}\right)}{i! j! \Gamma (2 \D_\phi )^2 \Gamma (-i-j+p-4 \D_\phi )}~.
\end{align}
Upon including all the prefactors, evaluating at the IR fixed point, and expanding in $\eps$, we get
\begin{align}\label{Sigmabulktwopt1loop}
	&\frac{\langle \mathcal{S}_n (x_1) \mathcal{S}_n (x_2)\rangle^{(1)}}{n(6n + N - 4)} = -\frac{\eps}{N + 8}{\Gamma \left(\frac{p}{2}\right)} \frac{ \log (\zeta) \, \zeta^{-\frac{p}{2}}\,\, _2F_1\left(\frac{1}{2},\frac{p}{2};1;-\frac{4}{\zeta}\right)}{|\xper|^{n p}  (2-2\heta+\zeta)^{\frac{1}{2} (n-1) p}}\nonumber\\
	&+\frac{2\eps}{N + 8} \sum_{k=0}^\infty\frac{(-1)^k \Gamma \left(2k\right)  \Gamma \left(k+\frac{p}{2}\right)}{k^2 \Gamma (k)^3 \Gamma \left(\frac{p}{2}\right)}\left[\psi\left(k + \frac{1}{2}\right) - 2\psi\left(k + 1\right) + \psi\left(k+\frac{p}{2}\right)+\log 4\right]\nonumber\\
	&\times |\xper|^{-n p} \zeta^{-k-\frac{p}{2}}\left (2-2 \heta+\zeta\right)^{-\frac{1}{2} (n-1) p} +O(\eps^2)~,
\end{align}
where we have set $|\xper_2|=|\xper_1|=\xper$ for simplicity, $\heta$ is defined in \eqref{hetadef} and $\zeta=|\tau_{12}|^2/|\xper|^2=\hchi-2=(1 - \hat{r})^2/\hat{r}$ ($\hchi$ is defined in \eqref{hetadef}).

From the coefficient of the logarithmic piece in the first line above we obtain the $O(\eps)$ correction to the scaling dimensions. This is because \eqref{expanded-defect-blocks}, with the defect blocks \eqref{two_points_def} plugged in, instructs us to look for the coefficient of $\log \hat{r}$ which is the same as the coefficient of $-\log \zeta$. As a warm-up, one can start with $n = 1$ which is especially simple. Eq.~\eqref{Sigmabulktwopt1loop} does not depend on the transverse angle for $n=1$, which means that it can be expanded into scalar defect blocks only. The anomalous dimensions of the defect operators, as well as their OPE coefficients, turn out to be
\begin{align}
	\hgamma_{m,s}^{(1)}=\frac{N+2}{N+8}\delta_{m,0}\delta_{s,0}~,\quad \omega^{(1)}_{m,s} =\hgamma_{0,0}^{(1)}\left[\psi \left ( \frac{p}{2} \right ) +\psi(1) \right]\delta_{m,0}\delta_{s,0}~,
\end{align}
which can now be applied to the long-range $O(N)$ model.\footnote{The bulk channel expansion is even simpler. It only features the bulk identity, as well as the operator $\mathcal{S}_2$ with coefficient
	\begin{align}
		a_{\mathcal{S}_2}\sqrt{2\frac{N + 2}{N}}=-\frac{\eps \hgamma_{0,0}^{(1)}}{2}\frac{ \Gamma \left(\frac{p}{2}\right)^2}{ \Gamma (p)}+O(\eps^2)~.
\end{align}}
This immediately shows why we had $\gamma^{(1)}_{n,\ell} \propto \delta_{n,0}$ for double-twist operators on the defect. Non-trivial powers of $\partial_i \partial^i$ cannot appear in operators that renormalize at one loop and these can be traded for $\partial_a \partial^a$ with Laplace's equation.

There is no reason to stop at $n = 1$. For general $\mathcal{S}_n$, this procedure will also yield non-trivial CFT data in the extension of the long-range $O(N)$ model which includes transverse spin. We find:
\begin{align}\label{anomalousdimsk}
	\hgamma_{m,s}^{(1)}&=  f_{s,m}\frac{(n-1) (6 n+N-4) \Gamma \left(\frac{1}{4} (2 n-1) p\right)\Gamma \left(\frac{n p}{2}+1\right) \Gamma \left(m+\frac{n p+s}{2}-\frac{p}{2}\right)}{2^{4 m} m!(N+8) \Gamma \left(m+\frac{1}{4} (2 n-1) p\right)\Gamma \left(\frac{n p-p}{2}+1\right) \Gamma \left(2 m+\frac{n p}{2}+s\right)}~,
\end{align}
where $f_{s,m}$ are complicated polynomials of $(n,p,s)$, which for $m\leq 3$ are found to be:
\footnotesize
\begin{align}\label{fskdef}
	f_{s,0}&=1~,\nonumber\\
	f_{s,1}&=4 n^2 p^2+n p (8 s+8-6 p)+p^2 (4 s+8-p)-12 p (s+1)~,\nonumber\\
	f_{s,2}&=32 n^4 p^4+32 n^3 p^3 (4 s+12-3 p)\nonumber\\
	&-8 n^2 p^2 \left[2 p^3-p^2 (8 s+33)+12 p (4 s+11)-16 (s^2+7s+11)\right]\nonumber\\
	&+8 n p (3 p-4 s-12) \left[p^3-4 p^2 (s+3)+6 p (2 s+5)-8 (s+2)\right]\nonumber\\
	&+p(p-4) \left[16 (p^2-4p+6) s^2-8 (p (p^2-12p+44)-60) s+(p-6)^2 (p-4)^2\right]~,\nonumber\\
	f_{s,3}&=384 n^6 p^6-576 n^5 p^5 (3 p-4 (s+5))\nonumber\\
	&-96 n^4 p^4 \left(3 p^3-3 p^2 (4 s+25)+36 p (3 s+14)-16 (3 s (s+11)+85)\right)\nonumber\\
	&+48 n^3 p^3 (3 p-4 (s+5)) \left(6 p^3-3 p^2 (8 s+35)+24 p (4 s+17)-16 (s (s+16)+45)\right)\nonumber\\
	&+12 n^2 p^2 \left[3 p^6-6 p^5 (4 s+25)+12 p^4 \left(4 s^2+70 s+231\right)-24 p^3 (s (44 s+423)+994)\right.\nonumber\\
	&+\left.96 p^2 (s (s (4 s+89)+553)+1035)-1152 p (s (s (s+18)+97)+162)+512 (3 s (s (s+14)+63)+274)\right]\nonumber\\
	&-6 n p (3 p-4 (s+5)) \left[3 p^6-24 p^5 (s+4)+12 p^4 (2 s+9) (2 s+11)\right.\nonumber\\
	&-\left.48 p^3 (2 s+9) (4 s+17)+96 p^2 (s (15 s+122)+245)-768 p (s (3 s+23)+43)+1024 (s+3) (s+4)\right]\nonumber\\
	&-(p-4) p \left[p^7-4 p^6 (3 s+11)+4 p^5 (3 s (4 s+33)+220)-16 p^4 (2 s (s (2 s+33)+181)+641)\right.\nonumber\\
	&+\left.16 p^3 \left(s \left(44 s^2+630 s+2941\right)+4483\right)-128 p^2 (s (s (31 s+390)+1625)+2248)\right.\nonumber\\
	&+\left.128 p (s+4) (s+5) (71 s+237)-9216 (s+3) (s+4) (s+5)\right]~.
\end{align}
\normalsize

From the last two lines of \eqref{Sigmabulktwopt1loop} we obtain $O(\eps)$ corrections to the (squared) tree-level bulk-defect OPE coefficients.
For the first few operators, these are
\begin{align}\label{dOPEcoeffsk}
	\omega^{(1)}_{0,s} &=\frac{2^s n (6 n+N-4) \left(\frac{1}{2} (n-1) p\right)_s}{(N+8) s!}[\psi(\tfrac{p}{2})+\psi(1)] ~,\nonumber\\
	\omega^{(1)}_{1,s} &=\frac{2^{s-1} n (6 n+N-4) \left(\frac{1}{2} (n-1) p\right)_{s+1}[\psi(\tfrac{p}{2})+\psi(1)]}{(N+8) s! (4 s+4 - p) (n p-p+2 s+2)}\left[f_{s,1} +\frac{g_{s,1}}{ (np-p+2s+2)[\psi(\tfrac{p}{2})+\psi(1)]}\right]~,\nonumber\\
	\omega^{(1)}_{2,s} &=\frac{2^{s-4} n(6 n+N-4) \left(\frac{1}{2} (n-1) p\right)_{s+2}[\psi(\tfrac{p}{2})+\psi(1)]}{(N+8) s! (p-4 (s+1)) (p-4 (s+2)) ((n-1) p+2 (s+2)) ((n-1) p+2 (s+3))}\nonumber\\
	&\times\left[f_{s,2} +\frac{g_{s,2}}{2 (n p-p+2 s+4) (n p-p+2 s+6)[\psi(\tfrac{p}{2})+\psi(1)]}\right]~,
\end{align}
with $f_{s,m}$ given in \eqref{fskdef} and
\begin{align}
	g_{s,1}&=(p-2) p (p-4 s-4) (n p-p+2 s+4)~,\nonumber\\
	g_{s,2}&=(p-2) p (p-4 s-8) (n p-p+2 s+8)\left[3 (n-1) p^4\right.\nonumber\\
	&\left.+2 p^3 (-n (8 n (2 n-5)+6 s+41)+9 s+24)\right.\nonumber\\
	&\left.-4 p^2 \left(16 n^2 (2 s+5)-2 n (27 s+64)+6 s^2+45 s+69\right)\right.\nonumber\\
	&\left.-16 p \left(n \left(8 s^2+42 s+58\right)-s (7 s+33)-42\right)-64 (s+3)^2\right]~.
\end{align}

\bibliography{bib}
\bibliographystyle{JHEP}

\end{document}